\documentclass[12pt]{article}
\usepackage{authblk}
\usepackage[utf8]{inputenc}
\usepackage[american]{babel}
\usepackage{csquotes}
\usepackage[section]{placeins}
\usepackage{appendix}
\usepackage{amsmath,amsfonts,amssymb}
\usepackage{caption}
\usepackage{subfigure}
\usepackage{booktabs}
\usepackage{multirow}
\usepackage{tabularx}
\usepackage{comment}
\usepackage{bm}
\usepackage{placeins}
\usepackage{graphicx}
\usepackage{verbatim}

\usepackage[hyphens]{url}
\usepackage[breaklinks]{hyperref}

\usepackage[rmargin=20pt,lmargin=20pt]{geometry}
\usepackage{xcolor}

\renewcommand{\l}{\left}
\renewcommand{\r}{\right}

\newcommand{\ppos }{\frac{p_{test} \sigma \lambda E}{p_{test} \sigma \lambda E+D}}

\title{Data-driven Optimized Control of the COVID-19 Epidemics}

\author[1,2]{Afroza Shirin}
\author[3]{Yen Ting Lin}
\author[1]{Francesco Sorrentino}
\affil[1]{Department of Mechanical Engineering, University of New Mexico, New Mexico, 87131}
\affil[2]{Electrical and Computer Engineering, University of New Mexico, New Mexico, 87131}
\affil[3]{Information Sciences Group, Computer, Computational and Statistical Sciences Division (CCS-3), Los Alamos National Laboratory, Los Alamos, New Mexico, 87544 USA} 

\begin{document}
	
	\maketitle

	\begin{abstract}
		Optimizing the impact on the economy of control strategies aiming at containing the spread of COVID-19 is a critical challenge. We use  daily new case counts of COVID-19 patients reported by local health administrations from different Metropolitan Statistical Areas (MSAs) within the US  to parametrize a  model that well describes the propagation of the disease in each area. We then introduce a time-varying control input that represents the level of social distancing imposed on the population of a given area and solve an optimal control problem with the goal of minimizing the impact  of social distancing on the economy
		in the presence of relevant constraints, such as a desired level of suppression for the epidemics at a terminal time. We find that with the exception of the initial time and of the final time, the optimal control input is well approximated by a constant,
		specific to each area, which contrasts with the  implemented system of reopening `in phases'.  For all the areas considered, this optimal level corresponds to stricter social distancing than the  level estimated from data. Proper selection of the time period for application of the control action optimally is important: depending on the particular MSA this period should be either short or long or intermediate. We also consider the case that the transmissibility increases in time (due e.g. to increasingly colder weather), for which we find that the optimal control solution yields progressively stricter measures of social distancing. {We finally compute the optimal control solution for a model modified to incorporate the effects of vaccinations on the population and we see that depending on a number of factors,  social distancing measures could be optimally reduced during the period over which vaccines are administered to the population.}
	\end{abstract}
	
	\section*{Introduction}
	
	The fast propagation of the COVID-19 pandemic has attracted unprecedented attention from both the public and the scientific community. {This has resulted in much research and funding getting redirected towards COVID-19 and stimulated  strong research collaboration between countries \cite{radanliev2020data}.} Due to the
	high fatality rate of SARS-CoV-2 \cite{richardson2020presenting}, governments throughout the world have adopted measures such as lock-down, stay-at-home, and shelter-in-place, which in turn have led to
	substantial economic losses, see e.g., \cite{GDP2ndQdrop}. In many countries control interventions have been articulated
	in phases, usually phase 1 to phase 3, with higher phases corresponding to progressively lower restrictions \cite{gottlieb2020national}. A fundamental challenge is to balance the need to suppress the spread of COVID-19 and the need to contain the economic impact of measures aiming at limiting the spread of the disease.  In this manuscript, we apply optimal control theory on a mathematical model for the propagation of the epidemics, parametrized by real-world data describing different regions, and compute control strategies which are optimal for each region from an economic standpoint.
	
	A number of papers and reports have focused on both modeling and controlling the pandemic. Flaxman et al.~\cite{flaxman2020report} looked at the effect of non-pharmaceutical interventions including school closures, banning of mass gathering, social distancing, etc. on the reproductive number $R_t$ of COVID-19. Sanche et al.~\cite{sanche2020novel} used a mathematical model with data on individual cases, real-time human travel, and infections, as well as estimated epidemiology parameters to compute $R_0$ and found that it is higher than initially estimated. 
	Chang et al.~\cite{chang2020modelling} adopted an agent-based model to determine the efficacy of several intervention strategies on the spread of COVID-19 in Australia. Anderson et al.~\cite{anderson2020estimating} performed a Bayesian analysis to estimate the impact of social distancing on number of reported cases and hospitalizations in British Columbia and found that when 78\% participation in social distancing has been accomplished the cases would decrease; it also noted that if the participation were below 45\%, an exponential growth would restart.   Morris et al.~\cite{morris2020optimal} explored COVID-19 intervention methods which are robust to implementation errors and found that these methods in conjunction with optimal time-limited methods derived from the standard SIR model can be used to mitigate the spread of the virus. Another study analyzed an open-loop optimal control policy updated weekly using real-world feedback, and found that this method is effective in reducing fatalities even if some measurements are inaccurate \cite{kohler2020robust}. A study published in March 2020 estimated the ICU occupancy and ventilator use from a statistical model under the conditions of social distancing and found that the demand for hospital beds and ventilators will exceed the supply \cite{covid2020forecasting}. 
	Another study explored optimal policies for decreasing economic cost and mortality rates from a multi-risk SIR model and found that strict lock-down policies which specifically target the elderly population were most effective in minimizing deaths and economic losses \cite{acemoglu2020multi}. Previous work has not computed optimal controls for data driven models.
	On the one hand, Refs.\ \cite{chang2020modelling}, \cite{morris2020optimal}, \cite{kohler2020robust}, \cite{acemoglu2020multi} explicitly compute optimal control strategies, but their models were stylized and not parametrized/calibrated by data; On the other hand, Refs.\ \cite{flaxman2020report} \cite{sanche2020novel}, \cite{anderson2020estimating}, \cite{covid2020forecasting} used data to parametrize the models, but the models do not have controllers which can be used to infer optimal control strategies.
	
	In what follows, we first construct a mathematical model, which is parametrized by historical and regional daily new case report. After parametrization, the data-driven model is capable to reproduce the regional progression of the COVID-19 epidemics up to the present. Then, we apply optimal control theory to the parametrized model to compute an optimal control strategy over the course of a pre-determined period into the future to suppress the epidemics to a desired level, while minimizing economic costs. This type of approach is suitable for long-term planning (i.e., over the course of several months) as opposed to short-term planning, which can be difficult to implement by the government and by businesses.
	
	{In most countries, distribution of the vaccine to the population is under way and will continue for most of 2021, which points out the need for planning interventions to contain the epidemics for several months while only a limited part of the population has received a vaccine.} Thus our proposed workflow aims to bridge the gap until the time $T_{vac}$ when an effective vaccine is massively manufactured, and administered to the majority of the population. 
	Another temporal consideration regards the time $T_{herd}$ at which a population achieves herd immunity in the absence of a vaccine and without overflowing the medical facilities. While herd immunity from COVID-19 is the subject of much ongoing discussion \cite{randolph2020herd}, in the Methods we provide a rough estimate of $T_{herd}$ from available data. In this paper we proceed under the assumption that the  inference period $T_{inf}$ and the control horizon $T_{cont}$ are such that $T_{inf}+T_{cont}<T_{vac}$ and $T_{inf}+T_{cont}<T_{herd}$.
	
	We set out the analysis by first introducing a compartmental model which describes key features of the COVID-19 epidemics. The whole population is divided into the following compartments. 
	The susceptible population compartment (S) includes the people who can contract the pathogen SARS-CoV-2. The exposed population (E) are those who have been infected but have not progressed long enough into the disease to transmit it to susceptible people. Those who can transmit the disease (`carriers') are divided into the asymptomatic group (A) who do not show symptoms and the infected symptomatic groups (I) who show symptoms. Both the symptomatic and asymptomatic groups can transmit the disease, but with different infectiousness---the asymptomatic people are less infectious. The infected symptomatic population $I$ is divided into three sub-compartments. The first sub-compartment $I_{sq}$ includes those who just self quarantine and do not get tested. The second sub-compartment $I_{tp}$ includes those who get tested, result positive, and get quarantined. In contrast to the previous two sub-compartments, the third sub-compartment $I_s$ includes the rest of the symptomatic population.  Those who were tested positive and those who decided to self-quarantine are moved into a quarantined compartment (Q), and stop interacting with other populations. The removed compartment (R) includes those who are completely recovered from the disease and have acquired immunity, and those who have died because of the disease. Both groups are not susceptible to reinfection and are removed from the system.

\begin{figure}[ht]
		\centering
		\includegraphics[width = 160mm]{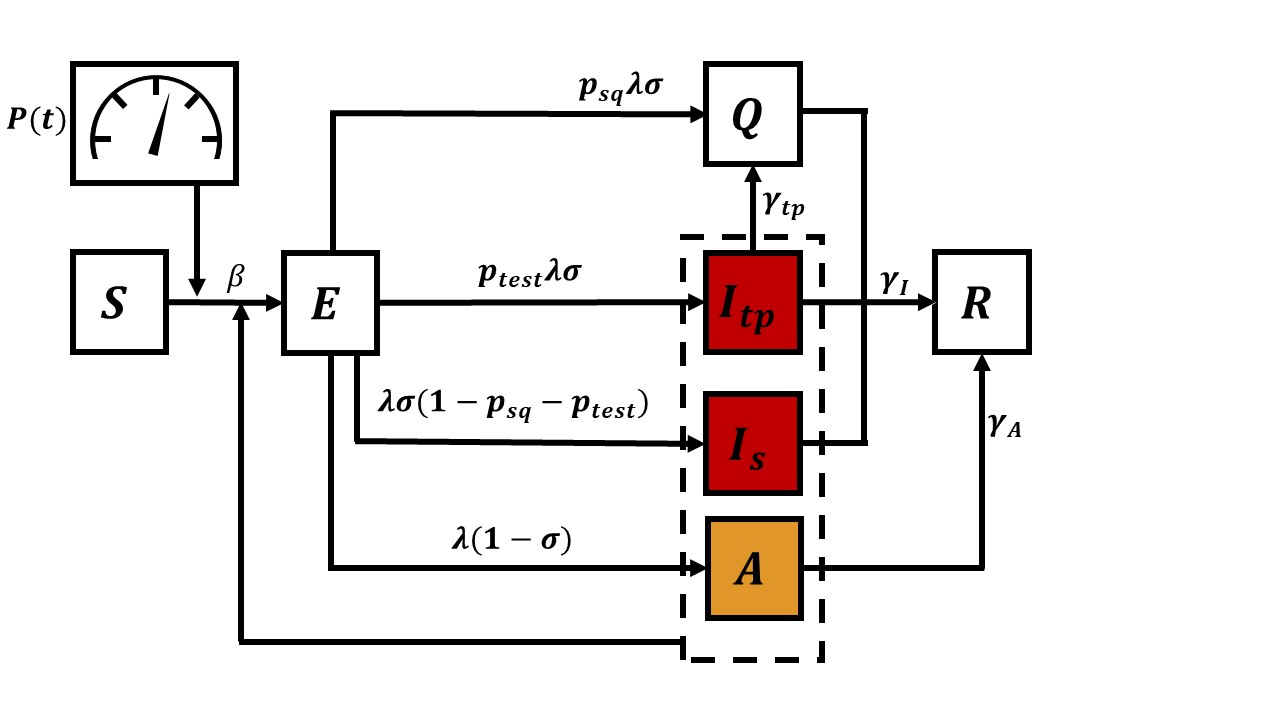}
		\caption{\textbf{Compartmental model corresponding to Eq.\ 2.} The  transition from the $S$ (susceptible) compartment to the $E$ (exposed) compartment is affected by the population densities $S(t)$, $I_{tp}(t)$, $I_s(t)$, $A(t)$ and by the time-varying control input $P(t)$. The dashed lines enclose the infectious populations.}
		\label{fig:schematicDiagram}
	\end{figure}

	Mathematically, the time evolution of the population density---defined as the compartmental population normalized by the total regional population---of each compartment is governed by the following set of coupled ordinary differential equations,
	\begin{subequations} \label{model-fullOrder}
		\begin{align} 
			\dot{S}\l(t\r)= {}&-\beta P^2\l(t\r) S(t)\l[I_s(t)+I_{sq}(t)+I_{tp}(t) +\mu A(t)\r] \\
			\dot{E}\l(t\r)= {}&\beta P^2\l(t\r) S(t) \l[I_s(t)+I_{sq}(t)+I_{tp}(t) +\mu A(t)\r]-\lambda E(t)\\
			\dot{A}\l(t\r)= {}&\lambda \l(1-\sigma\r) E(t)- \gamma_A  A(t) \\
			\dot{I}_{sq}\l(t\r)= {}& p_{sq} \lambda \sigma E(t)  -\l[\gamma_I + \gamma_{sq} \r]I_{sq}(t) \\
			\dot{I}_{tp}\l(t\r)= {}& p_{test} \lambda \sigma E(t)  -\l[\gamma_I + \gamma_{tp} \r]I_{tp}(t) \\
			\dot{I_s}\l(t\r)= {}& \left(1-p_{sq} - p_{test} \right) \lambda \sigma E(t)- \gamma_I I_s(t) \\
			\dot{Q}\l(t\r)={}&   \gamma_{tp} I_{tp}(t) +\gamma_{sq} I_{sq}(t) + p_{sq} \lambda \sigma E(t)  -\gamma_I Q(t) \\
			\dot{R}\l(t\r)= {}&  \gamma_A A(t) + \gamma_I \l[I_s(t)+I_{tp}(t)+Q(t)\r] ,
		\end{align}
	\end{subequations}
	where $\beta$ is the transmissibility, $\lambda$ is the transition rate from the exposed compartment to either the asymptomatic or symptomatic compartments, $\gamma_I$ is the transition rate from the infected compartment to the recovered compartment, $\gamma_A$ is the transition rate from the asymptomatic compartment to the recovered compartment, $\mu$ is the relative infectiousness of asymptomatic individuals (compared to symptomatic individuals), $\sigma$ is the fraction of exposed people who transition into the symptomatic compartments, 
	$p_{sq}$ is the fraction of symptomatic people who will self-quarantine, $\gamma_{sq}$ is the transition rate from from the $I_{sq}$ compartment into the quarantined compartment, $\gamma_{tp}$ is the transition rate from the $I_{tp}$ compartment into the quarantined compartment,  and $p_{test}$ is the fraction of infected who get tested (but only a fraction of them will be confirmed to be positive, $I_{tp}$). The model assumes that testing resources are not scarce, i.e., availability of a testing kit to every person in $I_{tp}$; we also assume positive people are identified as such with 100\% accuracy. The case with limited testing resources is discussed in the SI.
	Realistically, the time scale associated with $\gamma_{tp}$ can be several days, so we assume $\gamma_{tp}=0.5$ (2 days). We model {social distancing} by the control variable $0 \leq P(t) \leq 1$, which measures the reduction of contact probabilities between the susceptible and the infectious populations (which include both A and I). The model is structurally similar to the models analyzed in Refs.~\cite{anderson2020estimating,lin2020Daily}, but simplified to allow for efficient calculations of  optimal control solutions. Figure \ref{fig:schematicDiagram} illustrates a schematic diagram of the model.
	
	In order to reduce the dimensionality of the dynamical system, we introduce the following simplification:  we treat $\gamma_{sq}$ as a very large number; for $\gamma_{sq}\rightarrow\infty$, we assume that those becoming symptomatic immediately transition into the $Q$ compartment, yielding the following reduced-order model: 
	\begin{subequations} \label{model}
		\begin{align} 
			\dot{S}\l(t\r)= &-\beta P^2\l(t\r) S(t)\l[I_s(t)+I_{tp}(t) +\mu A(t)\r]\\
			\dot{E}\l(t\r)= &\beta P^2\l(t\r) S(t) \l[I_s(t)+I_{sq}(t)+I_{tp}(t) +\mu A(t)\r]-\lambda E(t)\\
			\dot{A}\l(t\r)= &\lambda \l(1-\sigma\r) E(t)- \gamma_A  A(t) \\
			\dot{I}_{tp}\l(t\r)= & p_{test} \lambda \sigma E(t)  -\l[\gamma_I + \gamma_{tp} \r]I_{tp}(t) \\
			\dot{I}_s\l(t\r)= & \left(1-p_{sq} - p_{test} \right) \lambda \sigma E(t)- \gamma_I I_s(t) \\
			\dot{Q}\l(t\r)=&   \gamma_{tp} I_{tp}(t) + p_{sq} \lambda \sigma E(t)  -\gamma_I Q(t) \\
			\dot{R}\l(t\r)= &  \gamma_A A(t) + \gamma_I \l[I_s(t)+I_{tp}(t)+Q(t)\r].
		\end{align}
	\end{subequations}
	
	The schematic diagram which fits the above model in Eq.\ \eqref{model} is shown in the Supplementary Note 1  \ref{fig:schematicDiagram}.
	
	\noindent To bridge the model and the data, we also integrate an auxiliary variable $C_I$ which counts the cumulative confirmed cases and evolves according to
	\begin{align}
		\dot{C}_{tp} \l(t\r) =  \gamma_{tp} I_{tp}(t),
	\end{align}
	We will fit $\Delta C_{tp}(t) := C_{tp}(t+1)-C_{tp}(t)$ to the new case counts reported on each day, detailed in the section on parametrization below.

	From an economic point of view, the measures of social distancing $P(t)$ and quarantining $Q(t)$  have very elevated costs for society. For example, the US real gross domestic product (GDP) dropped by roughly one third from year to year in the second quarter of 2020 {\cite{GDP2ndQdrop}}, due mainly to the COVID-19 pandemic, see e.g., \cite{nicola2020socio}. 
	
	We thus formulate the following optimal control problem,
	\begin{equation}\label{eq:cost}
		\min_{P} J=  c_p  \int_{t_i}^{t_f} \frac{1- P(t)}{ P(t)} dt + c_q  \int_{t_i}^{t_f} \frac{Q(t)}{ 1-Q(t)} dt, 
	\end{equation}
	$c_p,c_q \geq 0$, subject to Eq.\ \eqref{model}, the initial conditions, the terminal constraint 
	\begin{equation}\label{eq:const1}
		E(t_f)+I_s(t_f)+I_{tp}(t_f)+A(t_f)<\epsilon,
	\end{equation}
	and the following path constraint
	\begin{equation}\label{eq:const2}
		I_s(t)+I_{tp}(t)\leq I_{\max},
	\end{equation}
	where $I_{\max}$ is an upper limit on the number of infected people that can receive proper treatment in the hospitals. $t_i$ is the time at which we perform the inference and start optimizing the control action, $t_f$ is the final time of the optimization, the previously defined control horizon $T_{cont}=(t_f-t_i).$
	
	The objective function \eqref{eq:cost} takes into account an economic cost for social distancing with an appropriate coefficient $c_p>0$ and an economic cost for quarantining with an appropriate coefficient $c_q>0$.  
	We set the cost associated with social distancing per unit time to be modeled by $\left(1-P(t)\right)/P(t)$. By choosing this functional form, the cost is linearly dependent on the scale of the reduction ($1-P(t)$) when $1-P(t)\ll 1$, and nonlinearly diverges near total shut-down ($P(t)\ll 1$).  The economic cost increases with $1-P(t)$ to indicate that stricter measures of social distancing affect more and more `essential' workers, and so increasingly larger parts of the economy. For example, the cost of limiting large gatherings of people like concerts or sport events is lower than the cost of limiting customers' access to stores and restaurants. Similarly, imposing quarantine requires resources that are linearly proportional to the quarantined population when $Q$ is small, and diverge nonlinearly when the quarantined population is close to the entire population $Q=1$. We model such a cost by the functional form $Q(t)/\left(1-Q(t)\right)$.
	
	Both social distancing and quarantining have a cost associated with the lack of economic return generated by limiting person-to-person interactions. It is reasonable to assume $c_q \geq c_p$, as strict quarantining requires supervision costs as well as costs due to lowered productivity \cite{gupta2005economic} while social distancing only incurs costs due to lowered productivity \cite{koren2020business}.
	
	We will also consider the alternative objective function,
	\begin{equation}\label{eq:cost_alt}
		\min_{P} J^{\text{alt}}=  c_p  \int_{t_i}^{t_f} (1- P(t)) dt + c_q  \int_{t_i}^{t_f} {Q(t)} dt, 
	\end{equation}
	for which the integrand functions are linear in $P(t)$ and $Q(t)$. While the formulation \eqref{eq:cost_alt} is mathematically simpler, it does not take into account the fact that stricter measures of social distancing may result in progressively larger economic losses. The alternative objective function
	\eqref{eq:cost_alt} is here mainly introduced in order to compare the results to those obtained with \eqref{eq:cost}.
	
	The tunable parameter $\epsilon$ represents the desired suppression level for the epidemics at the final time $t_f$.  In general,  selected values for $\epsilon$ may depend on a number of factors, such as the time horizon over which optimization is performed and the particular stage of the epidemics (initial exponential growth, intermediate growth, or plateau.) A possibility is to require complete eradication of the epidemics, which corresponds to setting $\epsilon=1/N$, where $N$ is the number of individuals in the population. However, the high basic reproductive number of COVID-19 makes eradication unlikely; instead we set $\epsilon$ to a small number indicating suppression of the disease. 
	The other constraint given by $I_{\max}$ represents the need to contain the infected population below a given threshold at all times (or 'flatten the curve'.) {{
			In what follows, unless differently stated, we set the terminal constraint $\epsilon=10^{-5}$, which corresponds to imposing that the number of infected people is below one person per  $100000$. This number is derived from the guidelines of European countries about re-opening, see for example \cite{GermanOpening}, indicating that reopening occurred at about $2 \times 10^{-5}$. Also, European countries have official guidelines for reimposing stricter lock-down measures, see: \cite{franc24GermanyLockdown}, which sets a critical population equal to $50/100000=5 \times 10^{-4}$. The values of $I_{max}$ are set regionally based on the capacity of the ICU beds in different metropolitan areas and are summarized in Table 3. }}




	\newpage
	
	\section*{Parametrization of the model for different US metropolitan areas.}
	
	We partitioned the model parameters into two sets: a set of fixed parameters and a set of inferred parameters which are estimated by the daily case counts reported by regional health administration and registered in the repository curated by the New York Times \cite{NYTgithub}. 
	
	The fixed parameter sets includes $S_0$, $\lambda$, $\gamma_I$, $\gamma_A$, $\mu$, $\sigma$, $p_{test}$, and $p_{sq}$. $S_0$ is the regional total population, and we used the US Census Bureau-estimated regional population of each of the Metropolitan Statistical Areas (MSAs) or  `cities', which are delineated by the US Office of Management and Budget \cite{MSAdelineation}. Lauer et al.~\cite{lauer2020incubation} estimated the median of the incubation period to be about 5.2 days, however, there is evidence that patients become infectious roughly two days before the onset of symptoms \cite{nishiura2020serial}, which corresponds to approximately a three-day progression into \emph{contagiousness}. We thus set a rough estimation for $\lambda$ to be $1/3$ (d). In a more complex model with multiple stages of the disease progression \cite{lin2020Daily}, one could account for the fact that a patient can be both pre-symptomatic and infectious. 
	The coefficients $\gamma_I$ and $\gamma_A$, which are the transition rate to recovery of the symptomatic and asymptomatic populations, are estimated to be 0.12 (1/d) \cite{wolfel2020virological} and 0.26 (1/d) \cite{sakurai2020natural}. The relative infectiousness $\mu$ is set at $0.9$ \cite{nguyen2020natural}, and the fraction of the symptomatic population is set at 0.64 \cite{dimondPrincess,sakurai2020natural}. The parametrization of $\gamma_I$, $\gamma_A$, $\mu$, and $\sigma$ are consistent with a more complex model which was used to perform daily forecasts of the disease spread \cite{lin2020Daily}. We assume $p_{test}=0.25$ and $p_{sq}=0.4$, noting that these parameters were able to reproduce the infected population at the time of the inference (we estimated that about 15 to 20\% of the total population infected in the New York City MSA on 07-Jul-2020 when we parametrize the model.) 
	
	We used the data from 21-Jan-2020 to 07-Jul-2020 to infer the inferred parameters, which corresponds to the previously defined inference period $T_{inf}$.  We assume that the social distance function $P(t)$ before 07-Jul-2020 is piecewise-linear to avoid over-fitting (due to observation noise.) At the time when the analysis was performed, multiple MSAs had shown clear second-phase resurrection of the epidemics \cite{lin2020Daily,LANLgithub}. We found that a two-phase piecewise linear function is sufficient to reproduce the data of each of the MSAs:
	\begin{equation}
		P(t)= \left\{
		\begin{array}{ll}
			0, &  t\le t_0,  \nonumber\\
			1, & t_0 < t \le t_1 \\
			1+\frac{p_1-1}{t_2-t_1}\left(t-t_1\right), & t_1 < t \le t_2 \\
			p_1, & t_2 < t \le t_3 \\
			p_1+\frac{p_2-p_1}{t_4-t_3}\left(t-t_3\right), & t_3 < t \le t_4 \\
			p_2 & t_4 <t<t_i,\\
		\end{array}
		\right.
	\end{equation}
	where $t_0$ is the time when the disease was introduced into a specific MSA, the monotonic $t_1$, $t_2$, $t_3$, and $t_4$ ($t_j \le t_{j+1}$) are the times when the social-distancing behavior changes, and $p_1$ and $p_2$ are the two social-distancing strengths. {The two-phase model was selected by a model-selection procedure \cite{lin2020Daily} and is deemed the most evident model structure (v.~one- and three-phase).}  We define $\Delta t_j := t_j - t_{j-1}$ for $j=1, \ldots 4$. The time at which we perform the inference is $t_4$, which is also the time after which optimization of the control action begins. The two-phase piecewise linear model is the minimal model that we found capable to reproduce the data, and can be validated by the more rigorous model-selection procedure detailed in \cite{lin2020Daily}. 
	
	To fit the model by the noisy daily report new counts, we adopt a negative-binomial noise model. We brief the procedure here, noting that the procedure is similar to the inference method detailed in \cite{lin2020Daily}. Given a set of parameters $\theta$ (a stylized notation for the set of the inferred model parameters), the  model Eq.~\eqref{model} predicts a deterministic trajectory of the \emph{new} positive tested case on day $j$, $\Delta I_{tp}(j;\theta)$. This deterministic prediction is interpreted as the mean of a stochastic outcome, modeled by a negative binomial distribution $\text{NB}(r,p_j)$ where $r$ and $p_j$ are the parameter of the distribution, and $p_j$ is constrained by the deterministic model prediction
	\begin{equation}
		p_j = \frac{r}{r+ \Delta C_{tp}(j;\theta)}. 
	\end{equation}
	Here, $r$ is the \emph{dispersion} parameter which describes how disperse the noise distribution is; in the limit $r\rightarrow \infty$, the negative binomial statistics converges to Poissonian, and  in the limit $r\rightarrow 0$ the distribution looks closer to an exponential. The negative binomial noise model is phenomenological: it has the capability to capture a wide variety of single-modal distributions. With the negative binomial noise model, the likelihood function given a set of $N$ daily reported new case counts $\left\{\Delta C_{tp}(j)\right\}_{j=0}^{N-1}$ can be formulated \cite{lin2020Daily}:
	\begin{equation}
		\mathcal{L}\l(\theta; \left\{\Delta C_{tp}(j)\right\}_{j=0}^{N-1}\r) \equiv \prod_{j=0}^{N-1} \left(\begin{array}{c}\Delta C_{tp}(j) +r -1 \\ \Delta C_{tp}(j) -1 \end{array} \right) p_j^r  \left(1-p_j \right)^{\Delta C_{tp}(j) }. \label{eq:likelihood}
	\end{equation}
	
	In summary, the inferred parameters $\theta$ include the regional-specific disease transmissibility $\beta$, onset of the disease spread $t_0$, behavioral switching times $t_1$, $t_2$, $t_3$, $t_4$, the strengths of two social-distancing measures $p_1$, $p_2$, and a dispersion parameter $r$ of the negative binomial noise model. These parameters were inferred by the daily case reports from 21-Jan-2020 to 07-Jul-2020. With the formulated likelihood function \eqref{eq:likelihood}, we used the standard Markov chain Monte Carlo procedure (MCMC) with an adaptive sampler (\cite{andrieu2008tutorial}, detailed in \cite{lin2020Daily}) to estimate the maximum likelihood estimator of the parameters $\theta$ for each of the interested MSA's.  Figure 2 shows excellent agreement between the daily new case counts reported by local health administrations (plus signs) and the daily new cases obtained by integrating Eq.\ (2) after parametrization of the model (solid line.) 
	{Table 1 is a list of data-driven model parameters, with indication of whether they are free or fixed and of whether they are region-dependent. Table 2 summarizes the set of model parameters that were estimated for each MSA.}
	
	\begin{figure}[ht]
		\centering
		\includegraphics[width = 0.7\textwidth]{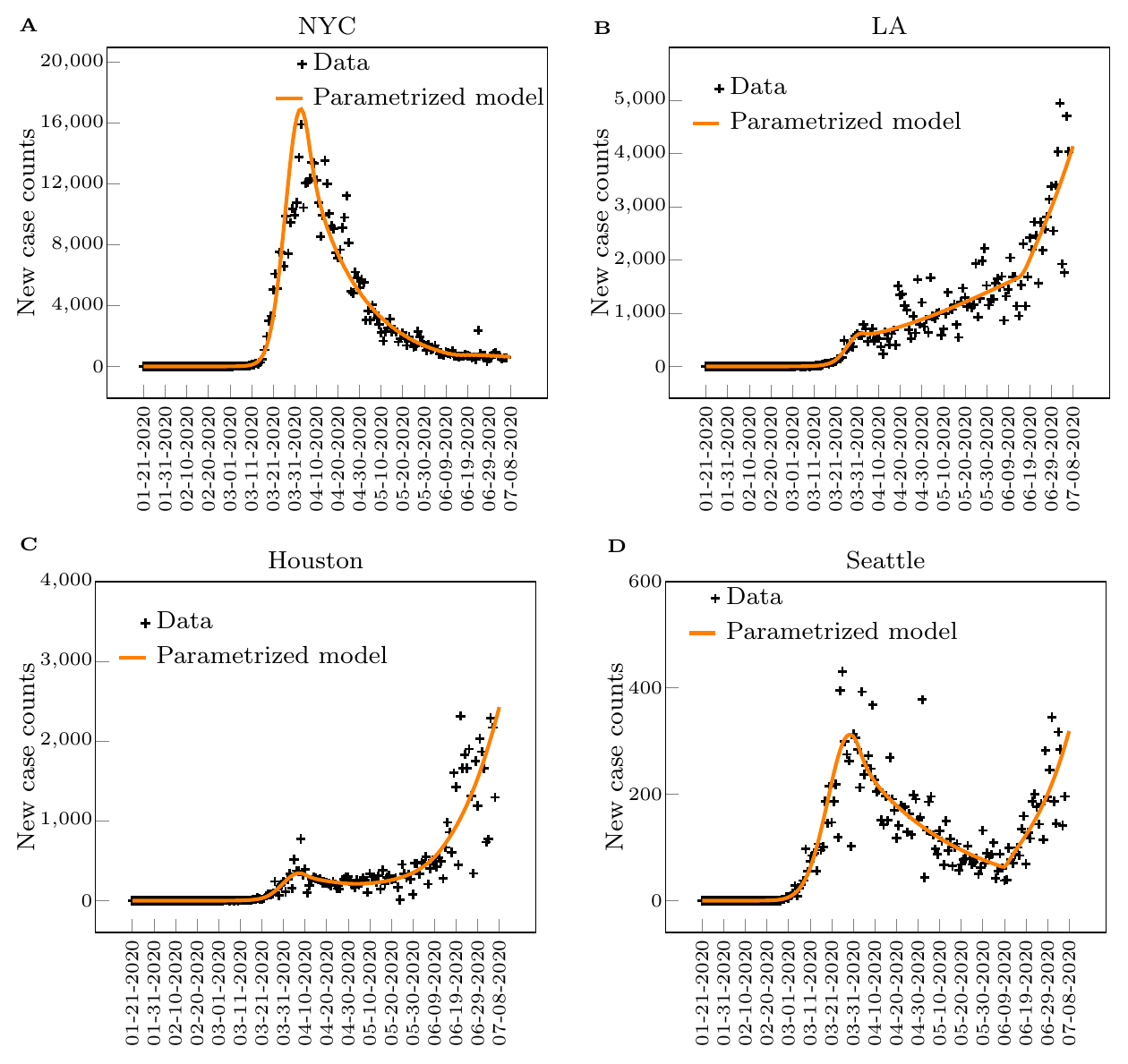}
		\caption{\textbf{New case counts from January 21, 2020 to July 8, 2020 in four Metropolitan Statistical Areas within the US.} Plus signs are daily new case counts reported by local health administrations. The solid line is the daily new cases obtained by integrating Eq.\ (2) after parametrization of the model{, which was deemed to be sufficient to reproduce the trends in new daily case report \cite{lin2020Daily}}}
		\label{fig:newcase}
	\end{figure}
	
	\begin{center}          
		\begin{table}[ht]
			\centering
			\caption{List of data-driven model parameters.}
			\begin{tabular}{clcc}
				Parameter & Description & Fixed/Free & Region-dependent \\ \hline\hline
				$S_{0}$ & Total regional population &  Fixed & Yes \\ 
				$\lambda$ & Transition rate constant from compartment $E$ to $I$ & Fixed & No  \\ 
				$\gamma_{I}$ & Recovery rate constant of the asymptomatic population & Fixed & No \\ 
				$\gamma_{A}$   &  Recovery rate constant of the asymptomatic population &  Fixed & No  \\ $\mu$   & Relative infectiousness of asymptomatic population & Fixed &	No  \\ 
				$\sigma$   & Fraction of exposed people who developed symptoms & Fixed & No  \\
				$p_{test}$   & Fraction of symptomatic people who took the test & Fixed & No \\
				$p_{sq}$   & Fraction of symptomatic people who self-quarantine & Fixed & No \\
				$r$   & Dispersion parameter of the negative-binomial noise & Free &	Yes  \\
				$\beta$  & Contact probability per person per time & Free &	Yes \\ 
				$t_{0}$ & Time when the disease was introduced to the region & Free & Yes\\ 
				$\Delta t_{1}$ & Transition time to the first social-distancing episode & Free & Yes \\
				$\Delta t_{2}$ & Duration of the first social-distancing episode &	Free & Yes \\ 
				$\Delta t_{3}$ & Transition time to the second social-distancing episode & Free & Yes \\
				$\Delta t_{4}$ & Duration of the second social-distancing episode & Free & Yes \\ 
				$P_1$ & Strength of the first episode of social-distancing & Free & Yes  \\ 
				$P_2$ &	Strength of the second episode of social-distancing & Free & Yes \\ 
				\hline\hline
			\end{tabular}
			\label{tab:paramList}
		\end{table}
	\end{center}
	
	\begin{center}          
		\begin{table}[ht]
			\centering
			\caption{Parameters by Metropolitan Statistical Area (MSA)}
			\begin{tabular}{|c|c|c|c|c|}
				\toprule \midrule
				Parameter & New York & Los Angeles & Houston & Seattle\\ \midrule \midrule
				$S_{0}$ & 1.92E+07 &  1.321E+07 &7.066E+06 & 3.979E0+6 \\ \midrule
				$\lambda$ & 3.333E-01 & 3.333E-01 & 3.333E-01 & 3.333E-01 \\ \midrule
				$\gamma_{I}$ & 1.200E-01 & 1.200E-01 & 1.200E-01 & 1.200E-01\\ \midrule
				$\gamma_{A}$   &  2.60E-01 &  2.60E-01 &	 2.60E-01 & 2.60E-01 \\ \midrule
				$\mu$   & 9.00E-01 & 9.00E-01 &	9.00E-01 &	9.00E-01 \\ \midrule
				$\sigma$   & 5.60E-01 & 5.60E-01 & 5.60E-01 & 5.60E-01 \\ \midrule
				$p_{test}$   & 2.50E-01 & 2.50E-01 & 2.50E-01 & 2.50E-01 \\ \midrule
				$p_{sq}$   & 4.00E-01 & 4.00E-01 & 4.00E-01 & 4.00E-01 \\ \midrule
				$r$   & 1.183E+01 & 1.100E+01 &	5.287 & 1.067E+01 \\ \midrule
				$\beta$  & 1.806 & 1.204 &	1.123 & 1.442\\ \midrule
				$t_{0}$ & 2.911E+01 & 2.865E+01 & 3.195E+01 & 2.339E+01\\ \midrule
				$\Delta t_{1}$ &	2.557E+01 & 3.237E+01 & 3.153E+01 & 1.388E+01 \\ \midrule
				$\Delta t_{2}$ & 2.148E+01 &	1.153+01 & 1.633E+01 & 3.549E+01\\ \midrule
				$\Delta t_{3}$ & 6.423E+01 & 7.171E+01 & 1.685E+01 & 6.614E+01 \\ \midrule
				$\Delta t_{4}$ & 8.800 & 4.289 & 4.658E+01 & 2.016\\ \midrule
				$P_1$ & 3.693E-01 & 5.371E-01 & 4.845E-01 & 4.160E-01 \\ \midrule
				$P_2$ &	4.403E-01 & 5.968E-01 &  6.391E-01 & 5.555E-01\\ \midrule
				\bottomrule
			\end{tabular}
			\label{tab:MSAparam}
		\end{table}
	\end{center}

	\section*{Results: Optimal Control Solutions}
	
	Following previous work by the authors \cite{AV,shirin2019optimal}, we use pseudo-spectral optimal control (see Supplementary Note 7) to compute solutions for the problem formulated in Eqs.\ \eqref{model},\eqref{eq:cost},\eqref{eq:const1},\eqref{eq:const2}. We set $t_i = 169$ (days) and $t_f = 259$, corresponding to a ninety day control horizon, $(t_f-t_i)=90$. {We wish to emphasize that our choice of $t_i$ is somehow arbitrary and coincides with the day when we initially computed the optimal control solutions. However, our procedure is pretty general and can be replicated for many possible choices of $t_i$.} We parametrize the solutions in $I_{\max}$, $\epsilon$ and $c_q$, after setting without loss of generality $c_p=1$. 
	
	We focus on four different US metropolitan statistical areas: New York City (NYC), Los Angeles (LA), Houston, and Seattle. NYC is the largest US metropolitan area; it emerged as the main early hotspot of the epidemics in the US in March and April, but since early June has achieved stable control of the epidemics. LA is the second largest metropolitan area in the US, it has seen a steady rise in the number of cases by the time when we performed the inference. Houston has seen a rapid increase of the cases in June and July.
	Seattle was also a very early hotspot, which has seen a decrease in the number of cases in April and May, followed by an increase in June and July. We chose these four MSA's to cover a wide range of different dynamics before the time of the inference.

	The solution of the optimal control problem  is the function of time $P^\ast(t)$ that minimizes the objective function \eqref{eq:cost} subject to the constraints (2), (5), and (6). Different from the observation period $t \le t_i$, for which we set $P(t)$ to be a piecewise linear function, in the optimization period $t \in [t_i,t_f]$ we let $P(t)$ be an arbitrary function of $t$, for the purpose of computing the optimal control solution.
	The optimal control solutions that we obtain for different MSAs are shown in Figs. \ref{fig:states_controls_Imax_mid}. These solutions are robust to parameter variations (such as different values of the coefficient $c_q$, see the Supplementary Note 2) and also qualitatively consistent for different MSA's. Robustness is also found with respect to the choice of the constraint $I_{max}$, see the Supplementary Note 2. Typically we first observe a quick drop in $P^{\ast}(t)$ (initial tightening), followed by a long nearly constant trend at $P^{\ast}_s$ (steady social distancing), and by another drop near the final time (final stranglehold). We remark that $P^{\ast}_s$ corresponds to the level of reduction needed to suppress the time-varying reproduction number $R_t\approx1$, and the cost per day associated with this level of reduction  is $\left(1-P^{\ast}_s\right)/P^{\ast}_s$. It follows that for each MSA, there is an almost constant value of $P^{\ast}_s$ which well describes the optimal solution, except for the initial time and the final time. The value of $P^{\ast}_s$ appears to be approximately the same for LA, Houston, and Seattle, while NYC has typically a slighter higher value of $P^{\ast}_s$. This also implies that the optimal cost function $J^{\ast}$ is lower for NYC than for the other cities.  In all the four metropolitan areas, $P^{\ast}_s$ is lower than the `current' value of $P(t)$ estimated from data, as can be seen from the initial dip in $P^{\ast}(t)$. NYC has the smallest initial drop, indicating that the control action at the time at which we performed the inference is the closest to optimal, followed by (in the order) Seattle, LA, and Houston. For each metropolitan area, $I_{\max}$ was estimated from available data about ICU beds for different US states, as shown in Table 3 in the Methods. 

	\begin{figure}
		\centering
		\includegraphics[width=0.9\textwidth]{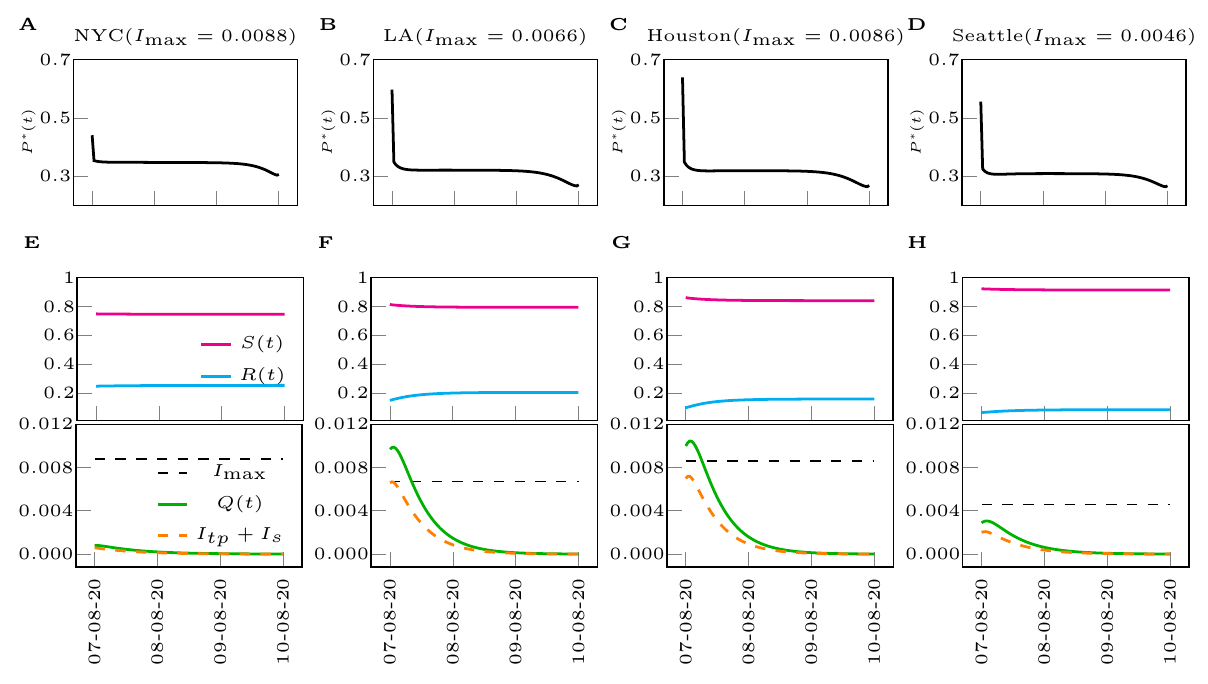}
		\caption{\textbf{Optimal control solutions for different Metropolitan Statistical Areas within the US.} (A-D) Optimal control inputs for the metropolitan cities of NYC, LA, Houston, Seattle (from left to right). (E-H) Time evolution of the states subject to the optimal control inputs. $I_{\max}$ are chosen as the minimum of the range in Table 3 ($\rho=2/3$) and $c_p=c_q = 1$. The legends in (F-H) are same as the legend in (E).  }
		\label{fig:states_controls_Imax_mid}
	\end{figure}

\begin{figure}
		\centering
		\includegraphics[width=0.9\textwidth]{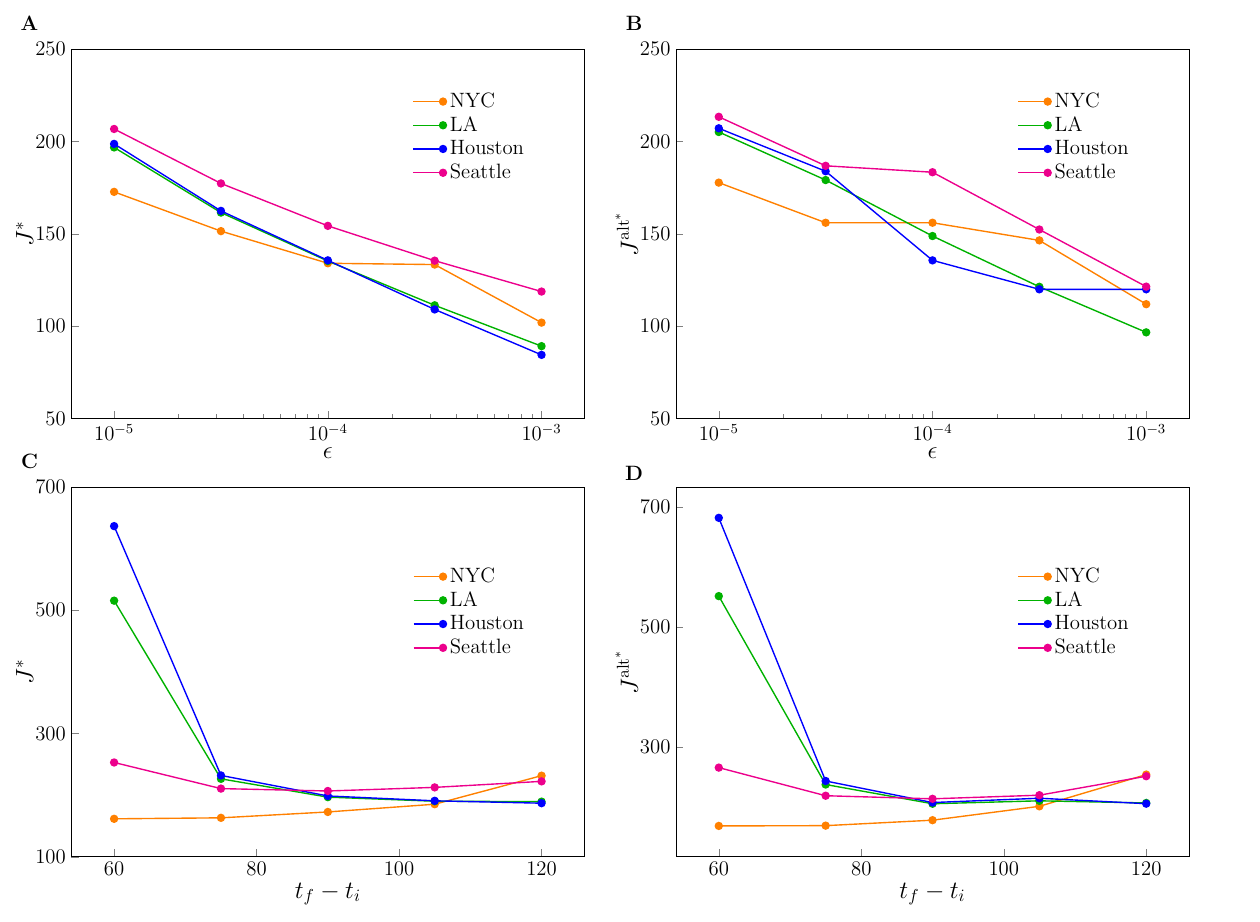}
		\caption{\textbf{Effects of varying the terminal suppression constraint $\epsilon$ and the control horizon $t_f-t_i$.} (A) The optimal cost $J^{\ast}$ obtained as the solution of the optimization problem \eqref{eq:cost} vs. the parameter $\epsilon$. (B) The optimal cost $J^{\text{alt}^{\ast}}$ obtained as the solution of the optimization problem \eqref{eq:cost_alt} vs. the parameter $\epsilon$. (C) The optimal cost $J^{\ast}$ obtained as the solution of the optimization problem vs. the control horizon $(t_f-t_i)$. (D) The optimal cost $J^{\text{alt}^{\ast}}$ obtained as the solution of the optimization problem \eqref{eq:cost_alt} vs. the control horizon $(t_f-t_i)$. The parameter $c_q$  and $c_p$ are both set to 1.}
		\label{fig:J_ep}
	\end{figure}

	{The most important parameter of the optimal control problem is the terminal suppression  constraint $\epsilon$, which describes the level at which one is trying to suppress the epidemic. The lower is $\epsilon$, the closer the goal is to eradication of the disease. The optimal value attained by the objective function $J^\ast$ versus $\epsilon$ is shown in Figs.\ \ref{fig:J_ep}(A) and (B).  It should be noted that for each value of $\epsilon$,  $J^\ast$ shown in Fig.\ \ref{fig:J_ep} is the lowest possible cost that can be attained. For all the US cities considered, this lowest possible cost increases dramatically as $\epsilon$ is reduced, which exemplifies the dilemma between saving human lives and protecting the economy. Again, here we are assuming that the measures of social distancing are optimally implemented, while the cost would be higher in case of non-optimal implementation (discussed in the Supplementary Note 7). For large values of the suppression constraint $\epsilon$, in all four cities considered, a different type of solution characterized by low cost emerges, which is discussed in more detail in what follows. Qualitatively similar results were obtained when the control input was chosen that minimizes the alternative objective function \eqref{eq:cost_alt}. A complete study of the effects of varying the suppression constraint parameter $\epsilon$ can be found in the Supplementary Note 3. There are deeper implications behind Figs.\  \ref{fig:J_ep} (A) and (B), namely setting a larger value of $\epsilon$ corresponds to delaying the economic cost in time, rather than removing it.  
		
		
		From Fig.\ 4 (C) and (D) we also see the effects of changing the control horizon $(t_f-t_i)$ of the optimal control problem, from a minimum of $60$ days to a maximum of $120$ days. We see here that different cities behave differently. For Houston and LA we see that a longer $t_f$ corresponds to a lower value of the optimal cost $J^\ast$, while for NYC the cost increases with $t_f$. A complete study of the effects of varying  $t_f$ can be found in the Supplementary Note 4. In particular, we see that increasing $t_f$ has two contrasting effects on the objective function. On the one hand the cost is integrated over a longer time period, on the other hand the longer time period can be exploited to allow for less stringent measures of social distancing at any time (larger values of $P^\ast$), which may result in a lower value for the objective function overall. Thus it appears that finely tuning the time horizon of the objective function may be used to critically and selectively affect different cities. The implications are particularly significant for those cities, Houston and LA, that seem to need a longer time period to suppress the epidemics. For Seattle we see that $J^\ast$ is minimized for intermediate values of $t_f$ in the interval $[255,285]$, indicating a specific advantage of choosing such a control horizon, see also Fig.\ 5. A remarkable observation is that by comparing the left plots and the right plots in Fig.\ 4, it is evident that the optimal control solution is quite independent of the particular form of the objective function (either $J$ or $J^{alt}$.) The reason for this is that when the integrand in \eqref{eq:cost} is nonlinear the optimal solution maintains $P^\ast(t)$ as close as possible to the linear regime (high values of $P^\ast$), which is why we do not see much difference with the linear case \eqref{eq:cost_alt}.  }

	In general, the main constraints of the problem are provided by $I_{\max}$ and $\epsilon$. However, for each case considered, typically either one of these two constraints is dominant. In all the simulations shown in  Figs. \ref{fig:states_controls_Imax_mid}, 
	the dominant constraint is provided by $\epsilon$, with $I_s(t)+I_{tp}(t)$ remaining well below $I_{max}$ over the entire period $[t_i,t_f]$.
	These solutions, which we will refer to as the type-1 solution, are characterized by strong measures of social distancing (low $P^\ast$) throughout the whole time interval considered, and the dominant constraint is to achieve suppression of the epidemics at the prescribed terminal time ($t_f$).  
	
	We have also seen the emergence of different solutions, which we refer to as the type-2 solution, when the dominant constraint is given by $I_{\max}$. In these solutions there is at least one time $t$ at which the constraint \eqref{eq:const2} is satisfied with the equal sign. Overall, type 2 solutions are less expensive economic-wise than type 1 solutions, i.e., the value of $J^\ast$ is lower.  There are also cases when the optimal solution is actively affected by both constraints. 
	In order to better understand the transition between type 1 and type 2 solutions, we have investigated the optimal control problem \eqref{eq:cost}, for the cases of LA, NYC, and Seattle, as both the suppression parameter $\epsilon$ and the control horizon $(t_f-t_i)$ are varied. The transition is characterized by a gradual change in $J^\ast$, high for type 1 solutions (in green) and low for type 2 solutions (in blue), with a transition area in between shown in yellow and red.  This is illustrated in Fig.\ 5 and in more detail in Fig.\  19 (LA), Fig.\ 20 (Seattle) and Fig.\ 21  (NYC) of
	the  Supplementary Note 5. Our results show that considerations about the timescale of the control action apply differently to different cities.
	As can be seen, the particular emergence of type-1 or type-2 solutions is affected by both the choice of $\epsilon$ and $(t_f-t_i)$. From Fig.\ 5 (A) for LA    we
	see that the most expensive control solutions are achieved when one is trying to suppress the epidemics to a low level in  a short time (small $\epsilon$ and short $t_f-t_i$.) This is opposite to what seen for NYC in Fig.\ 5 (B), where longer time horizons are usually associated with more expensive solutions. The most convenient solution arises when the suppression constant $\epsilon$ is large but the control horizon is short (area shown in blue.)
	Finally, Fig.\ 5 (C) for Seattle shows that in the case of small $\epsilon$,     expensive solutions are obtained when the control horizon is either short or long, while an intermediate control horizon is to be preferred. We also see that setting the suppression constraint to a larger value and the control horizon to be short can reduce the cost considerably (which is similar to NYC). 
	In general, 
	these results point 
	out the importance of carefully choosing the timescale over which one is seeking to suppress the epidemics, as well as the suppression level.
	{Intuitively, our results indicate that suppressing the epidemics in a short time is more costly when the epidemics is on the rise (LA) compared to cases when it is not (NYC.)}

	\begin{figure} 
		\centering
		\includegraphics[width=0.95\textwidth]{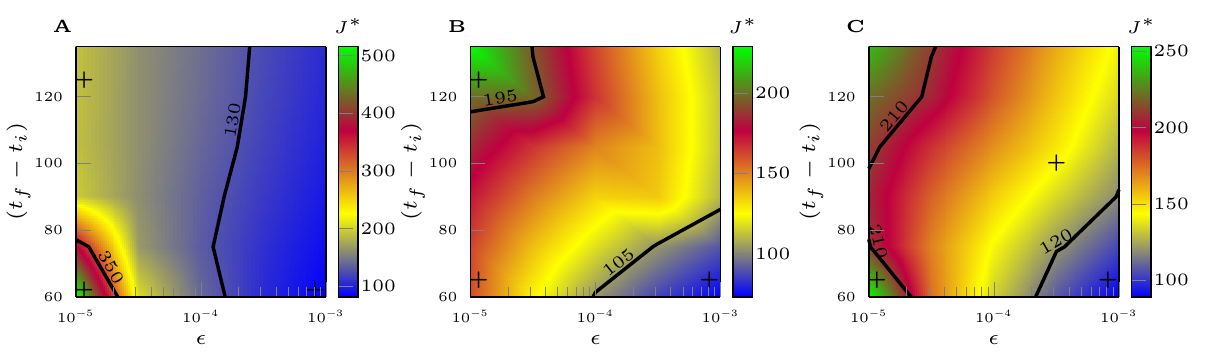}
		\caption{\textbf{The optimal cost $J^{\ast}$ in the $(t_f - t_i)$,  $\epsilon$ plane}.  (A) shows the case of the {Los Angeles Metropolitan Statistical Area}. (B) shows the case of the {NYC Metropolitan Statistical Area}. (C) shows the case of the {Seattle Metropolitan Statistical Area}. For each city, $I_{\max}$ is chosen as the maximum of the range in Table 3. Type 1 solutions (in green)
			are more expensive than type 2 solutions (in blue.) The regions in yellow/red correspond to the transition between the two types of solutions. The parameter $c_p$  and $c_q$ are both set to 1.}
		\label{latf}
	\end{figure}
	
	
	Next, we briefly consider the case that the transmissibility varies with time, i.e., we replace $\beta \rightarrow \beta(t)$ in Eq.\ (2). This is consistent e.g., with a situation in which the weather gets colder (such as during the fall season in the Northern emisphere), which has been associated with increased numbers of contagions. We then solve the optimal control problem (2) (4) (5) (6) by computing the optimal control $P^*(t)$ that minimizes the objective function $J$. 
	
\begin{figure}[h!] 
		\centering
		\includegraphics[width=0.7\textwidth]{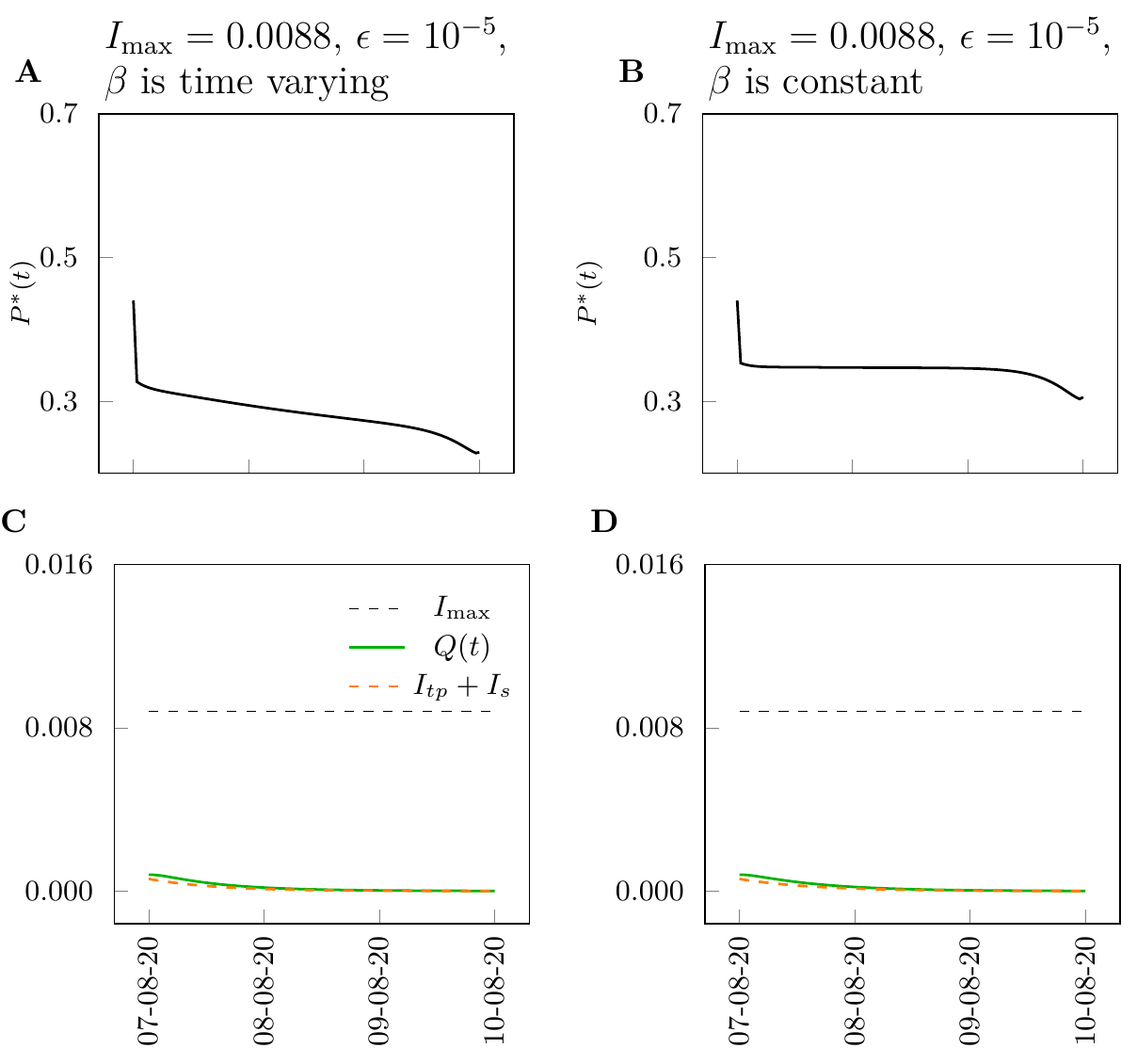}
		\caption{\textbf{Time-varying transmissibility}.  We consider the {NYC Metropolitan Statistical Area}. Left plots: time-invariant transmissibility. Right plots: the transmissibility increases linearly in time.  As can be seen, with the exception of the initial phase and of the final phase, the case of linearly increasing transmissibility corresponds to a linearly decreasing optimal control input $P^*(t)$.}
		\label{betavar}
	\end{figure}

	The results of our calculations are shown in Fig.\ \ref{betavar}, where we focus on the New York City Metropolitan Statistical Area and compare the two cases of constant transmissibility (previously studied) and time-varying transmissibility. For simplicity we take $\beta(t)$ to increase linearly in time, from the initial value in Table 1 $\beta(t_i)=1.806$ to the final value $\beta(t_f)=2 \times \beta(t_i)=3.612$. 
	As can be seen, the optimal control solution $P^*(t)$ for the case of linearly increasing $\beta$ differs substantially from the case of time-invariant $\beta$: in the central phase  $P^\ast(t)$ is now seen to decrease linearly in time, indicating that to a linear increase in transmissibility corresponds a linear decrease in the control parameter (an thus increasingly stricter measures of social distancing).  This can be used to plan long-term control interventions over periods of time over which the environmental conditions are subject to well characterized variations that affect the rate of transmissibility of the disease.

	{Finally, we include here computation of the optimal control solution for a model modified to incorporate the effects of the
		administration of an effective vaccine against COVID-19. When we initially started working on this paper, there was no certainty about the development and future availability of a vaccine.
		However, as of February 2021, we are now seeing several effective vaccines being developed, manufactured, and administered in different countries. An important observation is that the production and distribution of the vaccine is a process that takes time. For example, the United States is aiming to administer one million vaccine doses per day \cite{wang2021impacts}, which considering the entire US population, corresponds to a time-scale of roughly one year to achieve complete immunization. An important question is thus how the introduction of the vaccine is going to modify the optimal control solutions. To address this point, we (i) modify the model to include the effects of the vaccine being administered to the population, (ii) re-parametrize the model from real data and (iii) compute new optimal control solutions.  We briefly outline this process below.}
	
	We model the effects of administration of the vaccine as an additional flow from the $S$ compartment to the $R$ compartment.
	To accommodate for this, we keep using the set of equations (2), but replace Eq.\ (2a) with the following one,
	\begin{equation}  \nonumber
		\dot{S}\l(t\r)= -\beta P^2\l(t\r) S(t)\l[I_s(t)+I_{tp}(t) +\mu A(t)\r] - \kappa (t-t_v)
	\end{equation}
	and Eq.\ (2g) with
	\begin{equation} \nonumber
		\dot{R}\l(t\r)=   \gamma_A A(t) + \gamma_I \l[I_s(t)+I_{tp}(t)+Q(t)\r] + \kappa (t-t_v),
	\end{equation}
	where we take $t \geq t_v$ and $t_v$ is the time the vaccine administration started, which for the United States is 14-Dec-2020. Note that the model accounts for a rate of immunization linearly increasing with time, which is consistent with available data up to 2-Feb-2021, when we carried out this analysis.
	
	In our calculations we focused on the MSA of Seattle, for which we estimated $\kappa=0.000045$ (see the Supplementary Information Note 11 for more details.) We then used the same approach outlined before to parametrize the model. The parameters were inferred by the daily case reports over the extended period of time from 21-Jan-2020 to 14-Dec-2020. More information about the parametrization can also be found in the Supplementary Note 11.
	We then re-computed the optimal control solution by selecting  the time at which we start optimizing the control action $t_i=t_v+7$ and by varying the control horizon $T_{cont}=(t_f-t_i).$ As we already pointed out before, we found the choice of the control horizon to strongly affect the results of the optimization.

	Figure 7 shows the results of our computations for the city of Seattle by choosing two different values of the control horizon, $T_{cont}=(t_f-t_i)=130$ days (on the left) and $T_{cont}=(t_f-t_i)=150$ days (on the right.)  
	In both plots, we compared the two cases that the model includes the effects of vaccination (gray curve, $\kappa=0.000045$) or does not (black curve, $\kappa=0$.) The full black dot indicates the point in time at which the two curves depart from each other.  Both plots show the optimal control solution to deviate at an early time, roughly 15 to 25 days after the beginning of the administration of the vaccine (which corresponds to approximately $6\%$ to $8\%$ of the population getting vaccinated.)   This indicates that vaccinations can affect the optimal control solution, even when a relatively little percentage of the population is immunized. In all the simulations we have always seen the optimal control solution to deviate soon after the beginning of vaccinations.
	Moreover, the separation between the two curves appears to be larger over the longer control horizon. The plot on the right shows the parameter $P^*(t)$ approach $1$ at the end of the control horizon, which corresponds to complete removal of the social distancing measures. 
	
	Our results point out that in a realistic scenario the optimal level of social distancing is affected by the introduction of vaccinations, especially over longer periods of time.
	This is not surprising as the expectation is that the long-term effect of administration of the vaccines will be relaxation of the social distancing measures. However, the advantage of our analysis is that it can be used to assess  the most economically advantageous level of social distancing  
	\textit{while the vaccines are being administered to the population.} For example, Fig.\ 7B shows substantial relaxation of social distancing  roughly five months after the date in which administration of the vaccines began. {We remark that the derived timescale (150 days) is subject to the simple vaccination model, in which we optimistically assume that (1) all the doses are reserved for and administered to only the susceptible population, and (2) the production of the vaccines remains linear in this 150 days. Our simulation shows that, with these two assumptions, at the end of this period herd immunity is achieved and social distancing is removed.
		However, such a timescale can be treated as an optimistic forecast; the real timescale to reach  herd immunity  could be longer due to the violation of either of the assumptions.} It is important to point out once again that the optimization we perform is one for which the goal is to minimize the impact of social distancing on the economy, in the presence of relevant public health constraints.

	\begin{figure}[h!] 
		\centering
		\includegraphics[width=0.9\textwidth]{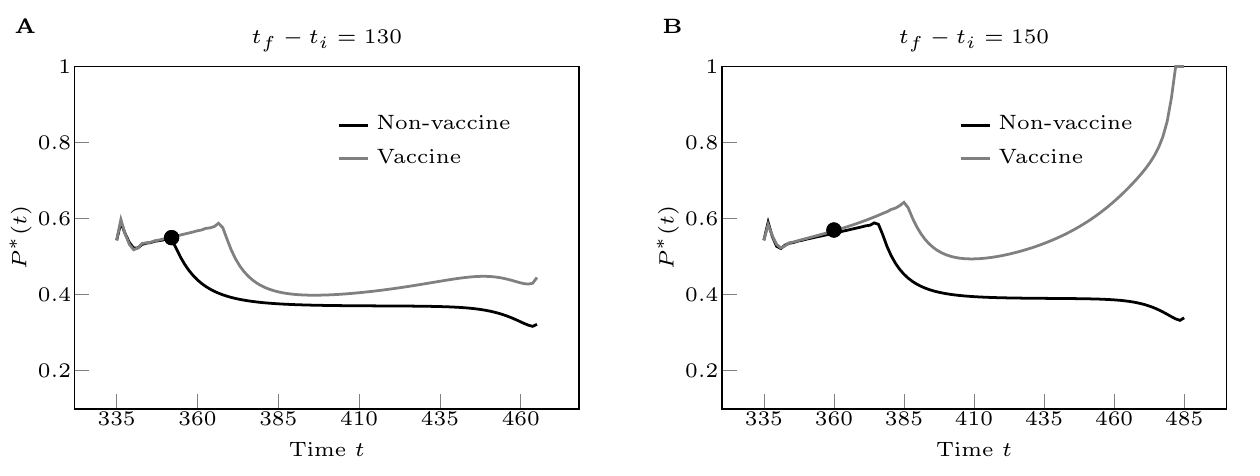}
		\caption{\textbf{Effects of vaccinations on the optimal control solution.}  We consider the {Seattle Metropolitan Statistical Area}. In A the control horizon $T_{cont}=(t_f-t_i)=130$ days and in B the control horizon $T_{cont}=(t_f-t_i)=150$ days.
			For both figures, we plot $P^*(t)$ by comparing the two cases that the model includes the effects of vaccination (gray curve) or does not (black curve.) The full black dot indicates the point in time at which the two curves depart from each other. As can be seen, the separation between the two curves appears to be much larger over a longer control horizon. The plot on the right shows the parameter $P^*(t)$ approach $1$ at the end of the control horizon, which corresponds to removal of the social distancing measures. Additional plots showing the time evolutions of the states corresponding to the optimal solutions in A and B can be found in the Supplementary Information, Note 11. }
		\label{vaccine}
	\end{figure}

	\color{black}

	\section*{Discussion}

	{Our analysis reveals that the cost of eradicating the disease---i.e., suppressing the number of infected individuals down to a certain critical threshold, such as $\ll 1$ person in the formulation of our stylized mathematical model, even in the optimal case, is significantly higher than the cost of `managing' the pandemic to avoid the saturation of regional medical resources. In light of the current progression of the pandemic in the US, our analysis brings  rigorous scientific and quantitative foundation for the latter strategy, which is adopted by many local administrations (e.g., State offices.) }
	
	{Our conclusion that the optimal control solution is well approximated by a constant level of social distancing contrasts with the implemented 
		system of reopening `in phases' 
		\cite{gottlieb2020national}. In many countries, the expected progression is from phase 1 to phase 2 and higher, but there have been several cases in which a premature reopening followed by a rise in COVID-19 numbers has led to folding back into phase 1 \cite{saul2020victoria,Italy2020TheGuardian,Spain2020TheGuardian,US2020NYT}. Based on our results, such alternating control interventions are suboptimal because the economic benefits of  momentarily relaxing the restrictions are lower than the costs associated with the successive tightening (such as e.g., a second lockdown.) Instead we have shown that in the presence of smoothly time-varying environmental conditions (e.g., affecting the transmissibility), the optimal control solution also varies smoothly in time. This allows us to plan long-term control interventions over periods of time over which the environmental conditions are subject to well characterized variations that affect the rate of transmissibility of the disease. 
	}
	
	{Several countries have adopted control strategies that are region-specific. For example, Italy has assigned to its regions color-coded restrictions (yellow, orange, and red)  \cite{pelagatti2020assessing}, which, based on a careful monitoring of the local progression of the disease, have been updated frequently. Our model focuses on individual MSA's (or regions) and as such neglects the larger-scale spatial resolution of the epidemic.  The need for frequent updates of the control interventions provides evidence of the inherent difficulty of `closing the control loop', due to many possible factors, such as partial and incorrect measurements, delays in obtaining information and implementing actions,  imperfect application of the controls, lack of resources, the effects of people traveling from area to area, and others. While all these factors would require separate consideration, our main recommendation that interventions should be maintained at a (nearly) constant level stands. Our work indicates that the best intervention for the economy is one that does not fluctuate in time,
		while alternating control actions are suboptimal.
		In addition, frequently changing interventions present other  disadvantages which have made them unpopular \cite{Angry}:  they are difficult to implement, and to follow, and they negatively affect business planning capability.}
	
	Occasionally, there have been 
	claims by policy makers and/or scientists that seeking herd immunity by exposure to the virus (in the absence of a vaccine) 
	may provide the best long-term path-forward \cite{orlowski2020four,kwok2020herd}. However, the timescales over which herd immunity can be achieved without violating the $I_{\max}$ constraint appear to be quite long (see Methods) in the absence of a vaccine. By setting a very large control horizon in our simulations, we have seen the emergence of these 
	`herd immunity solutions', shown in the Supplementary Note 6. However,
	the parameters of our realistic scenarios, estimated from real data, are far away from the point at which such solutions
	are
	optimal. In general, our study allows to quantify how `far' one is in the parameter space from the point at which a herd immunity solution becomes optimal. 

	\section*{Conclusion}
	
	In this paper, we have proposed an approach to optimize social distancing measures in time in order to contain the spread of the COVID-19 epidemics, while minimizing the impact on the economy. Our analysis has been applied to four different metropolitan {statistical} areas within the US, but can be directly applied to other geographical areas. Each MSA was shown to be well described by a stylized mathematical model whose parameters were {inferred} by daily new case counts reported by local health administrations.  Our approach based on modeling, inferring model parameters from real data, and computing optimal control solutions for the inferred model is general and may be applied to other 
	complex systems of interest.
	
	Our choice of the objective function is quite simplified, namely we assume that increasing levels of social distancing and quarantining result in progressively higher economical costs. {In addition, we did not account for the cost associated with  medical treatments, which is important yet significantly smaller than the cost of shutting down the economical activities of the entire society \cite{chen2020medical}.} Our approach can be easily generalized to other more complex, more realistic types of objective functions, see e.g., \cite{shirin2019optimal,AV}. Different from this study, these objective functions may also be specific to given regions or try to capture a particular socio-economical model of interest.

	We have found that the optimal control solutions are quite robust to the specific choice of the objective function and of its parameters, such as $c_q$. These control solutions tend to be qualitatively similar for different cities. However, they are affected by the choice of the constraints, in particular the constraint that we have associated to suppression of the epidemic, and by the time horizon of the control action. When these are varied, different cities behave differently, which points out the importance of our data-driven approach.  We have also seen that small deviations from the optimal solution can lead to dramatic violations of the constraints.

	It is possible to translate these optimal interventions in actual measures that can be imposed on the population, such as restricting the access to certain businesses or venues. {Available technology includes the usage of coupled digital non-contact healthcare systems \cite{radanliev2020covid}.}
	While implementation of a time-varying control may be challenging in practice,
	we found that the optimal solution is typically characterized by an initial drop (due essentially to the non-optimality of current control interventions), followed by a nearly constant control (specific to each city) for a long time, and by a final drop, which is needed to achieve the desired suppression of the epidemics at the final time. Thus the optimal control solution is almost constant except for the initial time and the final time, which significantly increases the applicability of this study. {The constant part of the solution could be practically achieved by dynamical regulation to control the time-varying reproduction number $R_t\approx 1$.} 
	A key observation is that the optimal solution $P^\ast$ is {generally} lower  than the value of $P$ {inferred by} data. The initial drop in $P^\ast$ may provide a measure of non-optimality of current interventions. This drop was seen to be smallest for NYC compared to other US cities, but was present in all the cities we have analyzed.

	The time-scale of control interventions, i.e., the control horizon, appears to play a fundamental role. Considerations about the control horizon
	may vary from area to area and may be affected by a number of factors, including the times at which a vaccine becomes available and is distributed to the population. It appears that cities that have seen an increase of cases during the inference period need a longer control horizon to suppress the epidemics optimally.
	In certain instances, the
	impact on the economy can be minimized by tuning the control horizon; for example, for the city of Seattle we found that an optimal control horizon was equal to roughly 90 days when
	the suppression constraint $\epsilon=10^{-5}$.
	We wish to emphasize that given the very large economical impact of social distancing measures, even a small improvement in the control strategy can lead to considerable economical benefits.

	We also computed the most economically advantageous level of social distancing during the time over which the vaccines are administered to the population. Our simulations show that the optimal control solution is affected by the introduction of the vaccine quite early on, which indicates the possibility of gradually relaxing measures of social distancing soon after the beginning of vaccinations 
	and strongly reduce them roughly five months later. While optimal control interventions computed during the vaccination period were found to be quite sensitive to the specific choice of model and control parameters, in all of our simulations we always saw that a gradual relaxation of social distancing measures was possible when roughly 10\% of the population got vaccinated. {We acknowledge that our study on the effects of the vaccines is limited, as we only focus on one region and we see that the optimal control solution is strongly affected by the choice of the control horizon. We also do not consider the emergence of new variants against which existing vaccines may be less effective. Our assumption of a linearly increasing rate of vaccination of the population should be validated by using region-specific data. Extending the analysis to other regions and incorporating more realistic data would require a major effort that is beyond the scope of this paper. Our conclusion that over longer time periods vaccination would allow substantial reductions in the level of social distancing, should be taken with the due precautions. One should be reminded that our objective in this paper is to minimize the impact of social distancing on the economy, in the presence of relevant public health constraints. Nonetheless, our approach based on modeling, model parametrization, and optimal control, could be easily adapted to different geographical areas  to provide region-specific recommendations.} 

	{We briefly discuss next other limitations of our work. Limited testing capabilities may affect some of our results. The model could benefit from incorporation of considerations about 
		spatial effects. Many regulations have been introduced to limit people's mobility during the pandemic; however essential travel has often remained in place, which has probably been responsible for much of the disease's propagation. One first step to incorporate spatial resolution in the model would be considering an extension of the model with two communicating regions. This could be the subject of future work. }

	One relevant question is whether policy-makers can assess whether a currently employed  control action is optimal or not. The so-called HAMVET procedure, initially proposed in \cite{ross2015primer} and presented in the Supplementary Note 9, can be used
	to validate a control strategy and evaluate its optimality.
	

%

\begin{thebibliography}{10}
		
		\bibitem{DailyVaccines}
		{"Daily COVID-19 vaccine doses administered". Our World in data}.
		\newblock
		\url{https://ourworldindata.org/grapher/cumulative-covid-vaccinations}.
		\newblock {February 2020}.
		
		\bibitem{LANLgithub}
		{LANL COVID-19 Prediction GitHub.}
		
		\bibitem{dimondPrincess}
		{Ministry of Health, Official report on the cruise ship Diamond Princess, May
			1, 2020.}
		
		\bibitem{NYTgithub}
		{New York Times repository of Covid-19 data in the United States.}
		
		\bibitem{MSAdelineation}
		{U.S. Office of Management and Budget Delineation Files}.
		
		\bibitem{acemoglu2020multi}
		Daron Acemoglu, Victor Chernozhukov, Iv{\'a}n Werning, and Michael~D Whinston.
		\newblock A multi-risk sir model with optimally targeted lockdown.
		\newblock Technical report, National Bureau of Economic Research, 2020.
		
		\bibitem{anderson2020estimating}
		Sean~C Anderson, Andrew~M Edwards, Madi Yerlanov, Nicola Mulberry, Jessica
		Stockdale, Sarafa~A Iyaniwura, Rebeca~C Falcao, Michael~C Otterstatter,
		Michael~A Irvine, Naveed~Z Janjua, et~al.
		\newblock Estimating the impact of covid-19 control measures using a bayesian
		model of physical distancing.
		\newblock {\em medRxiv}, 2020.
		
		\bibitem{andrieu2008tutorial}
		Christophe Andrieu and Johannes Thoms.
		\newblock A tutorial on adaptive mcmc.
		\newblock {\em Statistics and computing}, 18(4):343--373, 2008.
		
		\bibitem{Angry}
		Crispian~Balmer Balmer.
		\newblock Italian regions angry over government's covid-19 zones.
		\newblock {\em Reuters}, 2020.
		
		\bibitem{becerra2010solving}
		Victor~M Becerra.
		\newblock Solving complex optimal control problems at no cost with psopt.
		\newblock In {\em 2010 IEEE International Symposium on Computer-Aided Control
			System Design}, pages 1391--1396. IEEE, 2010.
		
		\bibitem{chang2020modelling}
		Sheryl~L Chang, Nathan Harding, Cameron Zachreson, Oliver~M Cliff, and Mikhail
		Prokopenko.
		\newblock Modelling transmission and control of the covid-19 pandemic in
		australia.
		\newblock {\em arXiv preprint arXiv:2003.10218}, 2020.
		
		\bibitem{chen2020medical}
		Jiangzhuo Chen, Anil Vullikanti, Stefan Hoops, Henning Mortveit, Bryan Lewis,
		Srinivasan Venkatramanan, Wen You, Stephen Eubank, Madhav Marathe, Chris
		Barrett, et~al.
		\newblock Medical costs of keeping the us economy open during covid-19.
		\newblock {\em Scientific reports}, 10(1):1--10, 2020.
		
		\bibitem{covid2020forecasting}
		IHME COVID, Christopher~JL Murray, et~al.
		\newblock Forecasting covid-19 impact on hospital bed-days, icu-days,
		ventilator-days and deaths by us state in the next 4 months.
		\newblock {\em MedRxiv}, 2020.
		
		\bibitem{flaxman2020report}
		Seth Flaxman, Swapnil Mishra, Axel Gandy, H~Unwin, Helen Coupland, T~Mellan,
		Harisson Zhu, Tresnia Berah, J~Eaton, P~Perez~Guzman, et~al.
		\newblock Report 13: Estimating the number of infections and the impact of
		non-pharmaceutical interventions on covid-19 in 11 european countries.
		\newblock 2020.
		
		\bibitem{franc24GermanyLockdown}
		{France 24. 2020}.
		\newblock {"Germany reimposes local lockdowns after regional coronavirus
			outbreak". France 24}.
		\newblock
		\url{https://www.france24.com/en/20200623-germany-reimposes-local-lockdown-after-coronavirus-outbreak}.
		\newblock {August 2020}.
		
		\bibitem{gottlieb2020national}
		Scott Gottlieb, Caitlin Rivers, Mark~B McClellan, Lauren Silvis, and Crystal
		Watson.
		\newblock National coronavirus response: a road map to reopening.
		\newblock {\em AEI Paper \& Studies}, 2020.
		
		\bibitem{gupta2005economic}
		Anu~G Gupta, Cheryl~A Moyer, and David~T Stern.
		\newblock The economic impact of quarantine: Sars in toronto as a case study.
		\newblock {\em Journal of Infection}, 50(5):386--393, 2005.
		
		\bibitem{MortalityRate}
		{Janet Christenbury. 2020}.
		\newblock {"Study in ICU Finds 30.9\% Mortality Rate From COVID-19". Futurity}.
		\newblock \url{https://www.futurity.org/covid-19-mortality-rate-2377362-2/}.
		\newblock {August 2020}.
		
		\bibitem{Spain2020TheGuardian}
		Ashifa Kassam.
		\newblock Spain warned of dire impact of second coronavirus lockdown.
		\newblock {\em The Guardian}, 2020.
		
		\bibitem{kirk2004optimal}
		Donald~E Kirk.
		\newblock {\em Optimal control theory: an introduction}.
		\newblock Courier Corporation, 2004.
		
		\bibitem{kohler2020robust}
		Johannes K{\"o}hler, Lukas Schwenkel, Anna Koch, Julian Berberich, Patricia
		Pauli, and Frank Allg{\"o}wer.
		\newblock Robust and optimal predictive control of the covid-19 outbreak.
		\newblock {\em arXiv preprint arXiv:2005.03580}, 2020.
		
		\bibitem{koren2020business}
		Mikl{\'o}s Koren and Rita Pet{\H{o}}.
		\newblock Business disruptions from social distancing.
		\newblock {\em arXiv preprint arXiv:2003.13983}, 2020.
		
		\bibitem{kwok2020herd}
		Kin~On Kwok, Florence Lai, Wan~In Wei, Samuel Yeung~Shan Wong, and Julian~WT
		Tang.
		\newblock Herd immunity--estimating the level required to halt the covid-19
		epidemics in affected countries.
		\newblock {\em Journal of Infection}, 80(6):e32--e33, 2020.
		
		\bibitem{lauer2020incubation}
		Stephen~A Lauer, Kyra~H Grantz, Qifang Bi, Forrest~K Jones, Qulu Zheng,
		Hannah~R Meredith, Andrew~S Azman, Nicholas~G Reich, and Justin Lessler.
		\newblock {The Incubation Period of Coronavirus Disease 2019 (COVID-19) From
			Publicly Reported Confirmed Cases: Estimation and Application.}
		\newblock {\em Annals of internal medicine}, 2019, 2020.
		
		\bibitem{US2020NYT}
		Jasmine~C. Lee, Sarah Mervosh, Yuriria Avila, Barbara Harvey, and Alex
		Leeds~Matthews.
		\newblock Spain warned of dire impact of second coronavirus lockdown.
		\newblock {\em The New York Times}, 2020.
		
		\bibitem{lin2020Daily}
		Yen~Ting Lin, Jacob Neumann, Ely~F Miller, Richard~G Posner, Abhishek Mallela,
		Cosmin Stafa, Jaideep Ray, Gautam Thakur, Supriya Chinthavali, and William~S
		Hlavacek.
		\newblock {Daily Forecasting of New Cases for Regional Epidemics of Coronavirus
			Disease 2019 with Bayesian Uncertainty Quantification}.
		\newblock {\em {\rm In press, }Emerging Infectious Diseases, {\rm preprint
				available at
				\url{https://www.medrxiv.org/content/10.1101/2020.07.20.20151506v2}}}, 2021.
		
		\bibitem{GDP2ndQdrop}
		{Lisa Mataloni, Dave Wasshausen, Erich Strassner, Jeannine Aversa. 2020}.
		\newblock {"Gross Domestic Product, Second Quarter 2020 (Advance Estimate) and
			Annual Update". U.S. Bureau of Economic Analysis}.
		\newblock
		\url{https://www.bea.gov/sites/default/files/2020-07/gdp2q20_adv_0.pdf}.
		\newblock {August 2020}.
		
		\bibitem{liu2020risk}
		Xiaofan Liu, Hong Zhou, Yilu Zhou, Xiaojun Wu, Yang Zhao, Yang Lu, Weijun Tan,
		Mingli Yuan, Xuhong Ding, Jinjing Zou, et~al.
		\newblock Risk factors associated with disease severity and length of hospital
		stay in covid-19 patients.
		\newblock {\em Journal of Infection}, 81(1):e95--e97, 2020.
		
		\bibitem{GermanOpening}
		{Lother Wieler, Ute Rexroth, Rene Gottschalk. 2020}.
		\newblock {"Emerging COVID-19 success story: Germany’s strong enabling
			environment". Our World in data}.
		\newblock \url{https://ourworldindata.org/covid-exemplar-germany}.
		\newblock {August 2020}.
		
		\bibitem{morris2020optimal}
		Dylan~H Morris, Fernando~W Rossine, Joshua~B Plotkin, and Simon~A Levin.
		\newblock Optimal, near-optimal, and robust epidemic control.
		\newblock {\em arXiv preprint arXiv:2004.02209}, 2020.
		
		\bibitem{nicola2020socio}
		Maria Nicola, Zaid Alsafi, Catrin Sohrabi, Ahmed Kerwan, Ahmed Al-Jabir,
		Christos Iosifidis, Maliha Agha, and Riaz Agha.
		\newblock The socio-economic implications of the coronavirus pandemic
		(covid-19): A review.
		\newblock {\em International journal of surgery (London, England)}, 78:185,
		2020.
		
		\bibitem{nishiura2020serial}
		Hiroshi Nishiura, Natalie~M Linton, and Andrei~R Akhmetzhanov.
		\newblock Serial interval of novel coronavirus (covid-19) infections.
		\newblock {\em International journal of infectious diseases}, 2020.
		
		\bibitem{nocedal2006numerical}
		Jorge Nocedal and Stephen Wright.
		\newblock {\em Numerical optimization}.
		\newblock Springer Science \& Business Media, 2006.
		
		\bibitem{orlowski2020four}
		Eric~JW Orlowski and David~JA Goldsmith.
		\newblock Four months into the covid-19 pandemic, sweden’s prized herd
		immunity is nowhere in sight.
		\newblock {\em Journal of the Royal Society of Medicine}, 113(8):292--298,
		2020.
		
		\bibitem{pelagatti2020assessing}
		Matteo~M Pelagatti.
		\newblock Assessing the effectiveness of the italian risk-zones policy during
		the second wave of covid-19.
		\newblock {\em University of Milan Bicocca Department of Economics, Management
			and Statistics Working Paper}, (457), 2020.
		
		\bibitem{perez2020serology}
		Javier Perez-Saez, Stephen~A Lauer, Laurent Kaiser, Simon Regard, Elisabeth
		Delaporte, Idris Guessous, Silvia Stringhini, Andrew~S Azman,
		Serocov-POP~Study Group, et~al.
		\newblock Serology-informed estimates of sars-cov-2 infection fatality risk in
		geneva, switzerland.
		\newblock {\em medRxiv}, 2020.
		
		\bibitem{quah2020mortality}
		Pipetius Quah, Andrew Li, and Jason Phua.
		\newblock Mortality rates of patients with covid-19 in the intensive care unit:
		a systematic review of the emerging literature.
		\newblock {\em Critical Care}, 24:1--4, 2020.
		
		\bibitem{radanliev2020data}
		Petar Radanliev, David De~Roure, and Rob Walton.
		\newblock Data mining and analysis of scientific research data records on
		covid-19 mortality, immunity, and vaccine development-in the first wave of
		the covid-19 pandemic.
		\newblock {\em Diabetes \& Metabolic Syndrome: Clinical Research \& Reviews},
		14(5):1121--1132, 2020.
		
		\bibitem{radanliev2020covid}
		Petar Radanliev, David De~Roure, Rob Walton, Max Van~Kleek, Rafael~Mantilla
		Montalvo, Omar Santos, Stacy Cannady, et~al.
		\newblock Covid-19 what have we learned? the rise of social machines and
		connected devices in pandemic management following the concepts of
		predictive, preventive and personalized medicine.
		\newblock {\em EPMA Journal}, pages 1--22, 2020.
		
		\bibitem{randolph2020herd}
		Haley~E Randolph and Luis~B Barreiro.
		\newblock Herd immunity: Understanding covid-19.
		\newblock {\em Immunity}, 52(5):737--741, 2020.
		
		\bibitem{rao2009survey}
		Anil~V Rao.
		\newblock A survey of numerical methods for optimal control.
		\newblock {\em Advances in the Astronautical Sciences}, 135(1):497--528, 2009.
		
		\bibitem{richardson2020presenting}
		Safiya Richardson, Jamie~S. Hirsch, Mangala Narasimhan, James~M. Crawford,
		Thomas McGinn, Karina~W. Davidson, , and the Northwell COVID-19
		Research~Consortium.
		\newblock {Presenting Characteristics, Comorbidities, and Outcomes Among 5700
			Patients Hospitalized With COVID-19 in the New York City Area}.
		\newblock {\em JAMA}, 323(20):2052--2059, 05 2020.
		
		\bibitem{ross2015primer}
		I~Michael Ross.
		\newblock {\em A primer on Pontryagin's principle in optimal control}.
		\newblock Collegiate publishers, 2015.
		
		\bibitem{ross2012review}
		I~Michael Ross and Mark Karpenko.
		\newblock A review of pseudospectral optimal control: From theory to flight.
		\newblock {\em Annual Reviews in Control}, 36(2):182--197, 2012.
		
		\bibitem{ruktanonchai2020assessing}
		Nick~Warren Ruktanonchai, JR~Floyd, Shengjie Lai, Corrine~Warren Ruktanonchai,
		Adam Sadilek, Pedro Rente-Lourenco, Xue Ben, Alessandra Carioli, Joshua
		Gwinn, JE~Steele, et~al.
		\newblock Assessing the impact of coordinated covid-19 exit strategies across
		europe.
		\newblock {\em Science}, 2020.
		
		\bibitem{sakurai2020natural}
		Aki Sakurai, Toshiharu Sasaki, Shigeo Kato, Masamichi Hayashi, Sei-ichiro
		Tsuzuki, Takuma Ishihara, Mitsunaga Iwata, Zenichi Morise, and Yohei Doi.
		\newblock {Natural History of Asymptomatic SARS-CoV-2 Infection}.
		\newblock {\em New England Journal of Medicine}, 0(0):null.
		
		\bibitem{sanche2020novel}
		Steven Sanche, Yen~Ting Lin, Chonggang Xu, Ethan Romero-Severson, Nicolas~W
		Hengartner, and Ruian Ke.
		\newblock The novel coronavirus, 2019-ncov, is highly contagious and more
		infectious than initially estimated.
		\newblock {\em arXiv preprint arXiv:2002.03268}, 2020.
		
		\bibitem{saul2020victoria}
		Allan Saul, Nick Scott, Brendan~S Crabb, Suman~S Majundar, Benjamin Coghlan,
		and Margaret~E Hellard.
		\newblock Victoria’s response to a resurgence of covid-19 has averted
		9,000-37,000 cases in july 2020.
		\newblock {\em The Medical Journal of Australia}, page~1, 2020.
		
		\bibitem{shirin2019optimal}
		Afroza Shirin, Fabio Della~Rossa, Isaac Klickstein, John Russell, and Francesco
		Sorrentino.
		\newblock Optimal regulation of blood glucose level in type i diabetes using
		insulin and glucagon.
		\newblock {\em PloS one}, 14(3):e0213665, 2019.
		
		\bibitem{AV}
		Afroza Shirin, Feng Song, Yen-Ting Lin, William~S Hlavacek, and Sorrentino S.
		\newblock Prediction of optimal drug schedules for controlling autophagy.
		\newblock {\em Scientific Reports}, 9(1428), 2019.
		
		\bibitem{Italy2020TheGuardian}
		Lorenzo Tondo.
		\newblock Italy at crossroads as fears grow of covid-19 second wave.
		\newblock {\em The Guardian}, 2020.
		
		\bibitem{nguyen2020natural}
		Nguyen {Van Vinh Chau}, Vo~Thanh Lam, Nguyen~Thanh Dung, Lam~Minh Yen, Ngo
		Ngoc~Quang Minh, Le~Manh Hung, Nghiem~My Ngoc, Nguyen~Tri Dung, Dinh
		Nguyen~Huy Man, Lam~Anh Nguyet, Le~Thanh~Hoang Nhat, Le~Nguyen~Truc Nhu,
		Nguyen Thi~Han Ny, Nguyen Thi~Thu Hong, Evelyne Kestelyn, Nguyen Thi~Phuong
		Dung, Tran~Chanh Xuan, Tran~Tinh Hien, Nguyen~Thanh Phong, Tran Nguyen~Hoang
		Tu, Ronald~B Geskus, Tran~Tan Thanh, Nguyen~Thanh Truong, Nguyen~Tan Binh,
		Tang~Chi Thuong, Guy Thwaites, Le~{Van Tan}, and Oxford University Clinical
		Research Unit COVID-19~Research Group.
		\newblock {The Natural History and Transmission Potential of Asymptomatic
			Severe Acute Respiratory Syndrome Coronavirus 2 Infection}.
		\newblock {\em Clinical Infectious Diseases}, 2020.
		
		\bibitem{wachter2006implementation}
		Andreas W{\"a}chter and Lorenz~T Biegler.
		\newblock On the implementation of an interior-point filter line-search
		algorithm for large-scale nonlinear programming.
		\newblock {\em Mathematical programming}, 106(1):25--57, 2006.
		
		\bibitem{wang2021impacts}
		Xutong Wang, Zhanwei Du, Kaitlyn Johnson, Spencer Fox, Michael Lachmann,
		Jason~S McLellan, and Lauren~Ancel Meyers.
		\newblock The impacts of covid-19 vaccine timing, number of doses, and risk
		prioritization on mortality in the us.
		\newblock {\em medRxiv}, 2021.
		
		\bibitem{wolfel2020virological}
		Roman W{\"o}lfel, Victor~M Corman, Wolfgang Guggemos, Michael Seilmaier, Sabine
		Zange, Marcel~A M{\"u}ller, Daniela Niemeyer, Terry~C Jones, Patrick Vollmar,
		Camilla Rothe, et~al.
		\newblock Virological assessment of hospitalized patients with covid-2019.
		\newblock {\em Nature}, 581(7809):465--469, 2020.
		
	\end{thebibliography}

\begin{thebibliography}{1}
	
	\bibitem{DailyVaccines}
	{"Daily COVID-19 vaccine doses administered". Our World in data}.
	\newblock
	\url{https://ourworldindata.org/grapher/cumulative-covid-vaccinations}.
	\newblock {February 2020}.
	
	\bibitem{becerra2010solving}
	Victor~M Becerra.
	\newblock Solving complex optimal control problems at no cost with psopt.
	\newblock In {\em 2010 IEEE International Symposium on Computer-Aided Control
		System Design}, pages 1391--1396. IEEE, 2010.
	
	\bibitem{kirk2004optimal}
	Donald~E Kirk.
	\newblock {\em Optimal control theory: an introduction}.
	\newblock Courier Corporation, 2004.
	
	\bibitem{nocedal2006numerical}
	Jorge Nocedal and Stephen Wright.
	\newblock {\em Numerical optimization}.
	\newblock Springer Science \& Business Media, 2006.
	
	\bibitem{rao2009survey}
	Anil~V Rao.
	\newblock A survey of numerical methods for optimal control.
	\newblock {\em Advances in the Astronautical Sciences}, 135(1):497--528, 2009.
	
	\bibitem{ross2015primer}
	I~Michael Ross.
	\newblock {\em A primer on Pontryagin's principle in optimal control}.
	\newblock Collegiate publishers, 2015.
	
	\bibitem{ross2012review}
	I~Michael Ross and Mark Karpenko.
	\newblock A review of pseudospectral optimal control: From theory to flight.
	\newblock {\em Annual Reviews in Control}, 36(2):182--197, 2012.
	
	\bibitem{ruktanonchai2020assessing}
	Nick~Warren Ruktanonchai, JR~Floyd, Shengjie Lai, Corrine~Warren Ruktanonchai,
	Adam Sadilek, Pedro Rente-Lourenco, Xue Ben, Alessandra Carioli, Joshua
	Gwinn, JE~Steele, et~al.
	\newblock Assessing the impact of coordinated covid-19 exit strategies across
	europe.
	\newblock {\em Science}, 2020.
	
	\bibitem{wachter2006implementation}
	Andreas W{\"a}chter and Lorenz~T Biegler.
	\newblock On the implementation of an interior-point filter line-search
	algorithm for large-scale nonlinear programming.
	\newblock {\em Mathematical programming}, 106(1):25--57, 2006.
	
\end{thebibliography}
	\newpage
	
	\section*{Methods}
	\subsection*{Estimation of $I_{\max}$}
	
	The $I_{\max}$ values can be approximated with simple assumptions as seen in Table \ref{tab:Imax_cities} for a selection of major U.S. cities. These $I_{\max}$ values are a ratio of ICU beds to the population of people which require them. Following \cite{lin2020Daily}, we estimate that the probability of death conditioned on symptomatic infection is equal to $f_H \times (1-f_R) = 0.01134$, where the two parameters $f_H$ and $f_R$ were independently computed in \cite{perez2020serology} and \cite{richardson2020presenting}, respectively. Data  \cite{quah2020mortality,MortalityRate} 
	shows that the mortality rate for patients sent to ICU is between $30 \%$ and $40 \%$, thus, it is reasonable to assume that an overall fraction of infected people equal to $0.01134/0.35=0.0324$ needs ICU beds. 
	In the hypothetical situation that the population of an entire state contracts COVID-19 $3.24\%$  will require an ICU bed. The $\rho$ term is a modifier which denotes how many ICU beds are available to COVID-19 patients as some beds could be used for other reasons. Reasonably, the value of $\rho$ ranges from 2/3 to 1 in Table \ref{tab:Imax_cities}. $I_{\max}$ is then calculated as the number of available ICU beds (including the $\rho$ assumption) divided by number people which contract COVID-19 and also require an ICU bed (3.24\% assumption).
	
	\begin{table}[htbp]
		\centering
		\caption{$I_{\max}$ values for U.S. Cities}
		\begin{tabular}{lll}
			City  & $I_{\max}$ ($\rho = 2/3$) & $I_{\max}$ ($\rho = 1$) \\
			\midrule
			New York, NY & 0.88\%(5704/648000) & 1.32\% (8556/648000) \\
			Los Angeles, CA & 0.64\% (8241/1280124) & 0.97\%(12361/1280124) \\
			Chicago, IL & 1.00\%(4094/410508) & 1.50\%(6141/410508) \\
			Dallas, TX & 0.86\%(8109/939600) & 1.29\%(12163/939600) \\
			Houston, TX & 0.86\%(8109/939600) & 1.29\%(12163/939600) \\
			Riverside, CA & 0.64\% (8241/1280124) & 0.97\%(12361/1280124) \\
			Miami, FL & 0.98\% (6829/695952) & 1.47\%(10244/695952) \\
			Philadelphia, PA & 1.047\%(4342/414720) & 1.57\%(6513/414720) \\
			Atlanta, GA & 0.65\%(2249/344088) & 0.98\%(3374/344088) \\
			Phoenix, AZ & 0.55\%(1309/2358450) & 0.83\%(1963/2358450) \\
			Boston, MA & 0.4\%(893/223330) & 0.60\%(1340/223330) \\
			San Francisco, CA & 0.64\% (8241/1280124) & 0.97\%(12361/1280124) \\
			Seattle, WA & 0.46\%(1143/246888) & 0.69\%(1714/24688) \\
			Indianapolis, IN & 0.83\%(1811/218120) & 1.25\%(2716/218120) \\
			Detroit, MI & 0.71\%(2315/32358) & 1.07\%(3472/32358) \\
			Baltimore, MD & 0.40\%(893/223330) & 0.60\%(1340/223330) \\
			Denver, CO & 0.56\%(1039/186590) & 0.84\%(1559/186590) \\
			Portland, OR & 0.56\%(768/136660) & 0.84\%(1152/136660 \\
			Las Vegas, NV & 0.75\%(752/99792) & 1.13\%(1129/99792) \\
		\end{tabular}%
		\label{tab:Imax_cities}%
	\end{table}%
	
	For  all our numerical experiments, we consider two values for $I_{\max}$, one corresponding to $\rho=2/3$ and another one corresponding to $\rho=1$.
	
	\subsection*{Timescale for $T_{\text{herd}}$}
	
	We now attempt to answer the following question. By enforcing satisfaction of the constraint with the equal sign  $I_s(t)+I_{tp}(t)=I_{\max}$ and in the absence of a vaccine, how long would it take before herd immunity is achieved? By assuming long-term immunity of those recovered from the virus, we can expect herd immunity to arise when roughly 80\% of the population has been exposed \cite{randolph2020herd}. Consider for example the case of NYC for which from Table 3 we see that $I_{\max}$ is between 0.88 \% and 1.32 \%. 
	We assume an average hospitalization of 20 days \cite{liu2020risk}. That means that the time to achieve herd immunity $T_{\text{herd}}$ varies between 606 days=$(80 \times 20/(1.32 \times 2))$ and 909 days=$(80 \times 12/(0.88 \times 2))$. The factor of $2$ accounts for the fact that roughly one exposed person out of two develops symptoms. Analogously, for Chicago, we estimate $T_{\text{herd}}$ to vary between 533 days=$(80 \times 20/(1.5 \times 2))$ and 800 days=$(80 \times 20/(1 \times 2))$.
	From these back-of-the-envelope calculations we see that
	the timescale over which herd immunity can be achieved without violating the $I_{\max}$ constraint appears to be quite long and definitely longer than the timescale over which vaccines have become available.

	\section*{Acknowledgements}
	The authors would like to thank Bill Hlavacek, Isaac Klickstein, Neethi Thevan T., and Enrico Del Frate for insightful discussions. Y.T.L. thanks the support from the Laboratory Directed Research and Development Program at Los
	Alamos National Laboratory (Project XX01) through the Center for Nonlinear Studies(CNLS). 
	
	\pagebreak
	\begin{center}
		\textbf{\large Supplemental Materials: Data-driven Optimized Control of the COVID-19 Epidemics}
	\end{center}

\setcounter{equation}{0}
\setcounter{figure}{0}
\setcounter{table}{0}
\setcounter{page}{1}

\makeatletter
\renewcommand{\theequation}{S\arabic{equation}}
\renewcommand{\thefigure}{S\arabic{figure}}

\newpage

\section*{Supplementary Note 1: Model Scheme}

The compartmental model described by Eq.\ 2 in the main text can be visualized in Figure 1 below.

\begin{figure}[h!]
	\centering
	\includegraphics[width = 170mm]{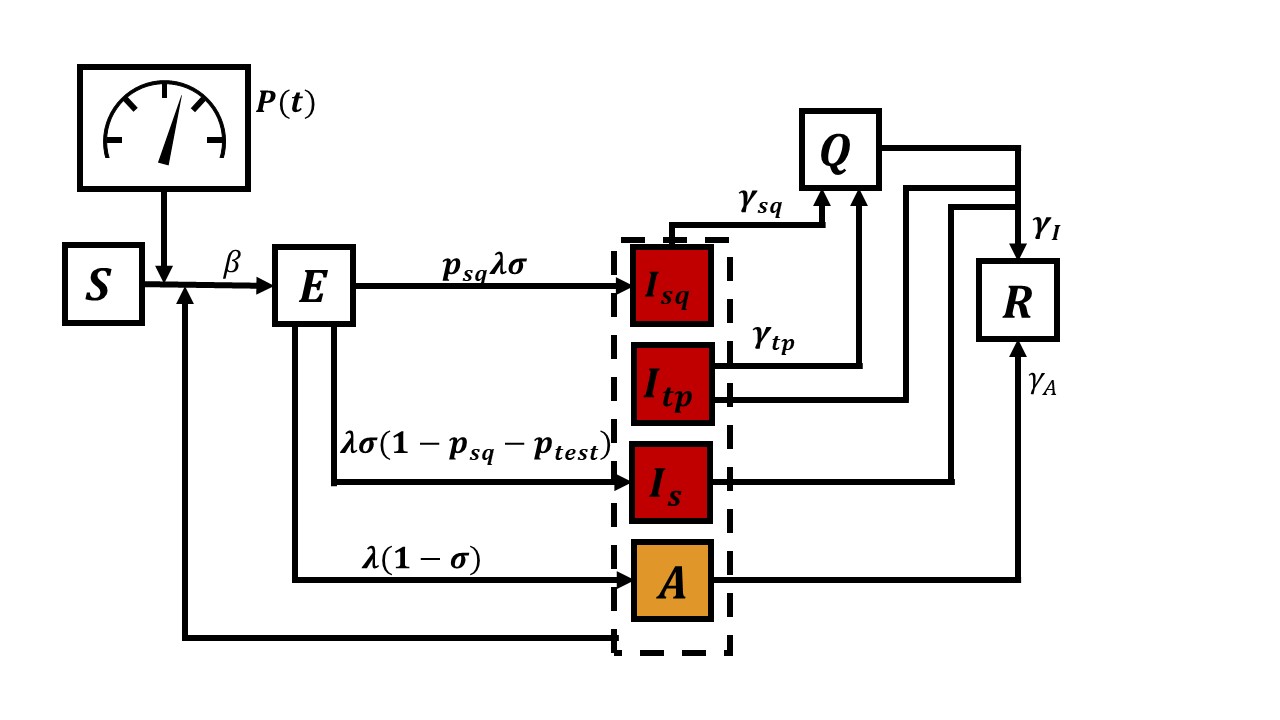}
	\caption{Compartmental model}
	\label{fig:schematicDiagramEQ2}
\end{figure}

\section*{Supplementary Note 2: Robustness of the optimal control with respect to variations of the parameters $c_q$ and $I_{\max}$}

\begin{figure}[h!]
	\centering
	\includegraphics[width=\textwidth]{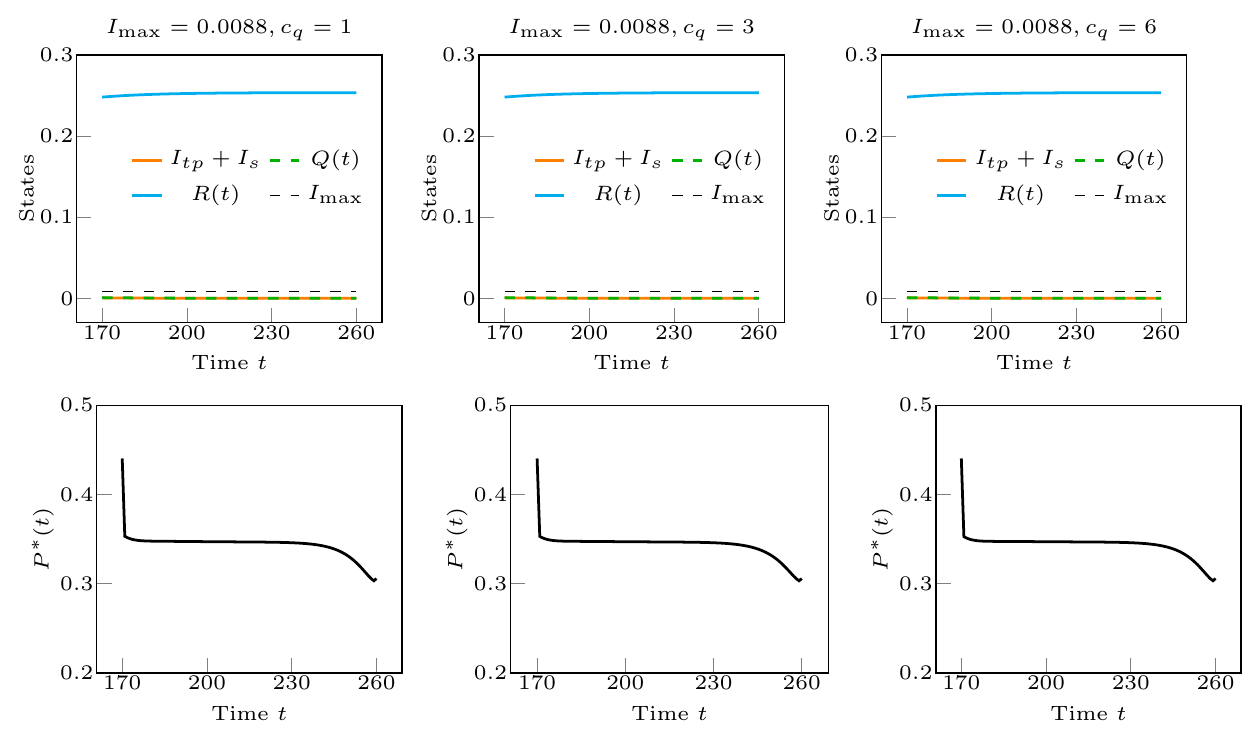}
	\caption{NYC Scenario. $I_{\max} = 0.0088$, minimum of the range.}
	\label{fig:NYC_Imax_0088}
\end{figure}

\begin{figure}
	\centering
	\includegraphics[width=\textwidth]{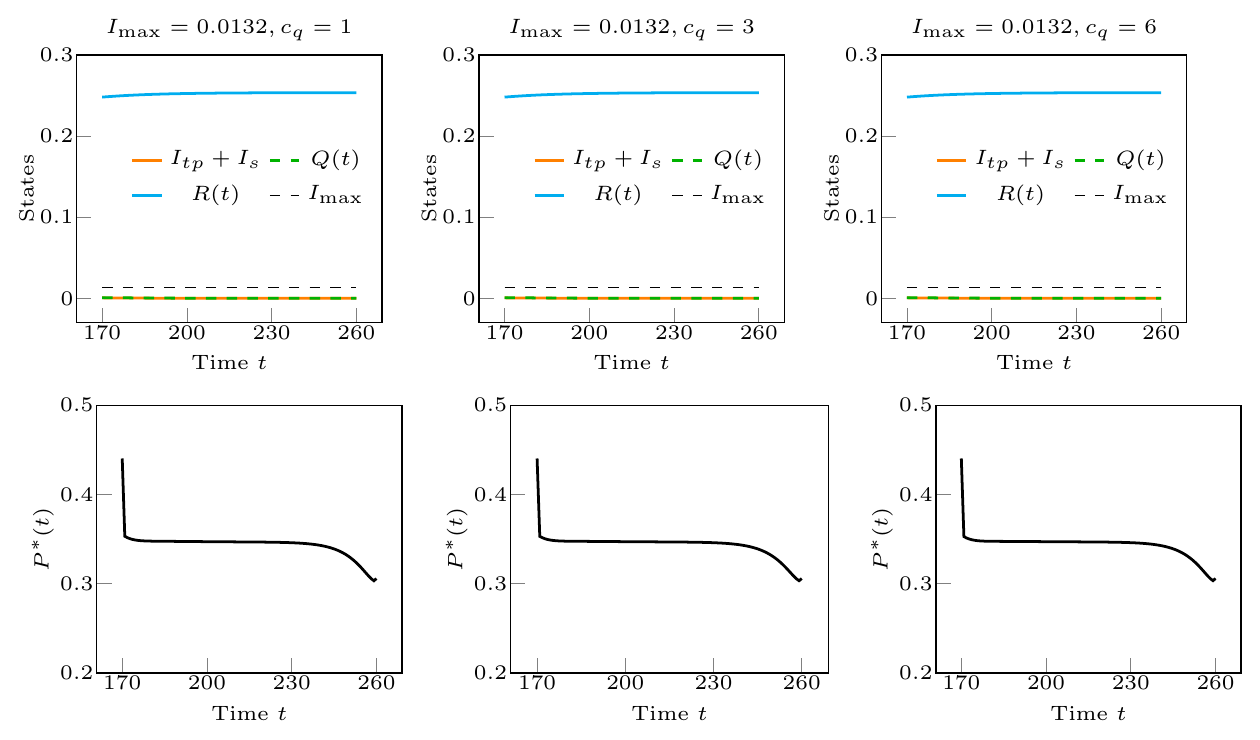}
	\caption{NYC Scenario. $I_{\max} = 0.0132$, maximum of the range.}
	\label{fig:NYC_Imax_0132}
\end{figure}

\begin{figure}
	\centering
	\includegraphics[width=\textwidth]{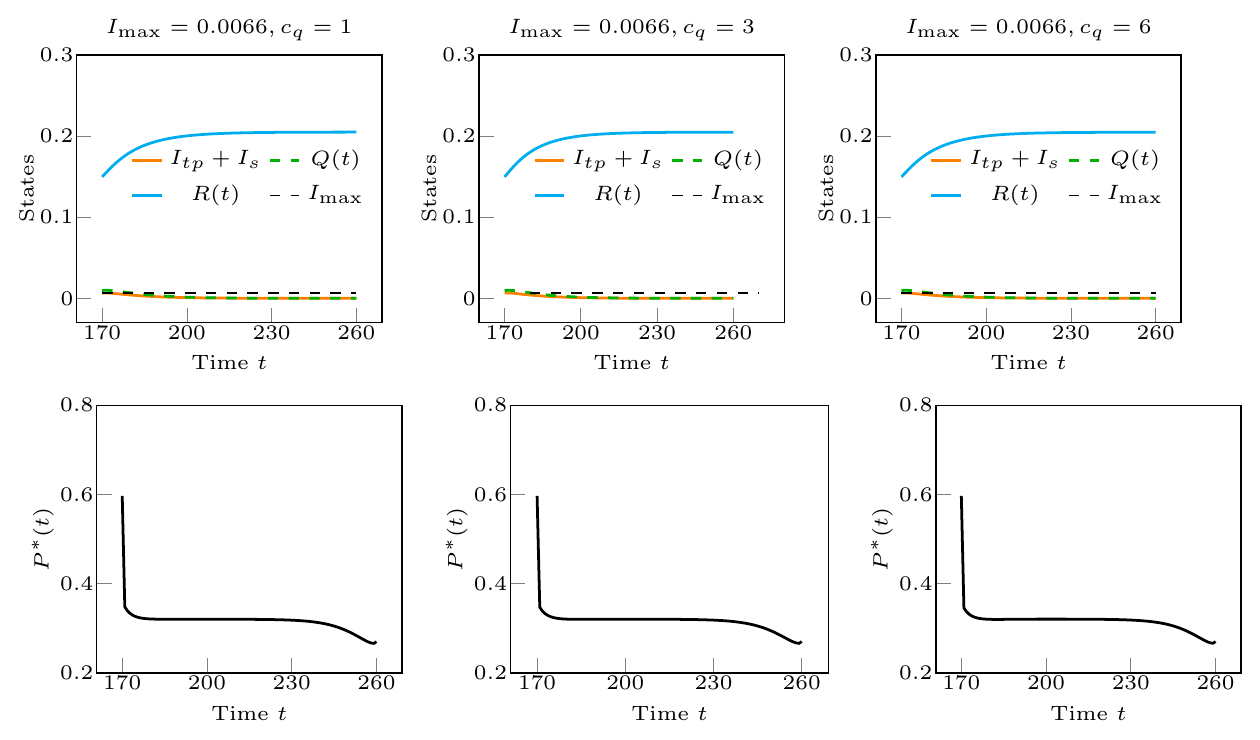}
	\caption{LA Scenario. $I_{\max} = 0.0066$, minimum of the range.}
	\label{fig:LA_Imax_0066}
\end{figure}

\begin{figure}
	\centering
	\includegraphics[width=\textwidth]{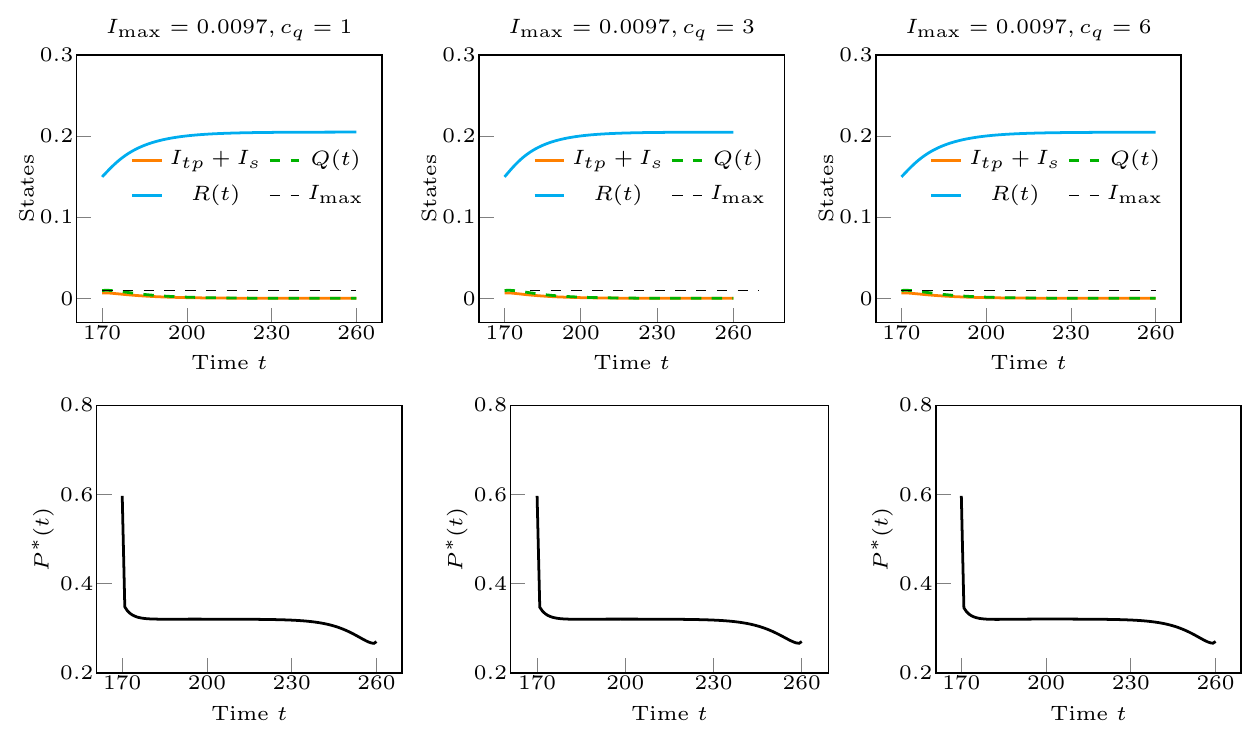}
	\caption{LA Scenario. $I_{\max} = 0.0097$, maximum of the range.}
	\label{fig:LA_Imax_0097}
\end{figure}

\begin{figure}
	\centering
	\includegraphics[width=\textwidth]{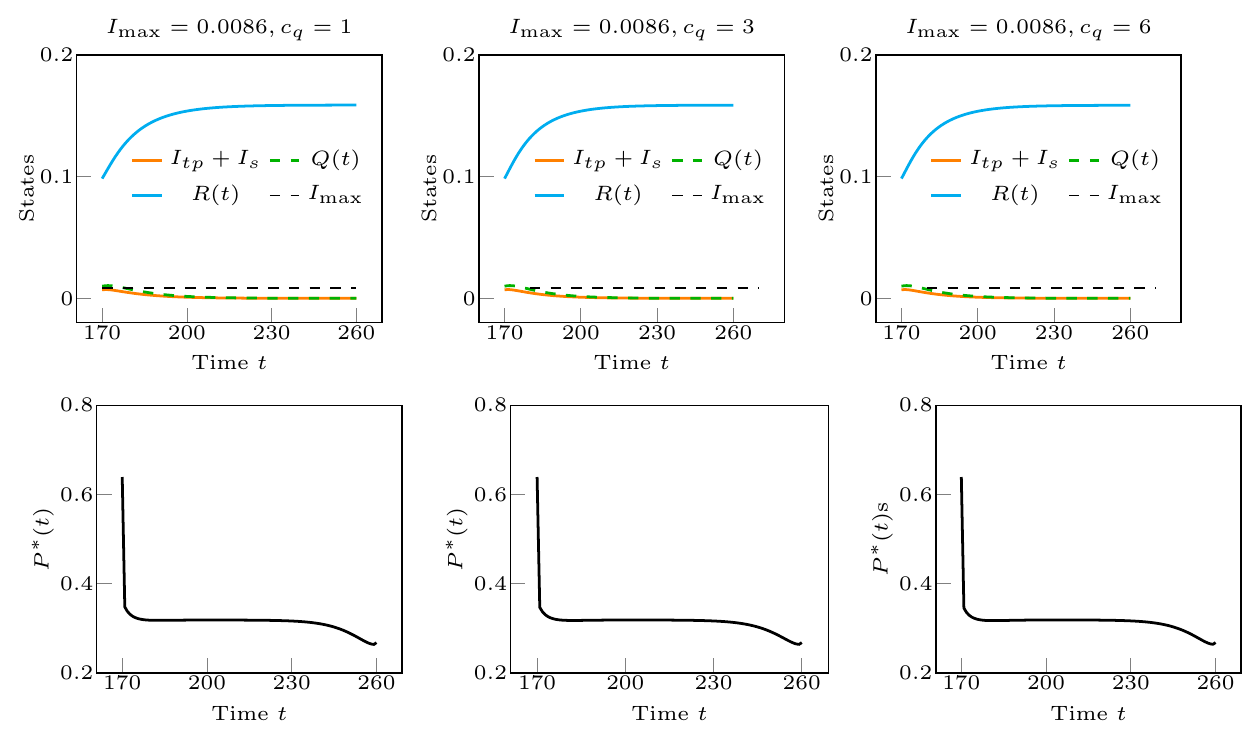}
	\caption{Houston Scenario. $I_{\max} = 0.0086$, minimum of the range.}
	\label{fig:Houston_Imax_0086}
\end{figure}

\begin{figure}
	\centering
	\includegraphics[width=\textwidth]{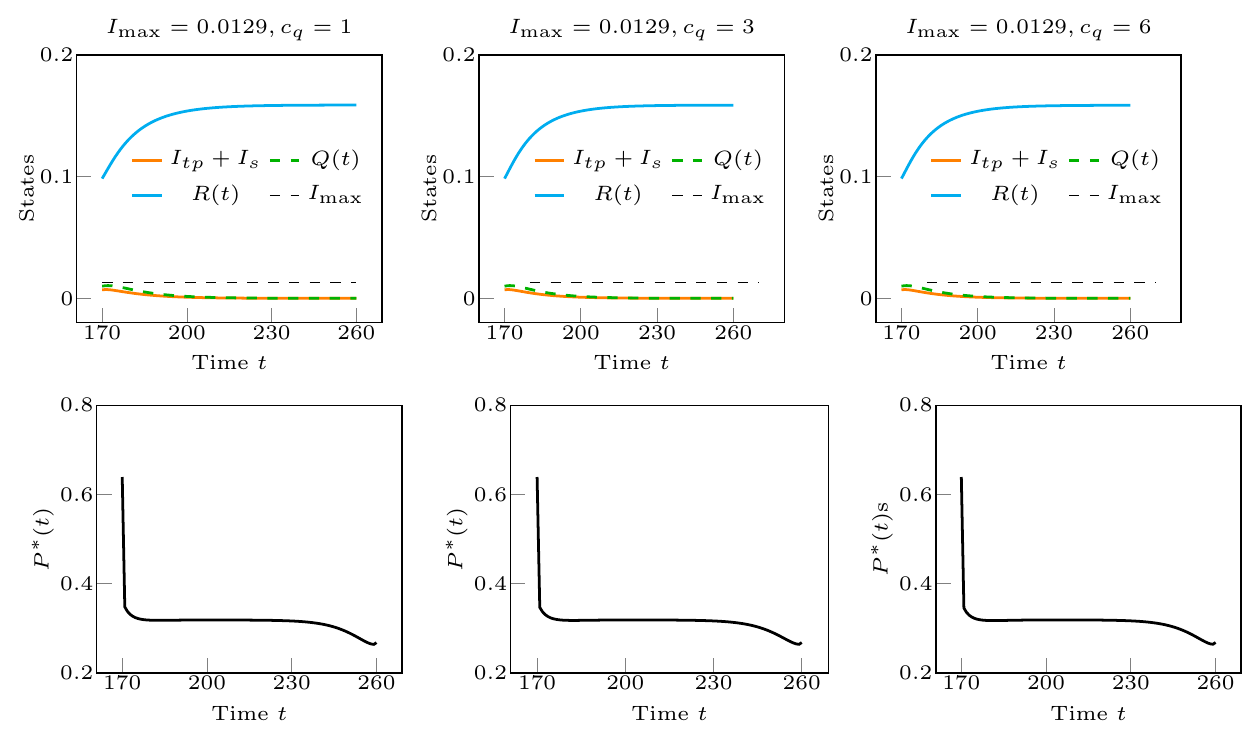}
	\caption{Houston Scenario. $I_{\max} = 0.0129$, maximum of the range.}
	\label{fig:Houstn_Imax_0129}
\end{figure}

\begin{figure}
	\centering
	\includegraphics[width=\textwidth]{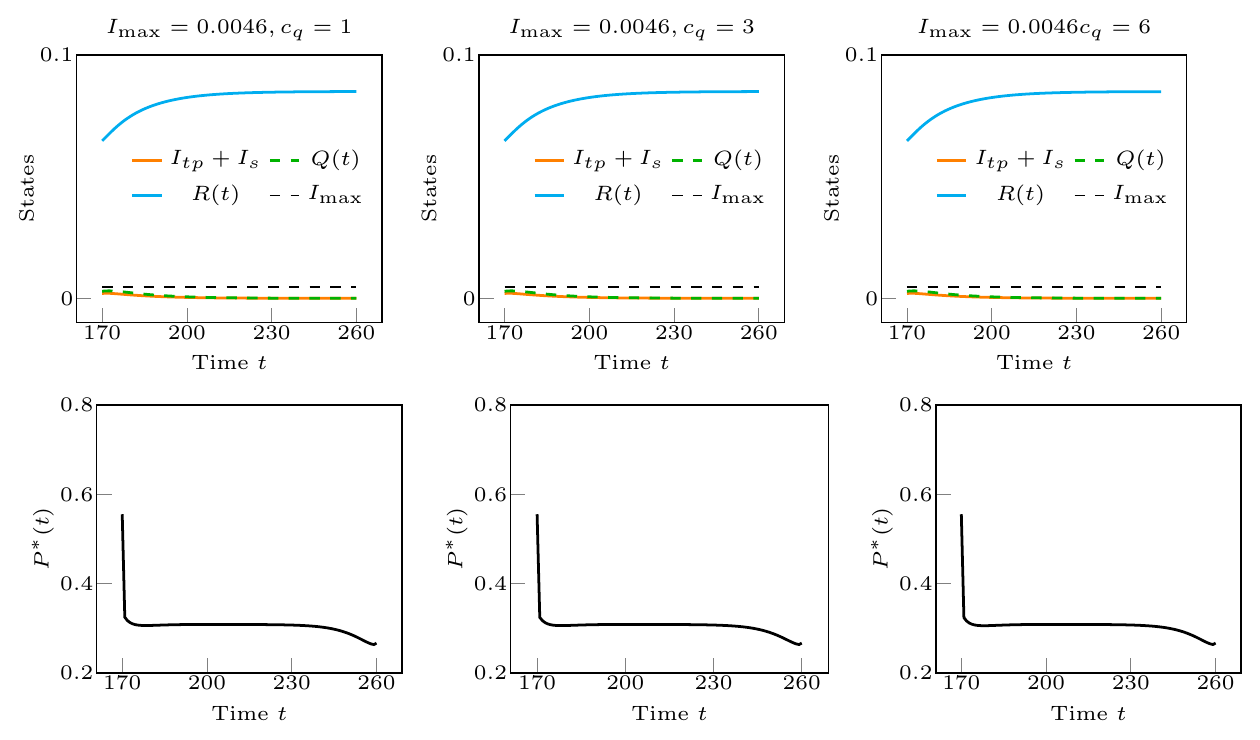}
	\caption{Seattle Scenario. $I_{\max} = 0.0046$, minimum of the range.}
	\label{fig:Seattle_Imax_0046}
\end{figure}

\begin{figure}
	\centering
	\includegraphics[width=\textwidth]{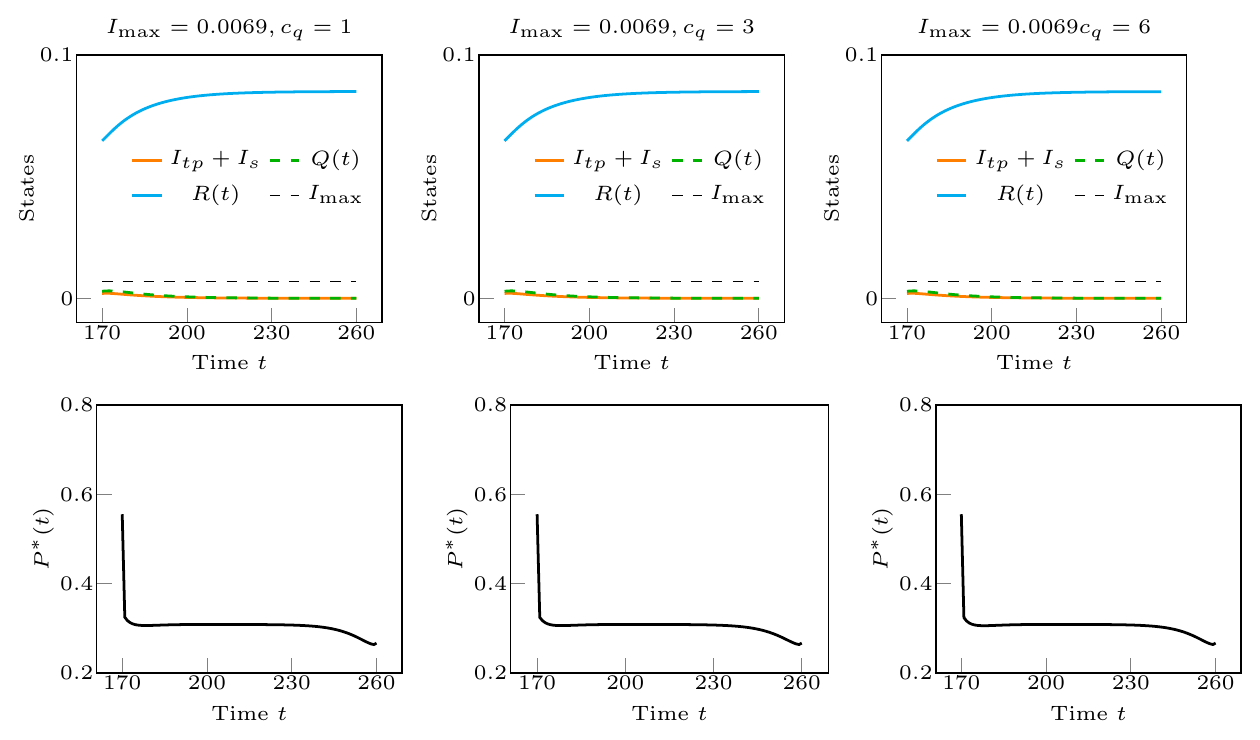}
	\caption{Seattle Scenario. $I_{\max} = 0.0069$, maximum of the range.}
	\label{fig:Seattle_Imax_0069}
\end{figure}

We now investigate the effect of varying the parameters $c_q$ and $I_{\max}$ on the optimal control solutions. Our main result is that we find the optimal control solutions to be robust to variations in both $c_q$ and $I_{\max}$. We choose two values of $I_{\max}$, a value corresponding to the lower value shown in Table 3 of the main manuscript ($\rho=2/3$) and a value corresponding to the larger value shown in Table 3 of the main manuscript ($\rho=1$).
In Figs.\ \ref{fig:NYC_Imax_0088} and  \ref{fig:NYC_Imax_0132}, the states and the optimal controls are plotted for the NYC scenario, for $I_{\max}$ = 0.0088 and 0.0132, respectively.
In Figs.\ \ref{fig:LA_Imax_0066} and  \ref{fig:LA_Imax_0097}, the states and the optimal controls are plotted for the LA scenario, for $I_{\max}$ = 0.0066 and 0.0097, respectively.
In Figs.\ \ref{fig:Houston_Imax_0086} and \ref{fig:Houstn_Imax_0129}, the states and the optimal controls are plotted for the Houston scenario, for $I_{\max}$ = 0.0088 and 0.0129, respectively.
In Figs.\ \ref{fig:Seattle_Imax_0046} and \ref{fig:Seattle_Imax_0069}, the states and the optimal controls are plotted for the Houston scenario, for $I_{\max}$ = 0.0046 and 0.0069, respectively.

All our results in Figs. \ref{fig:NYC_Imax_0088},  \ref{fig:NYC_Imax_0132}, \ref{fig:LA_Imax_0066},  \ref{fig:LA_Imax_0097},
\ref{fig:Houston_Imax_0086}, \ref{fig:Houstn_Imax_0129},
\ref{fig:Seattle_Imax_0046} and \ref{fig:Seattle_Imax_0069} show that the optimal control solutions are robust to variations in $c_q$.

\begin{figure}[h!]
	\centering
	\includegraphics[width=0.9\textwidth]{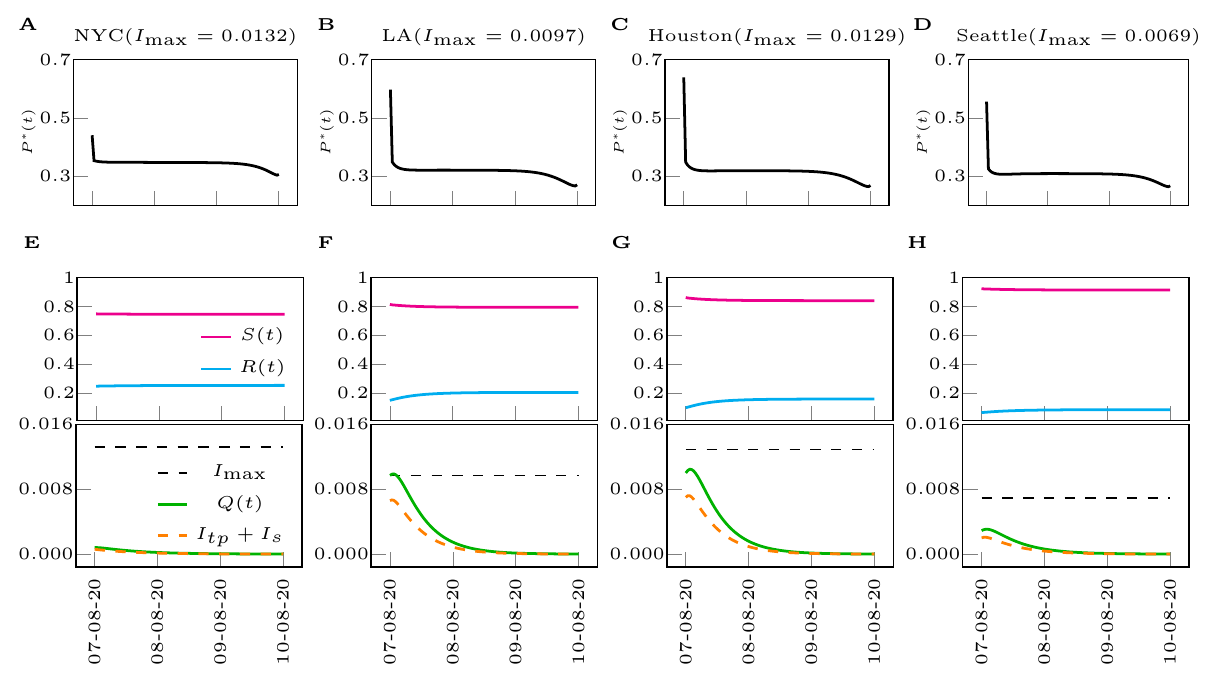}
	\caption{(A-D) Optimal control strategies for the metropolitan cities NYC, LA, Houston, Seattle, respectively. (E-F) Time evolution of the states subject to the optimal control inputs in (A-D). $I_{\max}$ are chosen as the maximum of the range in Table 3 of the main manuscript ($\rho=1$) and $c_q = 1$.  The legends in (F-H) are same as the legend in (E). }
	\label{fig:states_controls_Imax_max}
\end{figure}

Figure \eqref{fig:states_controls_Imax_max} is analogous to Figure 3 shown in the main manuscript but for the case that $I_{\max}$ are chosen as the maximum values in Table 3 of the main manuscript  ($\rho=1$.) As can be seen the solutions are the same as in Figure 3 of the main manuscript.

\section*{Supplementary Note 3: Effects of varying the terminal suppression constraint $\epsilon$}

Figures 11, 12, 13, and 14 show the effects of varying the final suppression constraint $\epsilon$ on the optimal control solutions, for the cases of the Metropolitan Statistical Areas of NYC, LA, Houston, and Seattle, respectively. For all cities, we see that for large enough $\epsilon$, solutions of type 2 emerge, for which $I_{tp}(t)+I_s(t)=I_{max}$ for certain times $t$. More specifically, this is seen for $\epsilon=3.16 \times 10^{-4}$ and $\epsilon= 10^{-3}$ in Figs.\ 11, 12, and 13 and for $\epsilon= 10^{-3}$ in Fig.\ 14.

\begin{figure}[h!]
	\centering
	\includegraphics[width=\textwidth]{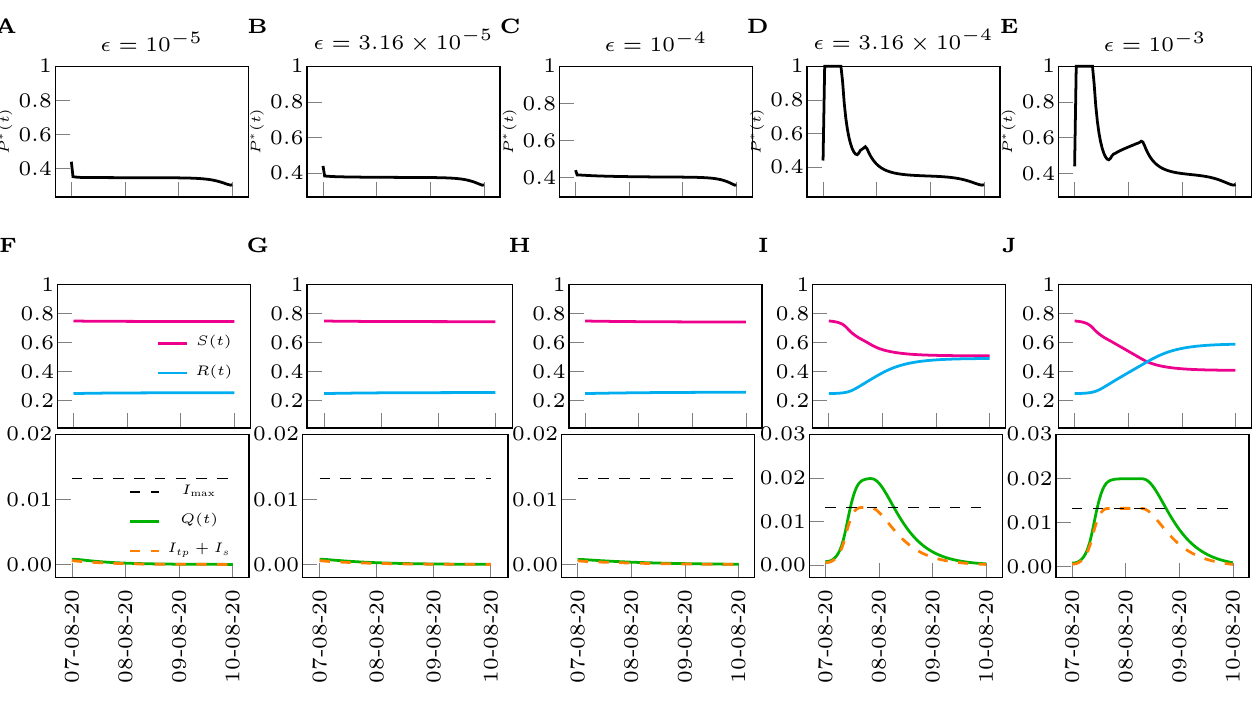}
	\caption{(A-E) Optimal control strategies for the NYC for different values of the parameter $\epsilon$. (F-J) Evolutions of the states subject to the optimal control inputs in (A-E). $I_{\max}$ are chosen from the maximum range of Table 3 of the main manuscript, $\rho=1$.}
	\label{fig:NYC_vary_ep}
\end{figure}

\begin{figure}
	\centering
	\includegraphics[width=\textwidth]{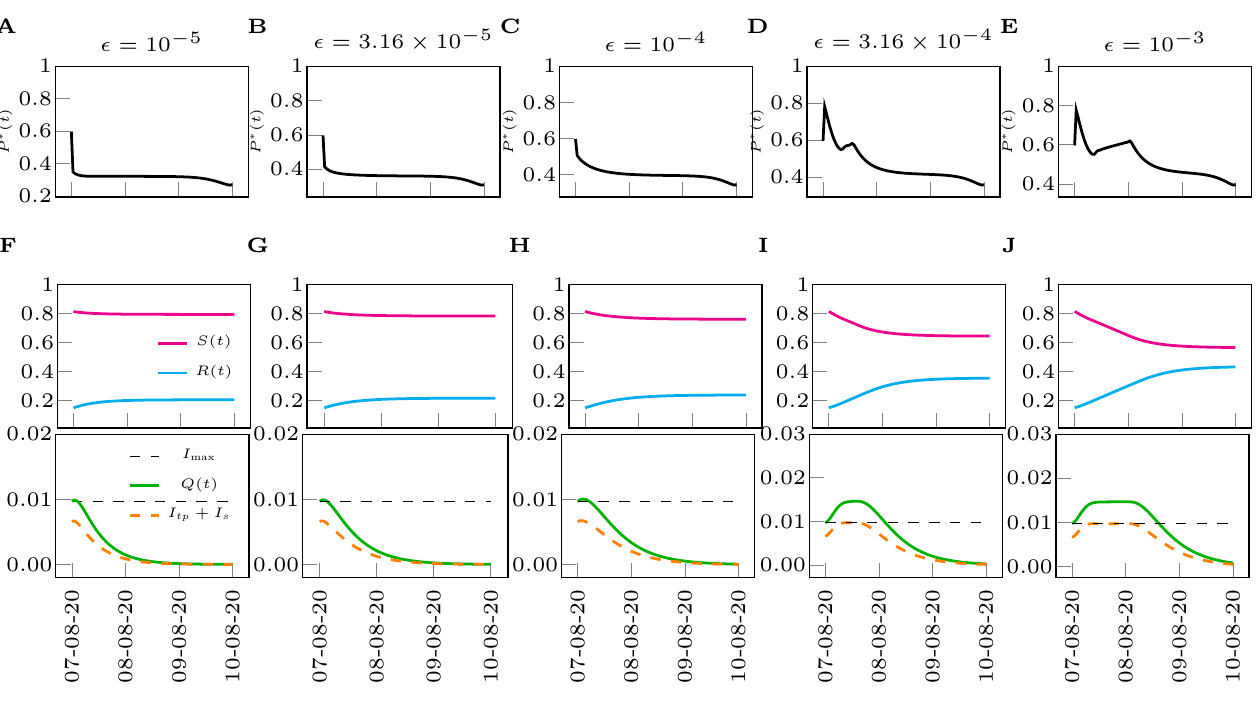}
	\caption{(A-E) Optimal control strategies for LA for different values of the parameter $\epsilon$. (F-J) Evolutions of the states subject to the optimal control inputs in (A-E). $I_{\max}$ are chosen from the maximum range of Table 3 of the main manuscript, $\rho=1$.}
	\label{fig:LA_vary_ep}
\end{figure}

\begin{figure}
	\centering
	\includegraphics[width=\textwidth]{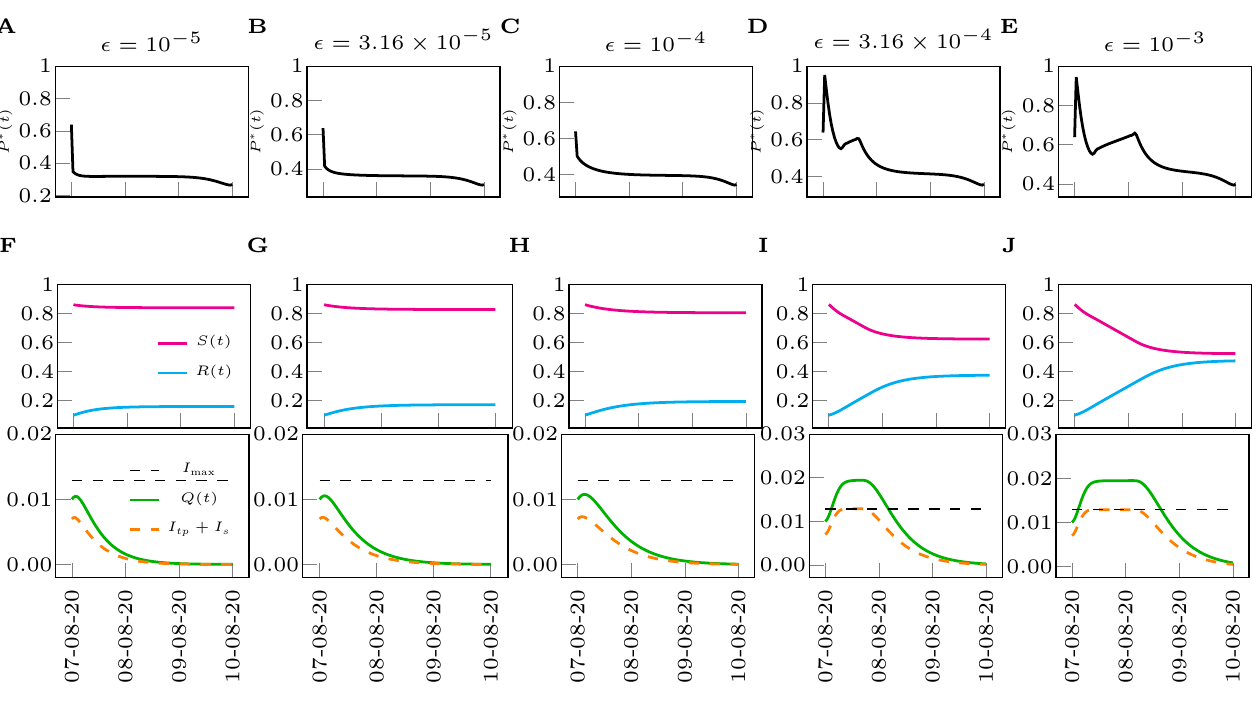}
	\caption{(A-E) Optimal control strategies for Houston for different values of the parameter $\epsilon$. (F-J) Evolutions of the states subject to the optimal control inputs in (A-E). $I_{\max}$ are chosen from the maximum range of Table 3 of the main manuscript, $\rho=1$.}
	\label{fig:Houston_vary_ep}
\end{figure}

\begin{figure}
	\centering
	\includegraphics[width=\textwidth]{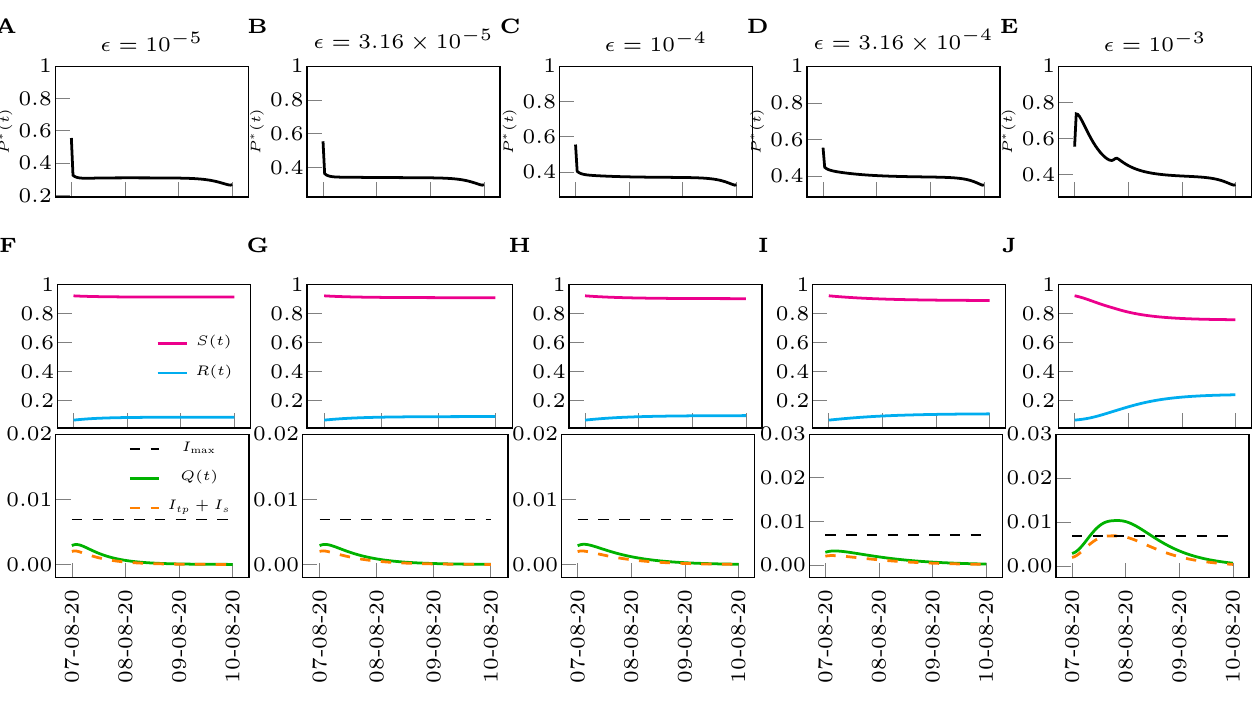}
	\caption{(A-E) Optimal control strategies for Seattle for values of the parameter $\epsilon$. (F-J) Evolutions of the states subject to the optimal control inputs in (A-E). $I_{\max}$ are chosen from the maximum range of Table 3 of the main manuscript, $\rho=1$.}
	\label{fig:Seattle_vary_ep}
\end{figure}

\newpage

\section*{Supplementary Note 4: Effects of varying $t_f$}

\begin{figure}[h!]
	\centering
	\includegraphics[width=\textwidth]{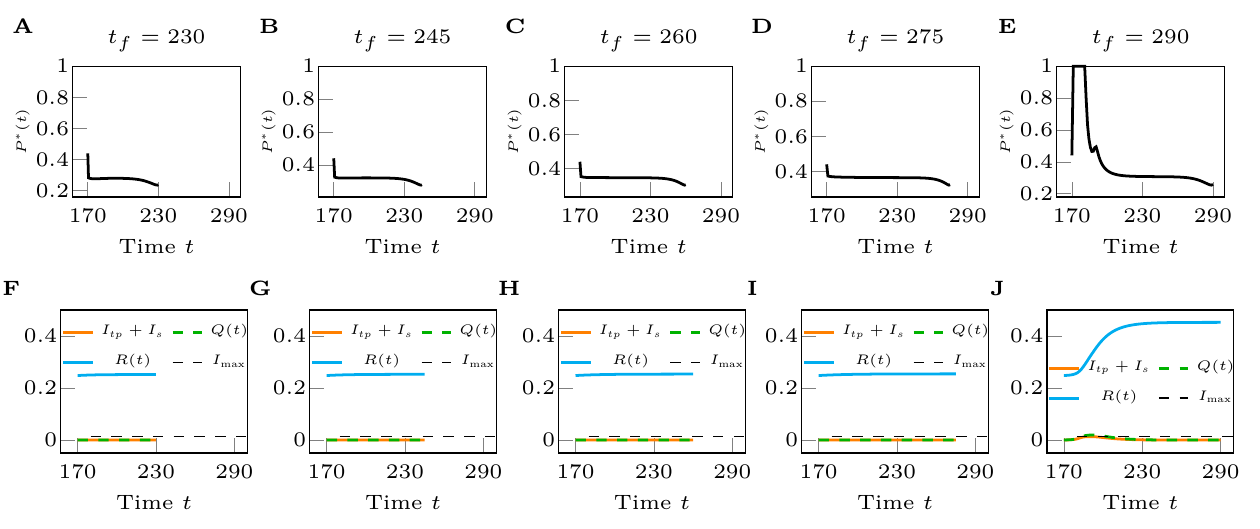}
	\caption{(A-E) Optimal control strategies for the NYC for different final times $t_f$. (F-J) Evolutions of the states subject to the optimal control inputs. $I_{\max}$ is chosen as the maximum of the range in Table 3 of the main manuscript ($\rho=1$).}
	\label{fig:NYC_vary_time}
\end{figure}

\begin{figure}
	\centering
	\includegraphics[width=\textwidth]{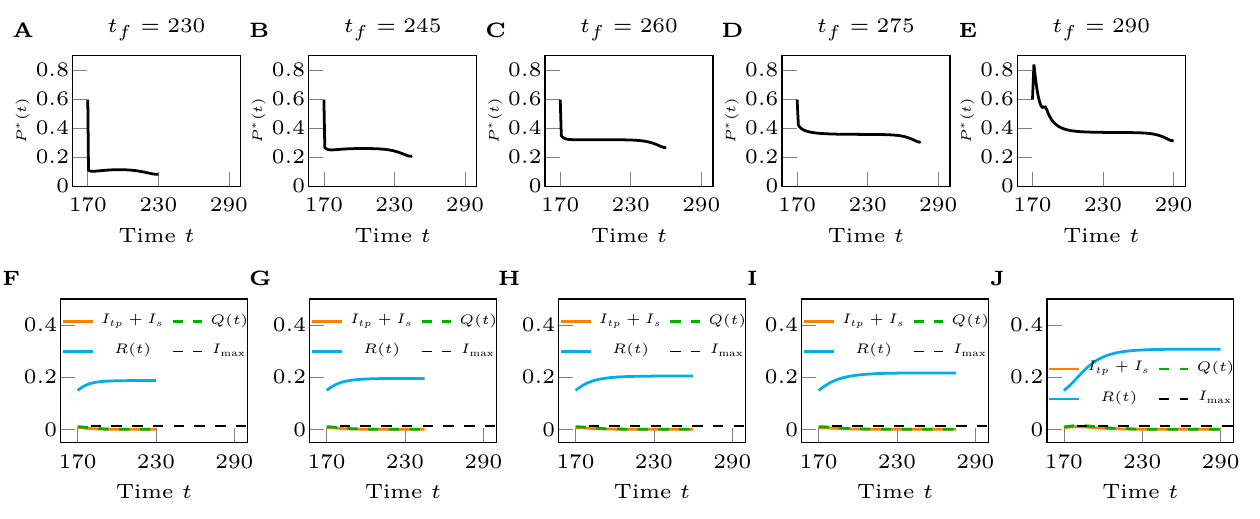}
	\caption{(A-E) Optimal control strategies for LA for different final times $t_f$. (F-J) Evolutions of the states subject to the optimal control inputs. $I_{\max}$ are chosen as the maximum of the range in Table 3 of the main manuscript ($\rho=1$) and  $\epsilon = 10^{-5}$.}
	\label{fig:LA_vary_time}
\end{figure}

\begin{figure}
	\centering
	\includegraphics[width=\textwidth]{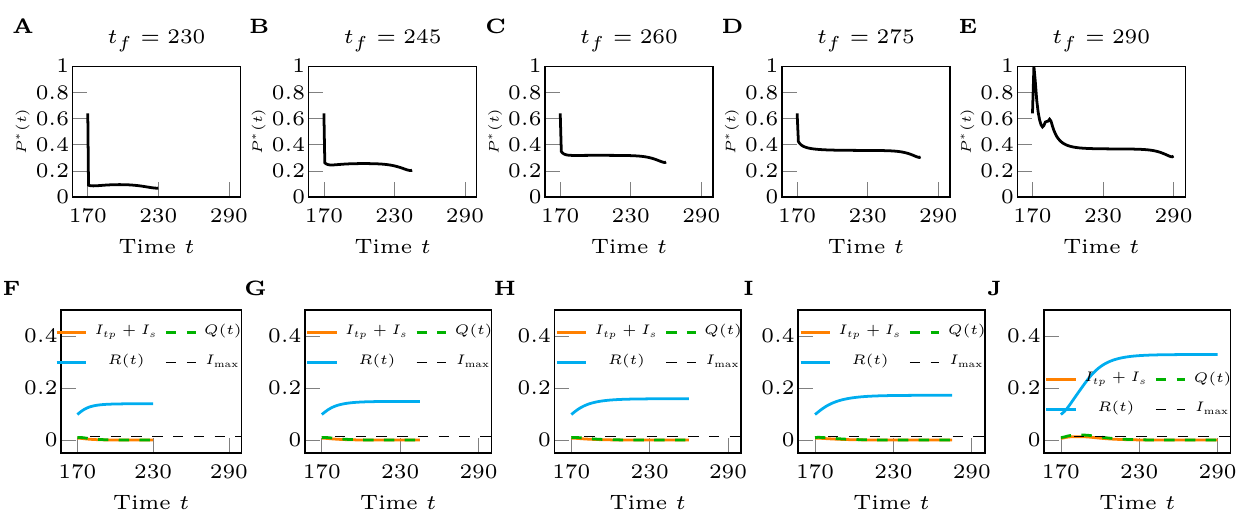}
	\caption{(A-E) Optimal control strategies for Houston for different final times $t_f$. (F-J) Evolutions of the states subject to the optimal control inputs. $I_{\max}$ are chosen from the maximum range of Table 3 of the main manuscript ($\rho=1$) and  $\epsilon = 10^{-5}$.}
	\label{fig:Houston_vary_time}
\end{figure}

\begin{figure}
	\centering
	\includegraphics[width=\textwidth]{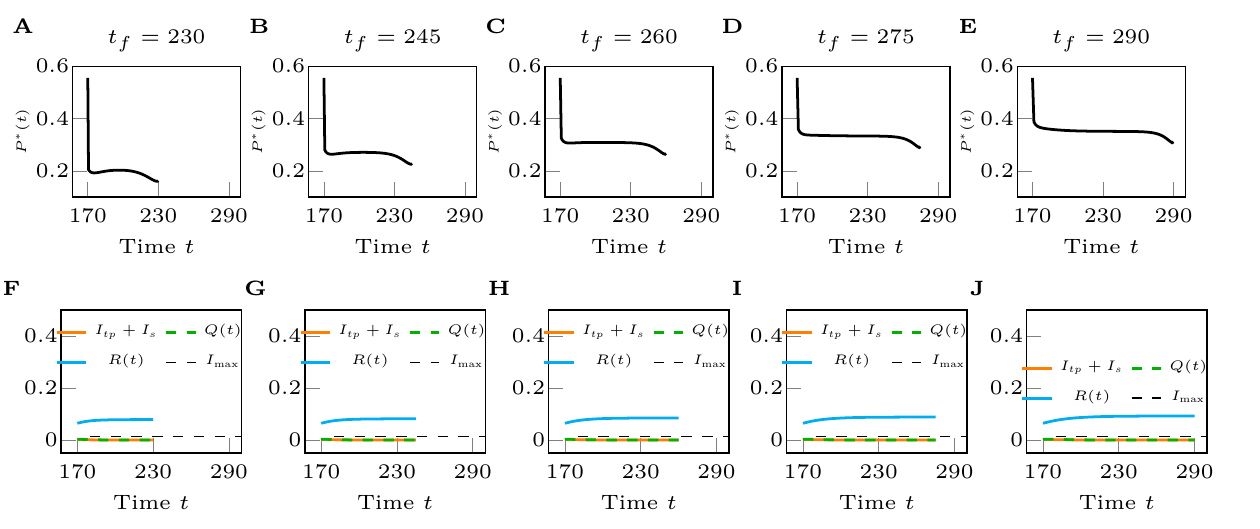}
	\caption{(A-E) Optimal control strategies for Seattle for different final times $t_f$. (F-J) Evolutions of the states subject to the optimal control inputs. $I_{\max}$ are chosen as the maximum of the range in Table 3 of the main manuscript  ($\rho=1$) and  $\epsilon = 10^{-5}$.}
	\label{fig:Seattle_vary_time}
\end{figure}

\newpage
\section*{Supplementary Note 5: Detailed comparison of different cities}

\begin{figure}[h!] \label{latf}
	\centering
	\includegraphics[width=0.9\textwidth]{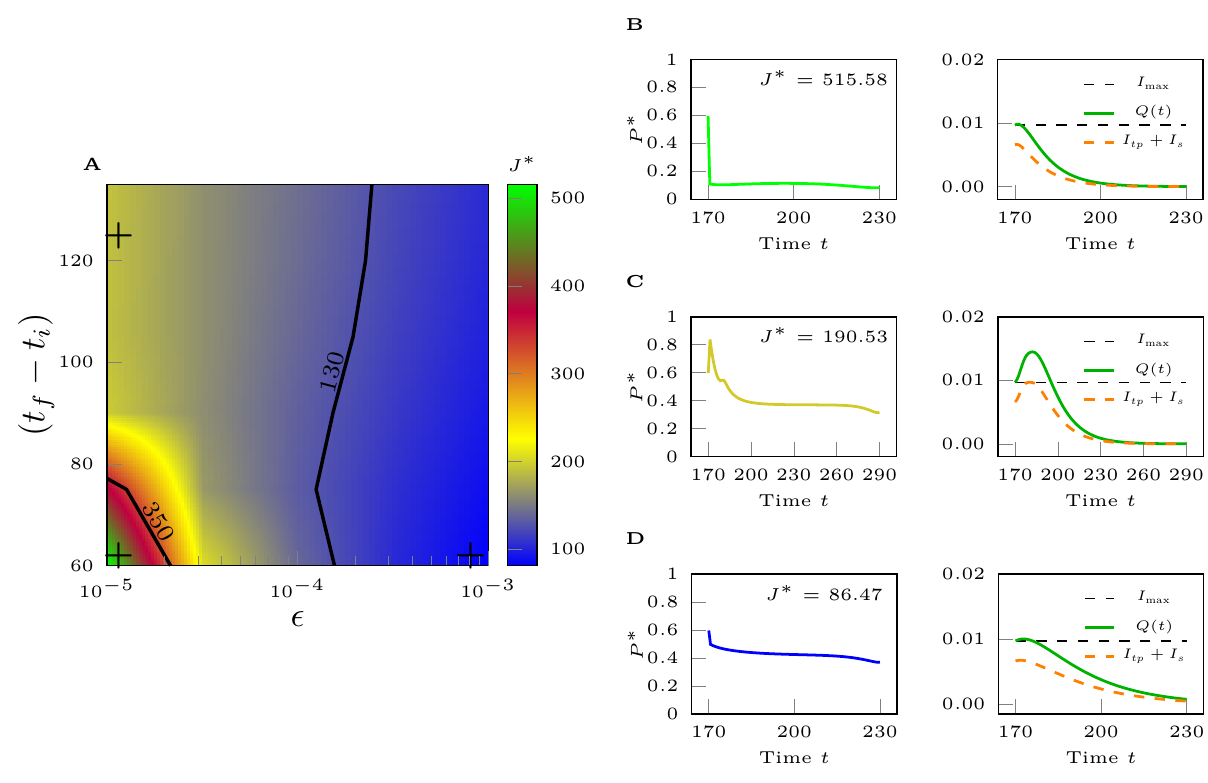}
	\caption{(A) The optimal cost $J^{\ast}$ in the $(t_f - t_i)$,  $\epsilon$ plane. The parameters correspond to the {Los Angeles Metropolitan Statistical Area}. $I_{\max}$ is chosen as the maximum of the range in Table 3 of the main manuscript . Type 1 solutions (in green)
		are more expensive than type 2 solutions (in blue.) The regions in yellow/red correspond to the transition between the two types of solutions. (B-D) Time evolutions of the optimal control inputs and states for three different points of the $(t_f - t_i)$-$\epsilon$ plane, points shown as
		plus signs in (A). The parameter $c_q$  and $c_p$ are both set to 1.}
\end{figure}

\begin{figure} \label{latf_seattle}
	\centering
	\includegraphics[width=0.9\textwidth]{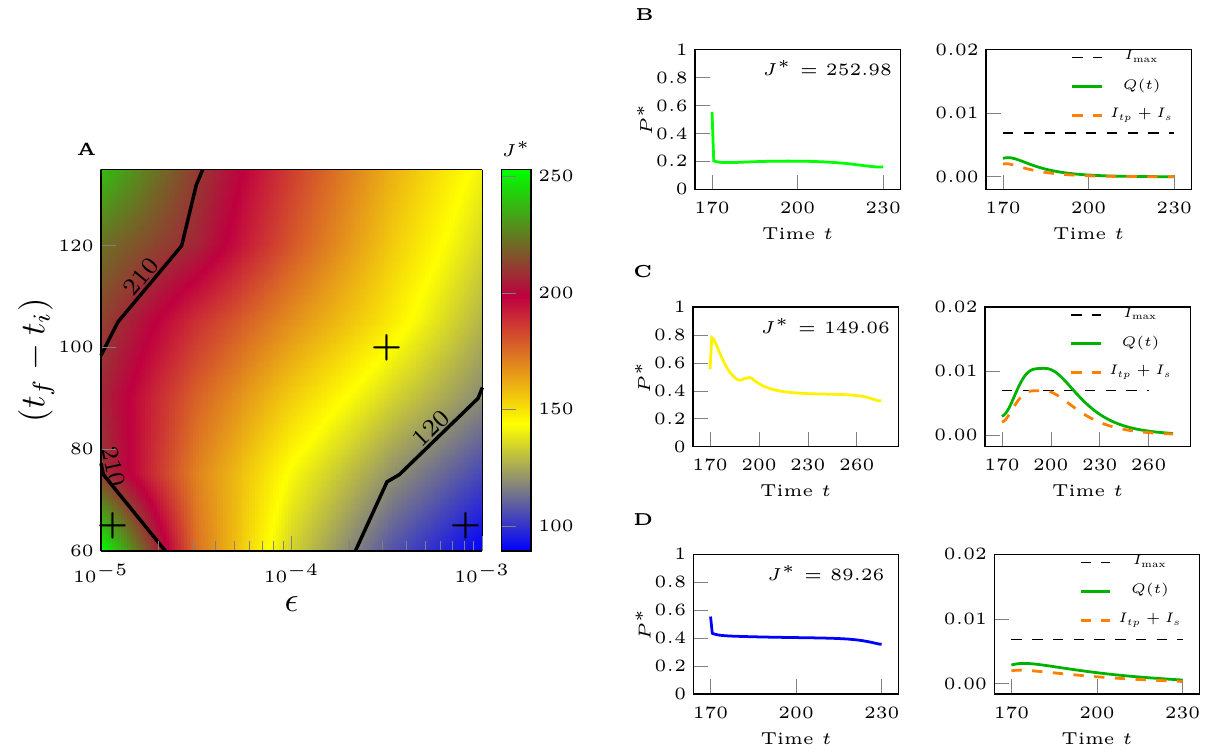}
	\caption{(A) The optimal cost $J^{\ast}$ in the $(t_f - t_i)$,  $\epsilon$ plane. The parameters correspond to the Seattle Metropolitan Statistical Area. $I_{\max}$ is chosen as the maximum of the range in Table 3 of the main manuscript . Type 1 solutions (in red) are more expensive than type 2 solutions (in blue.) The regions in yellow and red correspond to the transition between the two types of solutions. (B-D) Time evolutions of the optimal control inputs and states for three different points of the $(t_f - t_i)$-$\epsilon$ plane, points shown as plus signs in (A). The parameter $c_q$  and $c_p$ are both set to 1.}
\end{figure}

\begin{figure}
	\centering
	\includegraphics[width=0.9\textwidth]{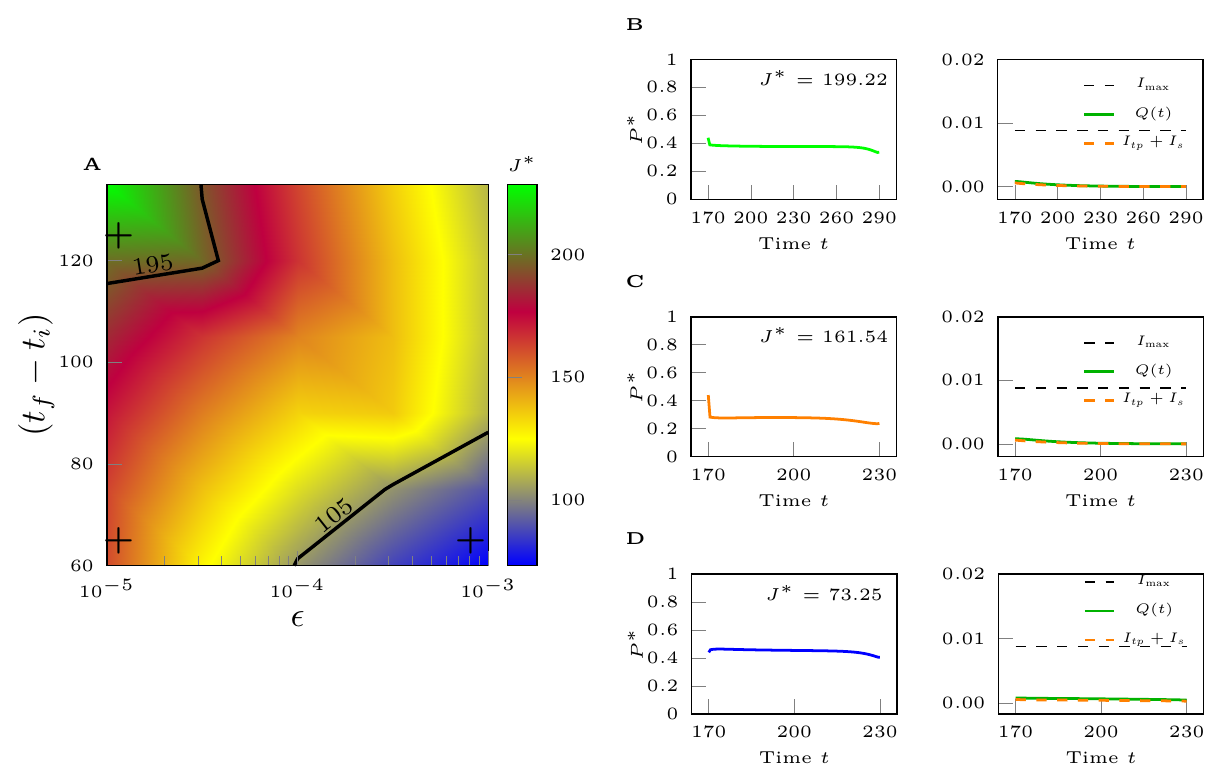}
	\caption{(A) The optimal cost $J^{\ast}$ in the $(t_f - t_1)$,  $\epsilon$ plane. The parameters correspond to the MSA of NYC. $I_{\max}$ is chosen as the maximum of the range in Table 3 of the main manuscript . Type 1 solutions (in red) are more expensive than type 2 solutions (in blue.) The regions in yellow and red correspond to the transition between the two types of solutions. (B-D) Time evolutions of the optimal control inputs and states for three different points  of the $(t_f - t_i)$-$\epsilon$ plane, points shown as plus signs in (A). The parameters $c_p$ and $c_q$ are both set to 1.}
	\label{fig:NYC_tf_ep}
\end{figure}

\section*{Supplementary Note 6: Herd Immunity Solutions}

\begin{figure}[h!] 
	\centering
	\includegraphics[width=0.85\textwidth]{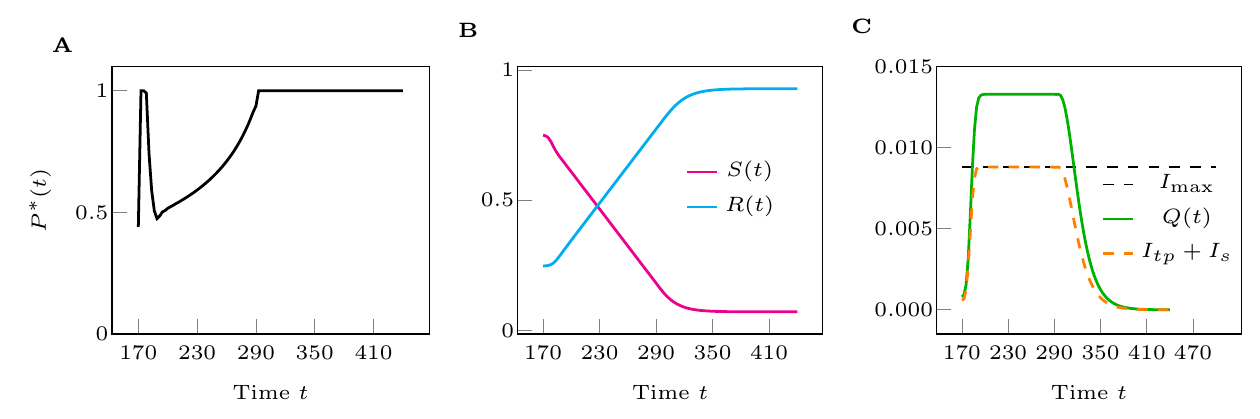}
	\caption{(A)  Optimal control strategy for NYC for $t_f = 440$. (B-C) Evaluations of the states to the optimal control input.  $I_{\max}$ is chosen as the minimum value in Table 3 of the main manuscript  $(\rho = 2/3)$ and  $\epsilon = 10^{-3}$. The parameters $c_p $ and $c_q $ are both set to 1.}
	\label{HI}
\end{figure}

As stated in the main manuscript, herd immunity solutions arise when the control horizon is very large. Figure \ref{HI} shows an example of such solution for the case of NYC, when the final time was set equal to $t_f = 440$.  These solutions are  characterized by three phases: (I)
$\dot{I} (t) > 0$ and $I (t) < I_{\max}$, $t \in [t_i,\tau_1)$, (II)
$\dot{I} (t) = 0$ and $I (t) = I_{\max}$, $t \in [\tau_1, \tau_2]$ and (III) $\dot{I} (t) < 0$ and $I (t) < I_{\max}$, $t \in (\tau_2, t_f ]$, $t_i \leq \tau_1 \leq \tau_2 \leq t_f$. The second phase has a natural interpretation: one of the objectives is to minimize the usage of social distancing while in the presence of the path constraint $I(t)\leq I_{\max}$, which results in setting $I=I_{\max}$ any time that the number of infected in the absence of controls would exceed $I_{\max}$. From simulations we see that this constant infection state corresponds to approximately setting $\dot{E}=\dot{A}=\dot{Q}=0$, $t \in [\tau_1, \tau_2]$. 
Then, the optimal $P^*(t)$ typically has a V-shape (see panel A of Fig.\ \ref{HI}), with stricter measures of social distancing only in the central phase.

\section*{Supplementary Note 7: Implementation of non-optimal control solutions}

We have provided an approach to robustly  minimize the effects on the economy of social distancing measures in the presence of relevant constraints. However, it is possible that a number of considerations may limit the implementation of such optimized interventions. 
We are interested in the effects of implementation of non-optimal controls solutions.
We thus consider application 
of a variation of the optimal solution 
\begin{equation}
	\tilde{P}(t)=\min\{1, (1+\alpha) P^\ast(t) \},
\end{equation}
$\alpha>0$, and analyze violations of the constraints $I_{\max}$ and $\epsilon$ as $\alpha$ is varied. Increasing values of $\alpha$ indicate stronger deviations of the control action from the optimal one. Table 1 summarizes application of such non-optimal interventions in all four cities, with $\alpha$ varying from $0.1$ to $0.5$. The letter `S' stands for constraint satisfied, otherwise we report the percentage by which the constraint is violated. We see that  the constraint on $I_{\max}$ remains always satisfied (this is expected as this constraint is not dominant) but strong violations of the suppression constraint $\epsilon$ are otherwise recorded.
From Table \ref{tab:non-optimal}
we see that for $\alpha=0.5$ the constraint on $\epsilon$ in the four areas of interest is violated by an amount that varies from roughly $4000\%$ for Houston to $ 80000\%$ for NYC (which corresponds to a fraction of infected people at the final time in the order of $10^{-2}$.) These
results are opposite to those observed previously. The city that is less resilient to variation in the control input is NYC, which was previously reported to have received closer to optimal control interventions. This is due to the higher $\beta$ for NYC (see Table 1) and to the fact that the optimal $P^\ast(t)$ for NYC is higher than for the other cities (corresponding to less strict social distancing required); as a result, increasing $P(t)$  further leads to the poorest outcome.  This also highlights the risk for resurgences of the epidemics after partial suppression has been achieved \cite{ruktanonchai2020assessing}, which is currently seen in different parts of the world.

\begin{table}
	\centering
	\caption{Violation in the constraints $I_{\max}$ and $\epsilon$ for the non-optimal solutions.}
	\begin{tabular}{cc}
		\includegraphics[scale = 0.95]{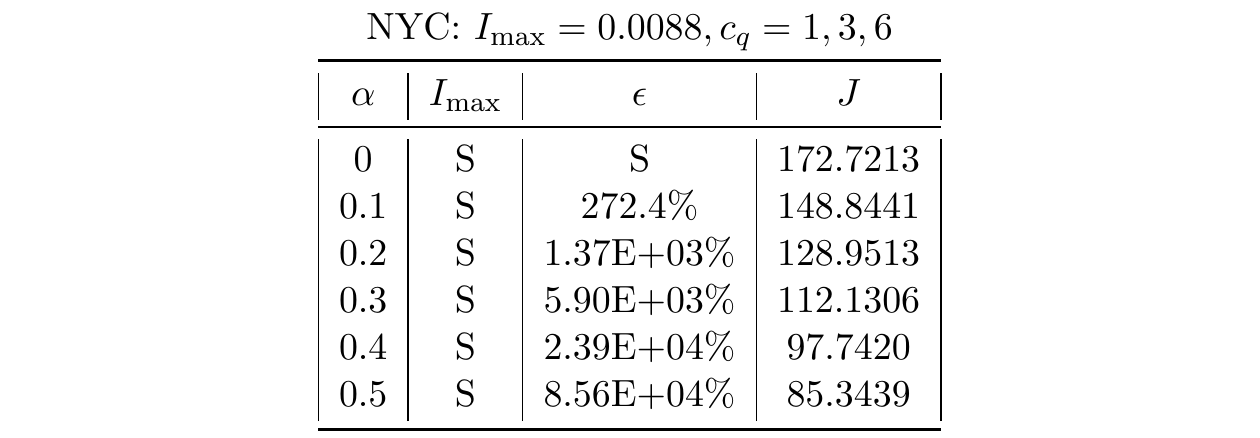}  & \hspace{-5 cm}\includegraphics[scale = 0.95]{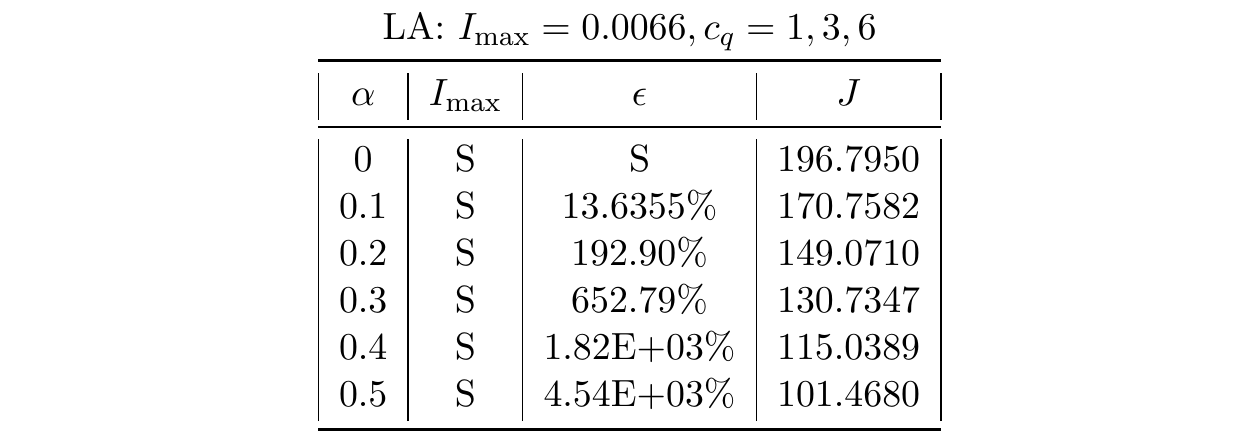} \\
		\includegraphics[scale = 0.95]{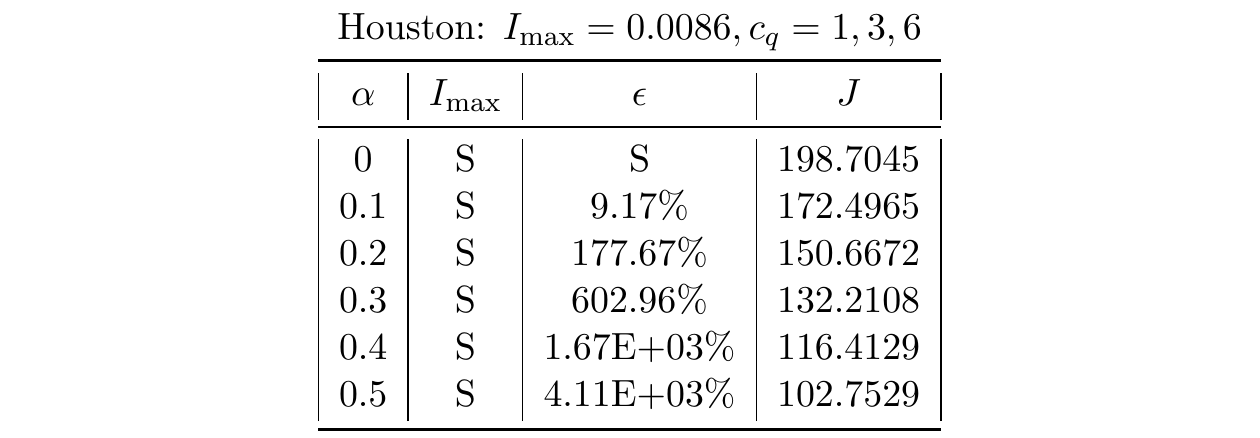}  & \hspace{-5 cm} \includegraphics[scale = 0.95]{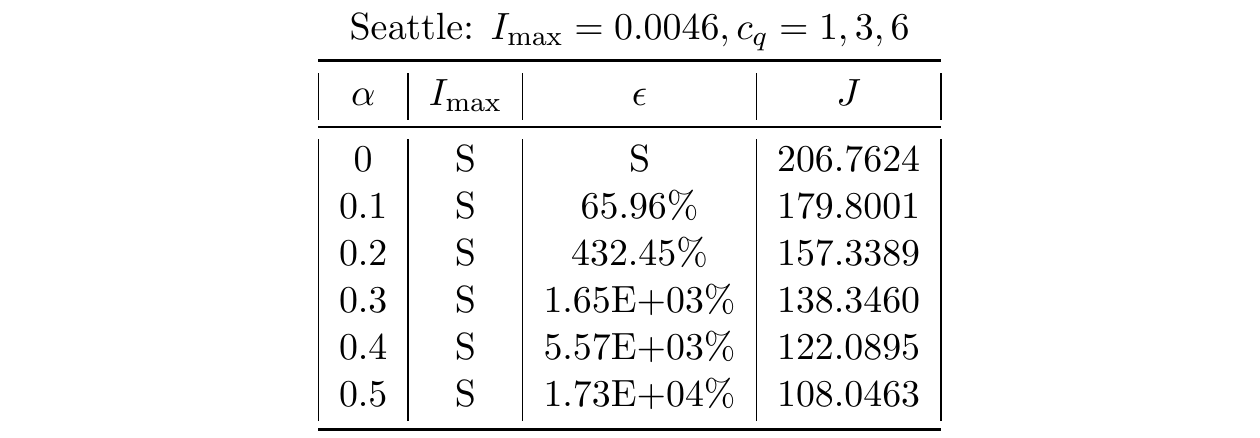}\\ 
	\end{tabular}
	\label{tab:non-optimal}
\end{table}

\section*{Supplementary Note 8: Pseudo-spectral Optimal Control}

Pseudo-Spectral Optimal Control (PSOC) is a computational method for solving optimal control problems.
PSOC \cite{rao2009survey,ross2012review} has provided a numerical tool to let scientists and engineers solve optimal control problems 
\begin{equation}\label{eq:4:OCP}
	\begin{aligned}
		\min_{\textbf{u}(t)} && &J(\textbf{x}(t),\textbf{u}(t),t) = E\left(\textbf{x}(t_i),\textbf{x}(t_f), t_i, t_f\right) + \int_{t_i}^{t_f} F\left( \textbf{x}(t), \textbf{u}(t), t\right) dt\\
		\text{s.t.} && &\dot{\textbf{x}}(t) = \textbf{f} ( \textbf{x}(t), \textbf{u}(t), t)\\
		&& &\textbf{e}^L \leq \textbf{e}(\textbf{x}(t_i), \textbf{x}(t_f), t_i, t_f) \leq \textbf{e}^U\\
		&& &\textbf{h}^L \leq \textbf{h}(\textbf{x}(t), \textbf{u}(t), t) \leq \textbf{h}^U\\
		&& &t \in [t_i,t_f]
	\end{aligned}
\end{equation}
reliably and efficiently in applications such as guiding autonomous vehicles and maneuvering the international space station \cite{ross2012review}.
PSOC is an approach by which an OCP can be discretized by approximating the integrals by quadratures and the time-varying states and control inputs with interpolating polynomials.
Here we summarize the main concepts behind the PSOC. 
We choose a set of $N+1$ discrete times $\{\tau_i\}$ $i = 0,1,\ldots,N$ where $\tau_0 = -1$ and $\tau_N = 1$ with a mapping between $t \in [t_i,t_f]$ and $\tau \in [-1,1]$. The discretization scheme includes the endpoints and is normalized by the mapping,
\begin{equation}
	t = \frac{t_f -t_i}{2} \tau + \frac{t_f+t_i}{2}
\end{equation}
The times $\{\tau_i\}$  are chosen as the roots of an $(N+1)$th order orthogonal polynomial such as Legendre polynomials or Chebyshev polynomials.
The choice of dicretization scheme is important to the convergence of the full discretized problem.
For instance, if we choose the roots of a Legendre polynomial as the discretization scheme, the associated quadrature weights can be found in the typical way for Gauss quadrature.
The time-varying states and control inputs are found by approximating them with Lagrange interpolating polynomials,
\begin{subequations}\label{eq:4:dxu}
	\begin{align}
		\hat{\textbf{x}}(\tau) & = \sum_{i=0}^N \hat{\textbf{x}}_i L_i(\tau)\\
		\hat{\textbf{u}}(\tau) & = \sum_{i=0}^N \hat{\textbf{u}}_i L_i(\tau),
	\end{align}
\end{subequations}
where $\hat{\textbf{x}}(\tau)$  and  $\hat{\textbf{u}}(\tau)$ are the approximations of $\textbf{x}(\tau)$ and $\textbf{u}(\tau)$, respectively, and $L_i(\tau)$ is the $i$th Lagrange interpolating polynomial. The Lagrange interpolating polynomials are defined as,
\begin{equation}
	L_i{\tau} = \prod_{j = 0, j \ne i}^{N} \frac{\tau - \tau_j}{\tau_i - \tau_j}
\end{equation}
The dynamical system is approximated by differentiating the approximation $\hat{\textbf{x}}(\tau) = \sum_{i=0}^N \hat{\textbf{x}}_i L_i(\tau)$ with respect to time.
\begin{equation}
	\begin{aligned}
		\frac{d \hat{\textbf{x}}}{d \tau} = \sum_{i=0}^N \hat{\textbf{x}}_i \frac{d L_i}{d\tau}
	\end{aligned}
\end{equation}
Let $D_{k,i} = \frac{d}{d\tau} L_i(\tau_k)$ which allows one to rewrite the original dynamical system constraints in \eqref{eq:4:OCP} as the following set of algebraic constraints.
\begin{equation}\label{eq:4:discstate}
	\begin{aligned}
		\sum_{i=0}^N D_{k,i} \hat{\textbf{x}}_i - \frac{t_f-t_i}{2} \textbf{f}(\hat{\textbf{x}}_k, \hat{\textbf{u}}_k, \tau_k) = \boldsymbol{0}_n, && k = 1,\ldots,N\\
		\hat{\textbf{x}}_N - \hat{\textbf{x}}_0 - \sum_{k=1}^N \sum_{i=0}^N w_k D_{k,i} \hat{\textbf{x}}_i = \boldsymbol{0}_n
	\end{aligned}
\end{equation}
The last set of algebraic constraints arise from the consistency condition $\int_{t_i}^{t_f} \dot{\textbf{x}}(t) dt = \textbf{x}(t_f) - \textbf{x}_0$.
Similarly to the consistency condition, the integral in the cost function is, 
\begin{equation}
	J = \int_{t_i}^{t_f} F(\textbf{x},\textbf{u},t) \approx \hat{J} = \frac{t_f-t_i}{2} \sum_{k=1}^N F(\hat{\textbf{x}}_k, \hat{\textbf{u}}_k, \tau_k)
\end{equation}
The original time-varying states, control inputs, the dynamical equations constrained and the cost function are now discretized approximation of the continuous NLP problem.
Thus the discretized approximation of the original OCP is compiled into the following nonlinear programming (NLP) problem.
\begin{equation}\label{eq:4:dOCP}
	\begin{aligned}
		\min_{\substack{\textbf{u}_i\\ i=0,\ldots,N}} && &\hat{J} = \frac{t_f-t_i}{2} \sum_{i=0}^N w_i f(\hat{\textbf{x}}_i,\hat{\textbf{u}}_i,\tau_i)\\
		\text{s.t.} && &\sum_{i=0}^N D_{k,i} \hat{\textbf{x}}_i - \frac{t_f-t_i}{2} \textbf{f}(\hat{\textbf{x}}_k,\hat{\textbf{u}}_k,\tau_k) = \boldsymbol{0},\ k = 0,\ldots,N\\
		&& &\hat{\textbf{x}}_N - \hat{\textbf{x}}_0 - \sum_{k=1}^N \sum_{i=0}^N w_k D_{k,i} \hat{\textbf{x}}_i = \boldsymbol{0}_n\\
		&& &\textbf{e}^L \leq \textbf{e}(\hat{\textbf{x}}_0,\hat{\textbf{x}}_N,\tau_0,\tau_N) \leq \textbf{e}^U\\
		&& &\textbf{h}^L \leq \textbf{h}(\hat{\textbf{x}}_k,\hat{\textbf{u}}_k, \tau_k) \leq \textbf{h}^U,\ k = 0,\ldots,N\\
		&& &t_i = \frac{t_f-t_i}{2}\tau_i + \frac{t_f+t_i}{2}
	\end{aligned}
\end{equation}
%


With the above results, we now present the application to the full multi-phase optimal control problem.
In general, let us assume there are $p>1$ phases where we set $p=2$ for simplicity.
Each phase is active within the interval $t \in [t_i^{(p)},t_f^{(p)}]$.
In each phase there is a cost function $J^{(p)}$, a dynamical system $\textbf{f}^{(p)}$, a set of endpoint constraints $\textbf{e}^{(p)}$, and a set of path constraints $\textbf{h}^{(p)}$.
If two phases, $p$ and $q$, are linked, then there also exists a set of linkage constraints $\Phi^{(p,q)}$.
\begin{equation}\label{eq:mpOCP}
	\begin{aligned}
		\min_{\textbf{u}^{(p)}} && &\sum_{p = 1}^P J^{(p)} = \sum_{p=1}^P \int_{t_i^{(p)}}^{t_f^{(p)}} F^{(p)}(\textbf{x}^{(p)},\textbf{u}^{(p)},t) d t\\
		\text{s.t.} && &\dot{\textbf{x}}^{(p)}(t) = \textbf{f}^{(p)}(\textbf{x}^{(p)}, \textbf{u}^{(p)}, t)\\
		&& &\textbf{h}^{L,(p)} \leq \textbf{h}^{(p)}(\textbf{x}^{(p)},\textbf{u}^{(p)},t) \leq \textbf{h}^{U,(p)}\\
		&& &\textbf{e}^{L,(p)} \leq \textbf{e}^{(p)}(\textbf{x}^{(p)}(t_i^{(p)}), \textbf{x}^{(p)}(t_f^{(p)}), t_i^{(p)}, t_f^{(p)}) \leq \textbf{e}^{U,(p)}\\
		&& &\Phi^{L,(p,q)} \leq \Phi^{(p,q)}(\textbf{x}^{(p)},\textbf{x}^{(q)},\textbf{u}^{(p)},\textbf{u}^{(q)}) \leq \Phi^{U,(p,q)}
	\end{aligned}
\end{equation}
Each phase is discretized with its own set of points, $\{\tau_i^{(p)}\}$ so that,
\begin{equation}
	\textbf{x}^{(p)}(\tau) \approx \hat{\textbf{x}}^{(p)}(\tau) = \sum_{i=1}^N \hat{\textbf{x}}_i^{(p)} L_i(\tau)
\end{equation}
so that the full multi-phase NLP is,
\begin{equation}\label{eq:dmpOCP}
	\begin{aligned}
		\min_{\textbf{u}_i^{(p)}} && &\sum_{p=1}^P \frac{t_f^{(p)} - t_i^{(p)}}{2} \sum_{k=1}^N F^{(p)}(\hat{\textbf{x}}_k^{(p)},\hat{\textbf{u}}_k^{(p)},\tau_k)\\
		\text{s.t.} && &\sum_{i=0}^N D_{k,i} \hat{\textbf{x}}_i^{(p)} - \frac{t_f^{(p)}- t_i^{(p)}}{2} \textbf{f}^{(p)} (\hat{\textbf{x}}^{(p)}_k, \hat{\textbf{u}}_k^{(p)},\tau_k) = \boldsymbol{0}_n, \quad p = 1,\ldots,P, \quad k = 1,\ldots,N\\
		&& &\hat{\textbf{x}}_N^{(p)} - \hat{\textbf{x}}_0^{(p)} - \frac{t_f^{(p)}-t_i^{(p)}}{2} \sum_{k=1}^N \sum_{i=0}^N w_k D_{k,i} \hat{\textbf{x}}_i = \boldsymbol{0}_n, \quad p = 1,\ldots,P\\
		&& &\textbf{e}^{L,(p)} \leq \textbf{e}^{(p)}(\hat{\textbf{x}}_0^{(p)},\hat{\textbf{x}}_N^{(p)},t_i^{(p)},t_f^{(p)}) \leq \textbf{e}^{U,(p)}, \quad p = 1,\ldots,P\\
		&& &\textbf{h}^{L,(p)} \leq \textbf{h}^{(p)} (\hat{\textbf{x}}^{(p)}_k, \hat{\textbf{u}}_k^{(p)}, \tau_k) \leq \textbf{h}^{U,(p)}, \quad k = 1, \ldots,N, \quad p = 1,\ldots P\\
		&& &\Phi^{L,(p,q)} \leq \Phi^{(p,q)}(\hat{\textbf{x}}_0^{(p)},\hat{\textbf{u}}^{(p)}_0, \hat{\textbf{x}}_N^{(q)}, \hat{\textbf{u}}_N^{(q)}) \leq \Phi^{U,(p,q)}, \quad p,q = 1,\ldots,P
	\end{aligned}
\end{equation}
To perform the discretization described in this subsection, we use the open-source C++ PSOC package $\mathcal{PSOPT}$ \cite{becerra2010solving}.

Next we show that Eq.\ \eqref{eq:dmpOCP} can be expressed in the typical NLP form \cite{nocedal2006numerical}.
Let $\textbf{z}^{(p)}$ contain all of the variables for phase $p$.
\begin{equation}
	\textbf{z}^{(p)} = \left[ \begin{array}{c}
		\hat{\textbf{x}}_0^{(p)} \\ \vdots \\ \hat{\textbf{x}}_N^{(p)} \\ \hat{\textbf{u}}_0^{(p)} \\ \vdots \\ \hat{\textbf{u}}_N^{(p)}
	\end{array} \right] \in \mathbb{R}^{(n+m)}
\end{equation}
Next, let $\textbf{z}$ contain the variables for every phase,
\begin{equation}
	\textbf{z} = \left[ \begin{array}{c}
		\textbf{z}^{(1)} \\ \vdots \\ \textbf{z}^{(P)}
	\end{array} \right] \in \mathbb{R}^{(N+1)(n+m)}
\end{equation}
With some algebraic manipulation, the entire discretized multi-phase OCP can be rewritten as an NLP in the typical form.
\begin{equation}\label{eq:4:NLPs}
	\begin{aligned}
		\min_{\textbf{z}} && &c(\textbf{z})\\
		\text{s.t.} && &\textbf{g}(\textbf{z}) = \boldsymbol{0}\\
		&& &\textbf{d}(\textbf{z}) \leq \boldsymbol{0}
	\end{aligned}
\end{equation}
To solve the large-scale NLP in Eq.\ \eqref{eq:4:NLPs} we employ an interior-point algorithm \cite{nocedal2006numerical}.
Specific details of the algorithm are outside the scope of this paper. We used the open-source C++ package IPOPT \cite{wachter2006implementation} to solve each instance of Eq.\ \eqref{eq:4:NLPs}.
We direct interested readers who would like to learn more about the technical detailed involved when solving Eq.\ \eqref{eq:4:NLPs} to the documentation provided with IPOPT.

The optimal solution returned, $\textbf{z}^\ast$, is separated into its component parts; first by splitting it into the phases $\textbf{z}^{(p)\ast}$, and second by reconstructing the discrete states and control inputs, $\hat{\textbf{x}}_i^\ast$ and $\hat{\textbf{u}}_i^\ast$.
The continuous time control inputs and states are then reconstructed using the Lagrange interpolating polynomials in Eq.\ \eqref{eq:4:dxu}.
With the continuous time states and control inputs, $\textbf{x}^\ast(t)$ and $\textbf{u}^\ast(t)$, we then verify that the necessary conditions are met to within an acceptable tolerance.

\section*{Supplementary Note 9: Necessary Conditions for PSOC Solutions}

{A closed form solution of Eq.\ \eqref{eq:4:OCP}  may not exist. Instead one  must typically turn to numerical methods, such as PSOC. Nonetheless, it is possible to derive a set of necessary conditions that any solution to Eq.\ \eqref{eq:4:OCP} must satisfy based on Pontryagin’s minimum principle \cite{kirk2004optimal}. Developing these necessary conditions allows one to construct a set of validation criteria with which we may test the quality of any solution returned by the numerical methods.}
In Ref.\  \cite{ross2015primer},  the so-called  HAMVET procedure has been proposed based on a slightly modified version of Pontryagin's principle to provide the necessary conditions for the general OCP in \eqref{eq:4:OCP}.

The HAMVET procedure is based on the following steps:

\begin{itemize}
	\item Construction of the Hamiltonian : (H)
	\item Adjoint equations : (A)
	\item Minimization of the Hamiltonian : (M)
	\item Evaluation of the Hamiltonian Value condition : (V)
	\item Evolution of the Hamiltonian : (E)
	\item Transversality conditions : (T)
	
\end{itemize} 

In what follows, we individually describe each one of the steps of the HAMVET procedure. A detailed analysis can be found in Ref.\ \cite{ross2015primer}.

\subsection*{Construction of the Hamiltonian}

The \textit{Hamiltonian} $H$ corresponding to the general OPC problem is

\begin{equation}\label{eq:4:H}
	H(\boldsymbol{\lambda},\textbf{x}, \textbf{u},t) = F(\textbf{x}, \textbf{u},t)+ \boldsymbol{\lambda}^T \textbf{f}(\textbf{x}, \textbf{u},t) 
\end{equation}
where $ \boldsymbol{\lambda}(t) \in \mathbb{R}^n$ is  the \textit{adjoint covector } which is a function of time $t$. The control input that minimizes the OCP satisfies the Hamiltonian Minimization Condition (HMC), that is,

\begin{equation}\label{eq:4:HMC}
	(HMC) \quad	\Bigg\{\begin{aligned}
		\min\limits_{u(t)}  &&  & H(\boldsymbol{\lambda},\textbf{x}, \textbf{u},t) \\
		\mbox{s.t.} &&  &\textbf{h}^L \le  \textbf{h}(\textbf{x},\textbf{u},t) \le \textbf{h}^U 
	\end{aligned}
\end{equation}

\subsection*{Adjoint equations}

The  Karush-Kuhn-Tucker (KKT) conditions can be used to solve the HMC. We define the \textit{Lagrangian of the Hamiltonian} $\bar{H}$ as

\begin{equation}\label{eq:4:LH}
	\bar{H}(\boldsymbol{\mu},\boldsymbol{\lambda},\textbf{x}, \textbf{u},t) =    H(\boldsymbol{\lambda},\textbf{x}, \textbf{u},t)  + \boldsymbol{\mu}^T \textbf{h}(\textbf{x},\textbf{u},t)
\end{equation}
where $ \boldsymbol{\mu}(t) \in \mathbb{R}^{h}$ is the \textit{path covector} which is a function of time $t$. Then the evolution of the adjoint covector $\boldsymbol{\lambda}(t)$ is given by,

\begin{equation}\label{eq:4:ADJ}
	- \dot{\boldsymbol{\lambda}} = \frac{\partial \bar{H}}{\partial \textbf{x}}
\end{equation}
Note that condition in \eqref{eq:4:ADJ} enforces the continuity but not differentiability of $\boldsymbol{\lambda}(t)$. So, the piecewise continuity of $\boldsymbol{\lambda}(t)$ is a necessary condition for an optimal control solution.
\\

\subsection*{Minimization of the Hamiltonian}

By the KKT condition, the minimization condition for the Hamiltonian yields 

\begin{equation}
	\frac{\partial \bar{H}}{\partial \textbf{u}} = \textbf{0}
\end{equation}
with the complementary conditions for path constraints,

\begin{equation}\label{eq:4:complementarypath1} 
	\left\{\begin{aligned}
		\mu_i & \le 0   &&  if &       h_i(\textbf{x},\textbf{u},t) = h_i^L \\[0.05in] 
		\mu_i & = 0   &&   if &    h_i^L  < h_i(\textbf{x},\textbf{u},t) < h_i^U  \\[0.05in] 
		\mu_i & \ge 0   &&   if &       h_i(\textbf{x},\textbf{u},t) = h_i^U  \\[0.05in]
		\mu_i & \mbox{ unrestricted } &&    if &       h_i^L = h_i^U  \\[0.05in]
	\end{aligned} \right.
\end{equation}
If there are path constraints, then one of the necessary conditions is

\begin{equation}\label{eq:4:complementarypath2}
	\mu_i(t)(h_i - h_i^L)(h_i - h_i^U) = 0
\end{equation}
\
\\
Along with the minimization of the Hamiltonian, there is an endpoint minimization condition (EMC) as well. The endpoint minimization problem is defined as
\begin{equation}\label{eq:4:EMC}
	(EMC) \quad	\Bigg\{\begin{aligned}
		\min  &&  & E(\textbf{x}(t_i),\textbf{x}(t_f),t_i,t_f) \\
		\mbox{s.t.} &&  &\textbf{e}^L \le  \textbf{e}(\textbf{x}(t_i),\textbf{x}(t_f),t_i,t_f) \le \textbf{e}^U 
	\end{aligned}
\end{equation}
To solve the EMC by KKT, we define the \textit{endpoint Lagrangian} $\bar{E}$ as 
\begin{equation}\label{eq:4:EL}
	\begin{aligned}
		\bar{E}(\boldsymbol{\nu},\textbf{x}(t_i),\textbf{x}(t_f),t_i,t_f) = & E(\textbf{x}(t_i),\textbf{x}(t_f),t_i,t_f)\\
		&+\boldsymbol{\nu}^T e(\textbf{x}(t_i),\textbf{x}(t_f),t_i,t_f)
	\end{aligned}
\end{equation}
where $\boldsymbol{\nu} \in \mathbb{R}^{e} $ is the \textit{endpoint covector}. Note that, $\boldsymbol{\nu}$ is a constant vector. The complementary conditions for event constraints are given by

\begin{equation}\label{eq:4:complementaryevent} 
	\left\{\begin{aligned}
		\nu_i & \le 0   &&  if &     e_i(\textbf{x}(t_i),\textbf{x}(t_f),t_i,t_f) = e_i^L  \\[0.05in] 
		\nu_i & = 0   &&  if &    e_i^L  < e_i(\textbf{x}(t_i),\textbf{x}(t_f),t_i,t_f) < e_i^U  \\[0.05in] 
		\nu_i & \ge 0   &&  if &       e_i(\textbf{x}(t_i),\textbf{x}(t_f),t_i,t_f) = e_i^U  \\[0.05in]
		\nu_i & \mbox{ unrestricted }   &&  if &       e_i^L = e_i^U  \\[0.05in]
	\end{aligned} \right.
\end{equation}
%

\subsection*{Hamiltonian Value condition}

The \textit{lower Hamiltonian} $\mathcal{H}$ is defined as the Hamiltonian evaluated at  $\textbf{u}(t)=\textbf{u}^\ast(t)$, the solution to the  HMC problem, i.e., 

\begin{equation}\label{eq:4:MH}
	\mathcal{H} = \min\limits_{\textbf{u}\in \mathbb{U}} H(\boldsymbol{\lambda},\textbf{x}, \textbf{u},t)
\end{equation}
where $\mathbb{U}$ is the set of feasible control inputs, i.e., they satisfy all of the constraints imposed by Eq.\ \eqref{eq:4:OCP}. The lower Hamiltonian  must satisfy the endpoint value conditions as a regular Hamiltonian

\begin{equation}
	\begin{aligned}
		\mathcal{H}(\boldsymbol{\lambda}(t_i),\textbf{x}(t_i),t_i) & =  \frac{\partial \bar{E}}{\partial t_i} \\
		\mathcal{H}(\boldsymbol{\lambda}(t_f),\textbf{x}(t_f),t_f) & = - \frac{\partial \bar{E}}{\partial t_f}
	\end{aligned}
\end{equation}
which provides another necessary conditions to check for the optimal control solution.

\subsection*{Time Evolution of the Hamiltonian}

As the lower Hamiltonian   $\mathcal{H}$ 
is obtained from the evaluation of the Hamiltonian at the $\textbf{u}^\ast(t)$, $\textbf{x}^\ast(t)$ and $\boldsymbol{\lambda}^\ast(t)$, 
where  $\textbf{x}^\ast(t)$ and $\boldsymbol{\lambda}^\ast(t)$ are the states and costates associated with the optimal control solution  $\textbf{u}^\ast(t)$,  $\mathcal{H}$ is a function  of time $t$ only. Thus the evolution of the lower Hamiltonian $\mathcal{H}$ can be defined as

\begin{equation}
	\dot{\mathcal{H}} = \frac{d\mathcal{H}}{dt} = \frac{\partial H}{\partial t}
\end{equation}
If $H$ in \eqref{eq:4:H} does not depend explicitly on time, then another necessary condition is

\begin{equation}\label{eq:4:EH}
	\dot{\mathcal{H}} = 0 \quad  \mbox{ or} \quad \mathcal{H} = \mbox{constant}
\end{equation}

\subsection*{Transversality conditions}
The endpoints of the adjoint covector $\boldsymbol{\lambda}(t)$ are related to the partial derivatives of the endpoint Lagrangian $\bar{E}$. The transversality conditions for the  adjoint covector $\boldsymbol{\lambda}(t)$ are
\begin{equation}\label{eq:4:trans}
	\boldsymbol{\lambda}(t_0) = - \frac{\partial \bar{E}}{\partial \textbf{x}(t_0)} \quad \mbox{and} \quad 
	\boldsymbol{\lambda}(t_f) =  \frac{\partial \bar{E}}{\partial \textbf{x}_f}
\end{equation}

\begin{figure}
	\centering
	\includegraphics[width= 0.75\textwidth ]{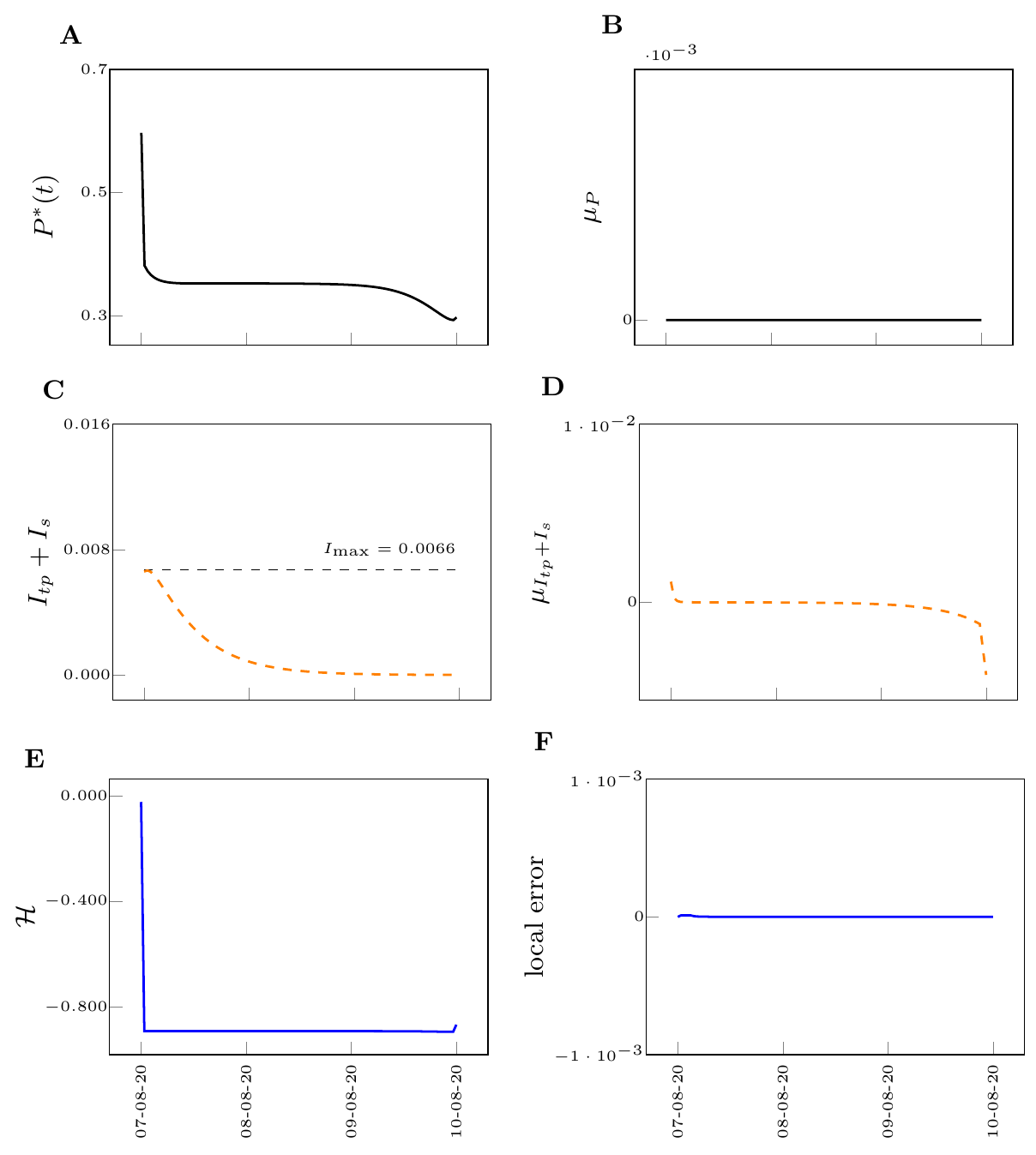}
	\caption{ 
		HAMVET procedure applied to our solution of the optimal control problem for the metropolitan city of LA. $I_{\max}$ and $\epsilon$ are set to 0.0066 and $10^{-5}$, respectively.
		A) The optimal time evolution of the control input $P^*(t)$. 
		B) The time evolution of the path covector $\mu_P$.
		C) The time evolution of $I_{tp}(t)+I_s(t)$.
		D) The time evolution of the path covector $\mu_{I_{tp}+I_s}(t)$.
		E) The time evolution of the lower Hamiltonian $\mathcal{H}$.
		F) The relative local discretization error at each time $t$.}
	\label{fig:VV}
\end{figure}

{
	\subsubsection*{Verification of the Necessary Conditions}
	We have presented the necessary conditions for the general OCP. Next we develop a set of necessary conditions for the solution of the OCP described in the main section. The OCP presented in the main text can be mapped to the general formulation presented in Eq.\ \eqref{eq:4:OCP} with the following definitions.
	\begin{itemize}
		\item The state variables $\textbf{x}(t) = \left[\begin{array}{ccccccc} S(t) & E(t) & A(t) & I_{tp}(t) & I_s(t) & Q(t) & R(t)\end{array} \right]^T \in \mathbb{R}^{7}$ and the control input $u(t)\equiv P(t) \in \mathbb{R}$.
		\item The cost function $J =   \int_{t_i}^{t_f} (1- P(t)) dt +   \int_{t_i}^{t_f} {Q(t)} dt$ (see Eq.\ (7) with $c_p = c_q =1$ in the main text) so that, from Eq.\ \eqref{eq:4:OCP}, $E \equiv 0$ and $F = 1- P(t)+Q(t)$. 
		\item The system dynamics, as presented in Eq.\ (1), are rewritten here,
		\begin{equation}
			\begin{aligned}
				\dot{\textbf{x}}(t) &= \left[ \begin{array}{c}
					\dot{S}(t)\\
					\dot{E}(t)\\
					\dot{A}(t)\\
					\dot{I}_{tp}(t)\\
					\dot{I}_s(t)\\
					\dot{Q}(t)\\
					\dot{R}(t)
				\end{array} \right] = \textbf{f}(\textbf{x}(t),\textbf{u}(t),t) \\
			\end{aligned}
		\end{equation}
		\item The only endpoint constraints are set at the initial time ($t_i$) and the terminal constraints in Eq.\ (4) in main text,
		\begin{equation}
			\begin{aligned}
				\textbf{e}(\textbf{x}(t_i),\textbf{x}(t_f),t_i,t_f) = \left[ \begin{array}{c}
					S(t_i) \\ E(t_i) \\ A(t_i) \\ I_{tp}(t_i) \\ I_s(t_i) \\  Q(t_i) \\ R(t_i) \\ P(t_i)\\ E(t_f)+I_s(t_f)+I_{tp}(t_f)+A(t_f)
				\end{array} \right], && \textbf{e}^L =  \left[ \begin{array}{c}
					\textbf{x}(t_i) \\ P_2 \\ 0
				\end{array} \right], && \textbf{e}^U = \left[ \begin{array}{c}
					\textbf{x}(t_i) \\ P_2 \\ \epsilon
				\end{array} \right]
			\end{aligned}  
		\end{equation}
		where  $ \textbf{x}(t_i) = \left[\begin{array}{ccccccc} S_{t_i} & E_{t_i} & A_{t_i} & I_{tp_{t_i}} & I_{s_{t_i}} &  Q_{t_i} & R_{t_i} \end{array} \right]^T$ are set to the values of the states at $t_i$.
		\item Finally, the path constraints consist of the bounds on $ I_{tp}(t)+ I_s(t)$ and possibly on $P(t)$
		\begin{equation}\label{eq:h}
			\begin{aligned}
				\textbf{h}(\textbf{x}(t),\textbf{u}(t),t) = \left[ \begin{array}{c}
					I_{tp}(t)+ I_s(t) \\ P(t) 
				\end{array} \right], && \textbf{h}^L = \left[ \begin{array}{c}
					0 \\ 0  
				\end{array} \right], &&
				\textbf{h}^U = \left[ \begin{array}{c}
					I_{\max} \\ 1 
				\end{array} \right]
			\end{aligned}
		\end{equation}
	\end{itemize}
	Developing the necessary conditions allows us to construct a set of validation criteria with which we may test the quality of any solution returned by our numerical methods. For this analysis, we consider the metropolitan city of LA. $I_{\max}$ and $\epsilon$ are set to 0.0066 and $10^{-5}$, respectively. 
	\\
	Define a vector of time-varying costates (or adjoint variables) as $\boldsymbol{\lambda}(t)$, 
	the Hamiltonian of the OCP in Eq.\ \eqref{eq:4:OCP} is defined as,
	\begin{equation}\label{eq:H}
		\begin{aligned}
			H(\boldsymbol{\lambda}, \textbf{x}, \textbf{u}, t) &= F(\textbf{x}, \textbf{u}, t) + \boldsymbol{\lambda}^T \textbf{f}(\textbf{x},\textbf{u},t)\\
			&= 1- P(t)+Q(t) +\boldsymbol{\lambda}^T \textbf{f}(\textbf{x},\textbf{u},t) 
		\end{aligned}
	\end{equation}
	where $\boldsymbol{\lambda}(t) \in \mathbb{R}^7$ are the costates (or adjoint variables).
	A solution to Eq.\ \eqref{eq:4:OCP} must also be a solution of the following minimization problem.
	\begin{equation}\label{eq:HMC}
		\begin{aligned}
			\min_{\textbf{u}(t)} && &H(\boldsymbol{\lambda},\textbf{x},\textbf{u},t)\\
			\text{s.t.} && &\textbf{h}^L \leq \textbf{h}(\textbf{x},\textbf{u},t) \leq \textbf{h}^U
		\end{aligned}
	\end{equation}
	To solve Eq.\ \eqref{eq:HMC}, we define the associated Lagrangian,
	\begin{equation}\label{eq:Hbar}
		\begin{aligned}
			\bar{H} (\boldsymbol{\mu},\boldsymbol{\lambda}, \textbf{x}, \textbf{u}, t) &= H(\boldsymbol{\lambda}, \textbf{x}, \textbf{u}, t) + \boldsymbol{\mu}^T \textbf{h}(\textbf{x},\textbf{u},t)\\
			&= 1- P(t)+Q(t) +\boldsymbol{\lambda}^T \textbf{f}(\textbf{x},\textbf{u},t) +  \mu_{I_{tp}+I_s} (I_{tp}+I_s) + \mu_P P(t)
		\end{aligned}
	\end{equation}
	where $\boldsymbol{\mu} = \left[ \begin{array}{cc} \mu_{I_{tp}+I_s} & \mu_P \end{array} \right]^T \in \mathbb{R}^h$ is the copath vector of the path constraints in Eq.\ \eqref{eq:h}.
	A solution to Eq.\ \eqref{eq:HMC}, and thus to our original OCP, must satisfy,
	\begin{equation}\label{eq:dHdu}
		\begin{aligned}
			\frac{\partial \bar{H}}{\partial P} =  \mu_P-1+ \left(\frac{\partial \textbf{f}}{\partial P}\right)^T\boldsymbol{\lambda}=0
		\end{aligned}
	\end{equation}
	where the costates evolve according to the dynamical equation,
	\begin{equation}
		\dot{\boldsymbol{\lambda}} = -\frac{\partial \bar{H}}{\partial \textbf{x}} = - \left( \frac{\partial \bar{\textbf{f}}}{\partial \textbf{x}} \right)^T \boldsymbol{\lambda} - \left[ \begin{array}{c}
			\boldsymbol{0}_3 \\ \mu_{I_{tp}+I_s} \\ \mu_{I_{tp}+I_s} \\ 1\\0
		\end{array} \right]
	\end{equation}
	The optimal control input $P^*(t)$, must satisfy the complementarity condition 
	\begin{equation}\label{eq:complement}
		\left\{ \begin{aligned}
			\mu_P(t) < 0  && \text{if} &&  P^*(t) = 0\\
			\mu_P(t) = 0 && \text{if} && 0<P^*(t) <1  \\
			\mu_P(t) > 0  && \text{if} && P^*(t) = 1
		\end{aligned} \right.
	\end{equation}
	\
	\\
	The path constraint $I_{tp}(t)+I_s(t)$ must satisfy the another complementary condition
	\begin{equation}
		\left\{ \begin{aligned}
			\mu_{I_{tp}+I_s} < 0  && \text{if} &&  I_{tp}(t)+I_s(t) = 0\\
			\mu_{I_{tp}+I_s} = 0 && \text{if} && 0<I_{tp}(t)+I_s(t) < I_{\max}  \\
			\mu_{I_{tp}+I_s} > 0  && \text{if} && I_{tp}(t)+I_s(t) = I_{\max}
		\end{aligned} \right.
	\end{equation}
	In Fig.\ \ref{fig:VV}(A) and (B), we plot $P^*(t)$ and $\mu_P(t)$, respectively. We note that $0<P^*(t) <1$ and $\mu_P(t)$ remains at zero over the the control time interval. In Fig.\ \ref{fig:VV}(C) and  (D), we plot $I_{tp}(t)+I_s(t)$ and $\mu_{I_{tp}+I_s}(t)$, respectively. Note that $\mu_{I_{tp}+I_s}(t) > 0$ when $I_{tp}(t)+I_s(t) = I_{\max}$, $\mu_{I_{tp}+I_s}(t) = 0$ when $0<I_{tp}(t)+I_s(t) < I_{\max}$, and when $\mu_{I_{tp}+I_s}(t) < 0$ when $I_{tp}(t)+I_s(t) \approx 0$. These necessary conditions are obtained from the complementary conditions while minimizing the Hamiltonian. 
	\
	\\
	Similarly, in order to minimize the endpoints constraints, the endpoint Lagrangian $\bar{E}$ in Eq.\ \eqref{eq:4:EL} becomes,
	\begin{equation}
		\bar{E} = \boldsymbol{\nu}^T \textbf{e}(\textbf{x}(t_0),\textbf{x}(t_f),t_0,t_f)
	\end{equation}
	\
	\\
	The endpoint covector must satisfy the conditions in Eq.\ \eqref{eq:4:complementaryevent}, but it is reasonable to verify at least $\nu_9$ associated with the terminal constraint $E_{tf}+I_s(t_f)+I_{tp}(t_f)+A(t_f)$. We obtained $\nu_9<0$ suggesting that $E_{tf}+I_s(t_f)+I_{tp}(t_f)+A(t_f) \approx 0$.
	\\
	Let us now assume that we have solved Eq.\ \eqref{eq:HMC}, that is,
	\begin{equation}
		\begin{aligned}
			\mathcal{H}(t) = \min_{P(t)} H(\boldsymbol{\lambda},\textbf{x},P^*,t)
		\end{aligned}
	\end{equation}
	The evolution of the Hamiltonian at the optimal solution can be written,
	\begin{equation}
		\frac{d \mathcal{H}}{d t} = \frac{\partial H}{\partial t}
	\end{equation}
	where, since in our OCP, $H$ does not explicitly depend on time, we expect that $d\mathcal{H} / d t = 0$ and so $\mathcal{H}$ should be constant. This is the next validation condition. 
	In Fig.\ \ref{fig:VV}(E), we plot  $\mathcal{H}(t)$ and it is clear that $\mathcal{H}(t)$ is constant over $t_i<t<t_f$.  The transversality conditions in Eq.\ \eqref{eq:4:trans} are also verified from the numerical solution of $\boldsymbol{\lambda}(t)$. In Fig.\ \ref{fig:VV}(F), we plot the relative discretization error at each time $t$. From Fig.\ \ref{fig:VV}, we conclude that the necessary condition derived for the optimal control input are satisfied.}

{
	\section*{Supplementary Note 10: Sensitivity Analysis}
	It is important to check the sensitivity of the optimal control solution to small perturbations of the system parameters. To this aim, we perturb the parameter $\beta$, i.e., the most important parameter of the model, and compute how much the optimal control input changes. The other parameters remain unchanged.  We define $\xi$ as the average normalized variation in the optimal control input when $\beta$ is varied by a quantity equal to $ \delta\%$,
	\begin{equation}
		\xi = \left\langle\frac{\vline P^{\ast}(t)-P^{\ast}_{ \delta}(t)\vline}{P^{\ast}(t)}\right\rangle
	\end{equation}
	where $\left\langle.\right\rangle$ indicates a time average over the control time, $P^{\ast}(t)$ is the optimal input for the original value of $\beta$ and $P^{\ast}_{\delta}(t)$ is the optimal control input for $(1 + \delta\%) \beta$. For this analysis, we consider the metropolitan city of Seattle. $I_{\max}$ and $\epsilon$ are set to 0.0046 and $10^{-5}$, respectively. In Fig.\ \ref{fig:sensitivity}, we plot $\xi$ vs. $\delta\%$.
	\begin{figure}
		\centering
		\includegraphics[width= 0.6\textwidth ]{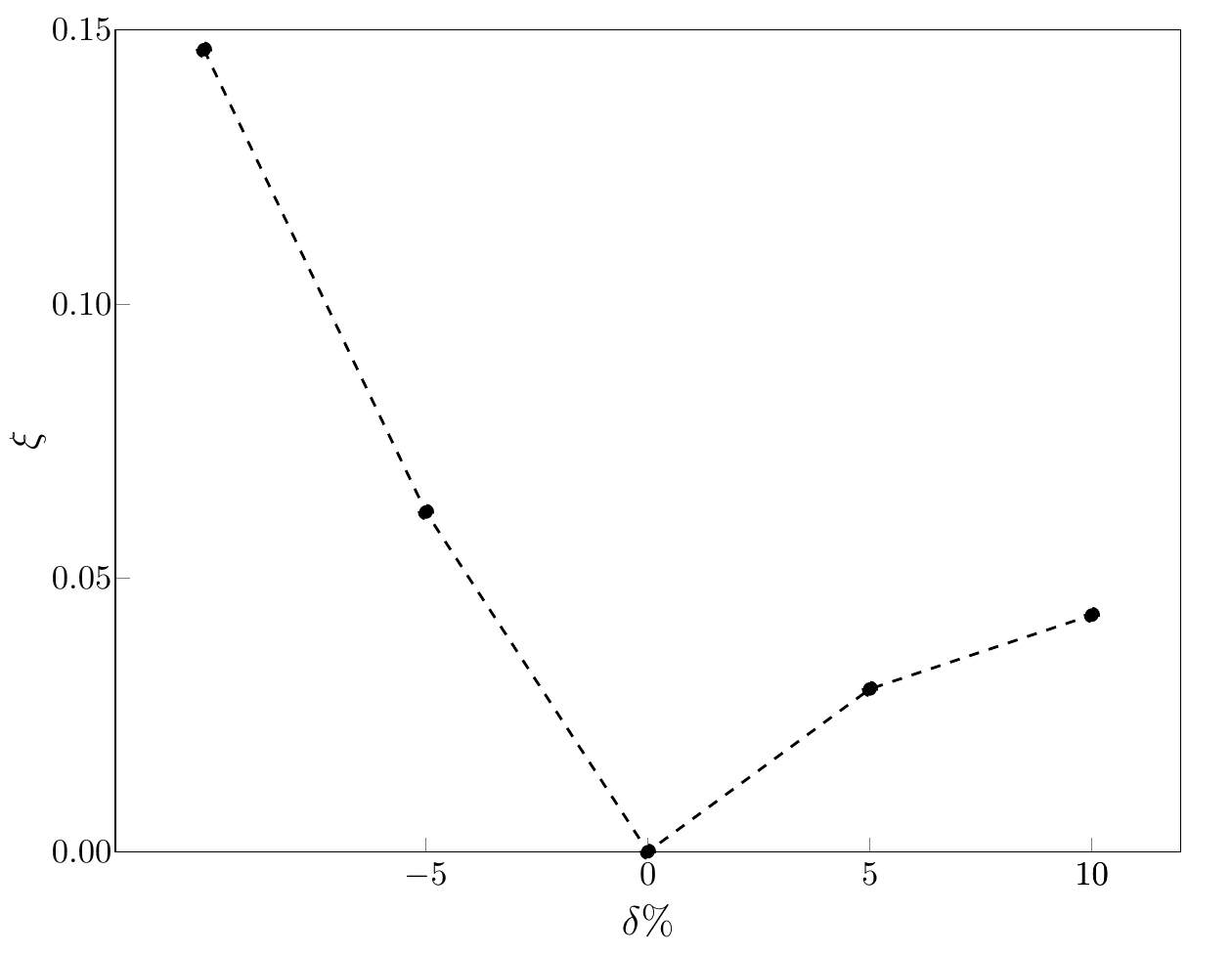}
		\caption{
			Sensitivity analysis of the optimal control input to a perturbation in the parameter $\beta$ for the metropolitan city of Seattle. The average variation $\xi$ in the optimal control input  vs. the change $\delta$ in the parameter $\beta$. $I_{\max}$ and $\epsilon$ are set to 0.0046 and $10^{-5}$, respectively.}
		\label{fig:sensitivity}
	\end{figure}
}

\section*{Supplementary Note 11: Incorporation of Vaccinations in the Model and Effects on the Optimal Control Solutions}

Figure 25 shows very good
agreement between the daily new case counts reported by local health administrations for the city of Seattle over the extended period from January 21, 2020 to December 14, 2020 (blue symbols) and the daily
new cases obtained by integrating Eq. (2) of the main text after parametrization of a 4-phase model (four social-distancing episodes with piece wise-linear transition of $P(t)$; red solid line). 

\begin{figure}[ht]
	\centering
	\includegraphics[width = 1\textwidth]{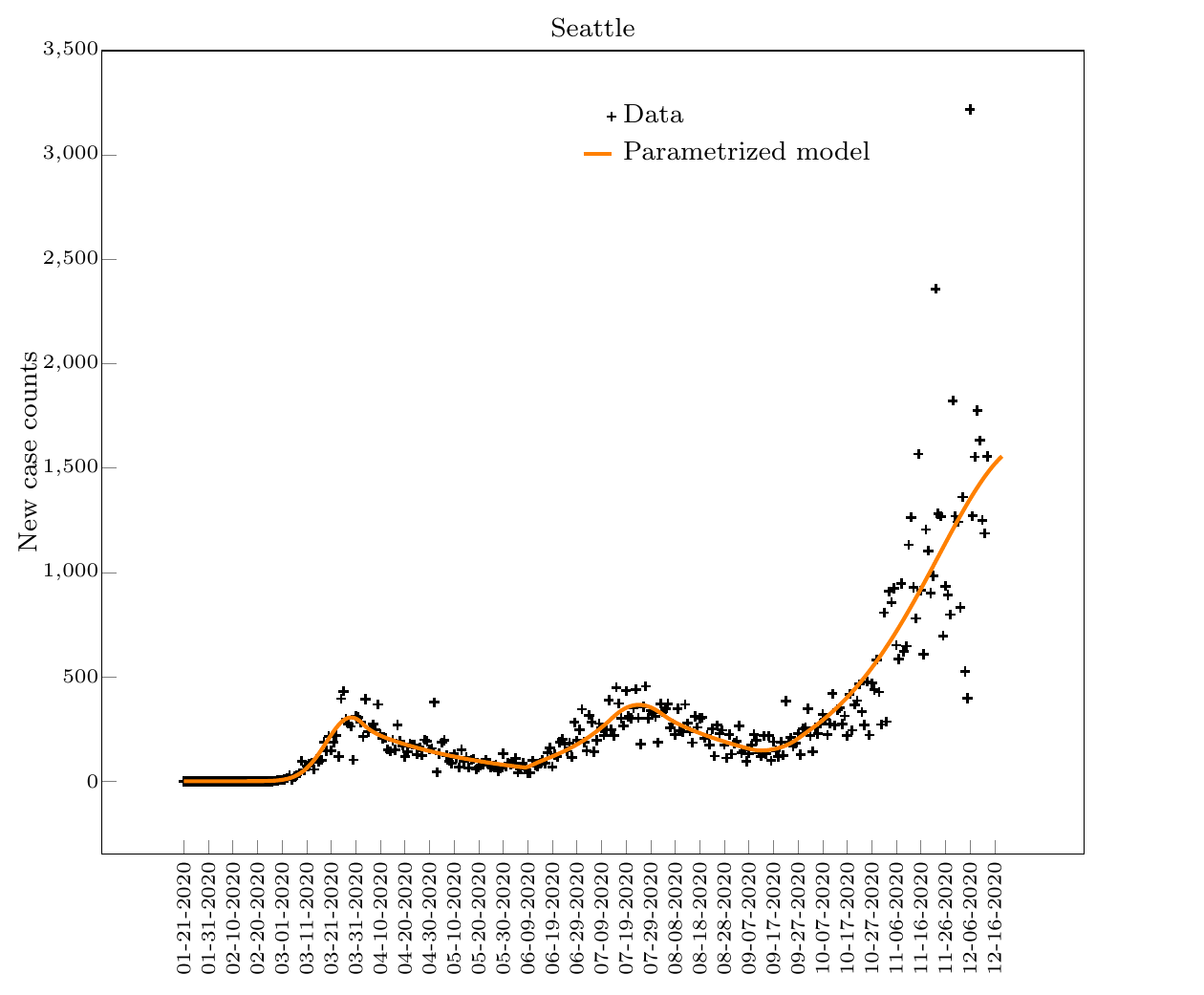}
	\caption{{New case counts from January 21, 2020 to December 14, 2020 for the Metropolitan Statistical Areas of Seattle.} Blue symbols are daily new case counts reported by the local health administration. The solid line is the daily new cases obtained by integrating Eq.\ (2) after parametrization of the model.}
	\label{fig:newcase}
\end{figure}

According to \cite{DailyVaccines} a total of $1.36$ million doses have been administered in the US as of February 1 2021, with the number of doses per day increasing roughly linearly with time. Thus we model the number of doses administered per day as $\kappa \times (t-t_v)$, where we set the time $t_v$ to be Dec-14-2020, the date in which vaccinations began in the US.
It becomes important to estimate the linear rate of increase 
$\kappa$.  There is a total of $49$ days from Dec-14-2020 to Feb-1-2021.
Assuming each person gets $2$ vaccine doses, we have 
$\kappa \times 49 = 1.36M/2$ leading to $\kappa=13878$ people/day nationally.
The MSA of Seattle has population of roughly $4M$, which is about $1.29 \%$ of the total US population.  
This means the $\kappa$ for Seattle should be equal to roughly $13878 \times 0.0129 = 180$ (person/day.)
In the non-dimensional case, this corresponds to  $\kappa=0.000045$ (1/day.)

Figure 26 shows the state time evolutions corresponding to the optimal control inputs shown in Figure 7 of the main text. The left plots are for a shorter control horizon, $T_{cont}=(t_f-t_i)=130$ days, while the right plots are for a longer control horizon, $T_{cont}=(t_f-t_i)=150$ days.  We compare the two cases that $\kappa=0$, i.e., we do not model the effects of vaccinations (in A and B), and that $\kappa=0.000045$, i.e., we model the effects of vaccinations (in C and D.)

\begin{figure}
	\centering
	\includegraphics[width=0.9\textwidth]{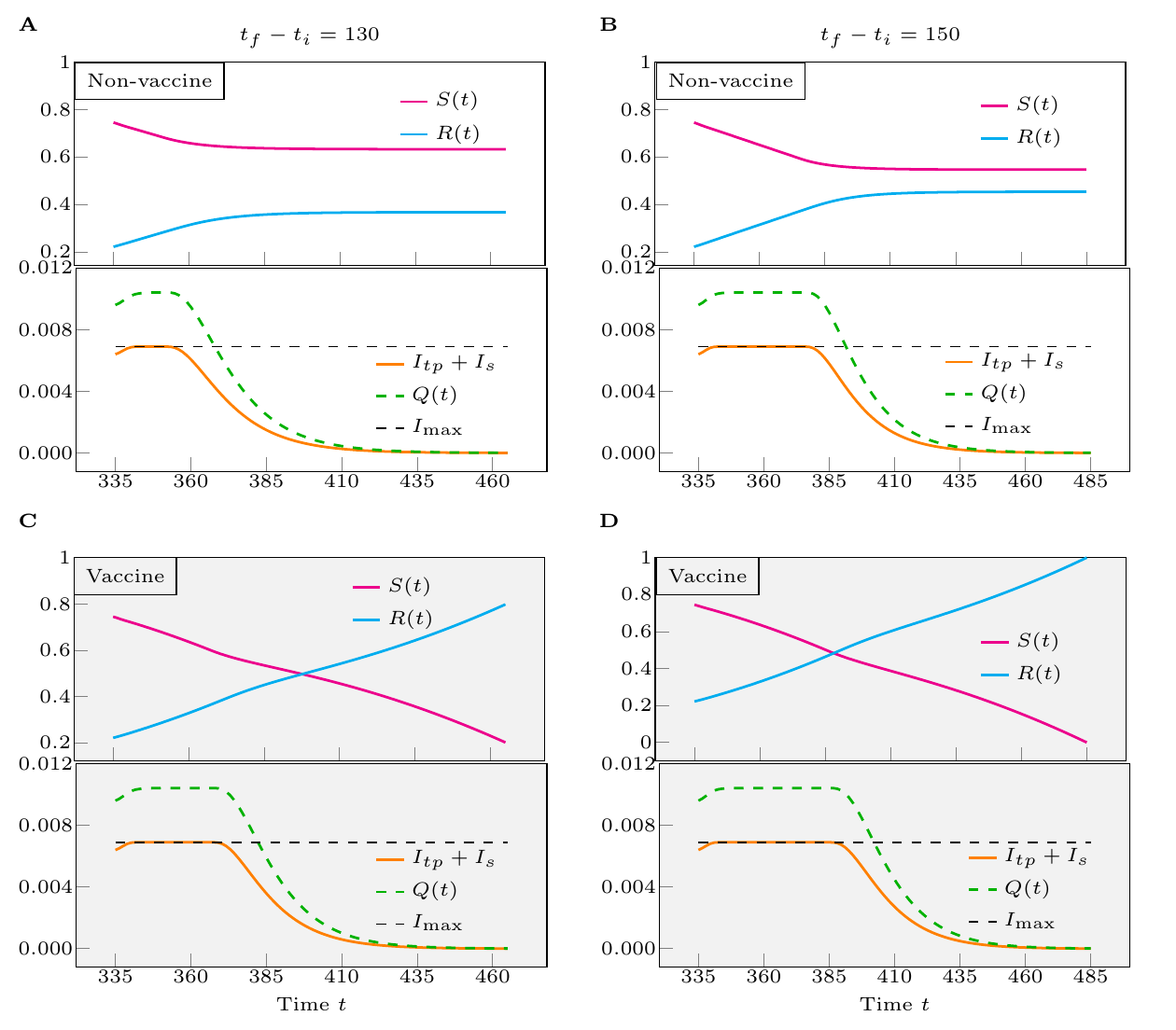}
	\caption{Time evolution of the states subject to the optimal control inputs shown in Fig.\ 7 of the main text for the city of Seattle. A and B are for the case that vaccinations are not modeled ($\kappa=0$), while C and D are for the case that vaccinations are modeled ($\kappa=0.000045$). Plots on the left are for a shorter control horizon $(t_f-t_i)=130$ days while the plots on the right are for a longer control horizon $(t_f-t_i)=150$ days. $I_{\max}$ are chosen as the maximum of the range in Table 3 ($\rho=1$) and $c_p=c_q = 1$.   }
	\label{fig:states_controls_Imax_mid}
\end{figure}

\color{black}

\section*{Supplementary Note 12: Non-Normalized Equations}

Consider the non-normalized quantities in the variables $\hat{S}(t)=N S(t)$, $\hat{E}(t)=N E(t)$, $\hat{A}(t)=N A(t)$, $\hat{I}(t)=N I(t)$, $\hat{I}_{tp}(t)=N I_{tp}(t)$, $\hat{Q}(t)=N Q(t)$, $\hat{R}(t)=N R(t)$, where $N$ is the total number of individuals in the population, such as $\hat{S}(t)+\hat{E}(t)+\hat{A}(t)+\hat{I}(t)++\hat{I}_{tp}(t)+\hat{Q}(t)+\hat{R}(t)=N$ at any time. The non-normalized quantities evolve based to  following equations,

\begin{subequations}
	\begin{align}
		\dot{\hat S}\l(t\r)= &-\hat \beta P^2\l(t\r) {\hat S}(t)\l[{\hat I}(t)+{\hat I}_{tp}(t) +\mu {\hat A}(t)\r]\\
		\dot{\hat E}\l(t\r)= &\hat \beta P^2\l(t\r) {\hat S}(t) \l[{\hat I}(t)+ \hat I_{sq}(t)+ \hat I_{tp}(t) +\mu \hat  A(t)\r]-\lambda \hat  E(t)\\
		\dot{\hat A}\l(t\r)= &\lambda \l(1-\sigma\r) \hat  E(t)- \gamma_A \hat  A(t) \\
		\dot{\hat I}_{tp}\l(t\r)= & p_{test} \lambda \sigma \hat E(t)  -\l[\gamma_I + \gamma_{tp} \r] \hat  I_{tp}(t) \\
		\dot{\hat  I}\l(t\r)= & \left(1-p_{sq} - p_{test} \right) \lambda \sigma \hat  E(t)- \gamma_I \hat I(t) \\
		\dot{\hat Q}\l(t\r)=&   \gamma_{tp} \hat  I_{tp}(t) + p_{sq} \lambda \sigma \hat  E(t)  -\gamma_I \hat  Q(t) \\
		\dot{\hat  R}\l(t\r)= &  \gamma_A \hat A(t) + \gamma_I \l[\hat I(t)+ \hat I_{tp}(t)+ \hat Q(t)\r].
	\end{align}
\end{subequations}
{where $\hat {\beta}=N^{-1} \beta$}. 
All other parameters are the same as defined in Eq.\ 2 of the main manuscript.

\section*{Supplementary Note 13: Model incorporating limited testing}

An extended version of the model that incorporates limited testing is the following.

\begin{subequations}
	\begin{align}
		\dot{S}\l(t\r)= &-\beta P^2\l(t\r) S(t)\l[I(t)+I_{tp}(t) +\mu A(t)\r]\\
		\dot{E}\l(t\r)= &\beta P^2\l(t\r) S(t) \l[I(t)+I_{tp}(t) +\mu A(t)\r]-\lambda E(t)\\
		\dot{A}\l(t\r)= &\lambda \l(1-\sigma\r) E(t)- \gamma_A  A(t) \\
		\dot{I}_{sq}\l(t\r)= & p_{sq} \lambda \sigma E(t)-\l[\gamma_I + \gamma_{sq} \r] I_{sq}(t) \\
		\dot{I}_{tp}\l(t\r)= & T  \ppos \wedge p_{test} \sigma \lambda E -\l(\gamma_I + \gamma_{tp} \r)I_{tp}(t) \\
		\dot{I}\l(t\r)= & \left(1-p_{sq} \right) \lambda \sigma E(t) - T \ppos \wedge \sigma p_{test} \lambda E - \gamma_I I(t) \\
		\dot{Q}\l(t\r)=&   \gamma_{tp} I_{tp}(t) + p_{sq} \lambda \sigma E -\gamma_I Q(t) \\
		\dot{R}\l(t\r)= & \gamma_A A(t) + \gamma_I \l[I(t)+I_{tp}(t)+Q(t)\r].
	\end{align}
\end{subequations}
\newpage
\pagebreak
\noindent

The flux of population that will be tested is  $p_{test} \sigma \lambda E$, and the positive detection probability is 
\begin{equation}
	\frac{p_{test} \sigma \lambda E}{p_{test} \sigma \lambda E+D}.
\end{equation}
This is to assume that the number of testing kits is less than the total people needed to be tested (scarcity) which can be described as: number of people which require testing $ \approx \l( p_{test} \sigma \lambda E+D \r) \Delta t$ in a small $\Delta t \ll 1$ time. Otherwise, we will assume a probability of 1 of detecting all those with COVID-19 while ignoring the false negatives. The variable $T$ is the flux of testing kits which are generated (so unit is number per day). We use the standard notation $a\wedge b = \min\{a,b \}.$ The testing flux $T$ is most-likely lower than $p_{test}\lambda E$, but if we take a temporally averaged value of $T$, it is likely that we can get everyone a kit at the early stage. The above general model considers the possibility that testing kits are scarce (i.e., $T< \l(p_{test} \sigma \lambda E + D\r) $).
However, for NY, at this moment (subdued pandemic), it is likely that the resource is not scarce. 

	\newcommand{\noop}[1]{}

\end{document}